\documentclass{iopart}
\usepackage{amstext}
\usepackage{iopams} 
\usepackage{fancyhdr}
\usepackage{bbm}
\usepackage[a4paper]{geometry}
 \usepackage{fullpage}
\usepackage{placeins}
\usepackage[square,sort&compress,comma,numbers]{natbib}

\usepackage{color}
\usepackage{graphicx}
\usepackage{hyperref}
\newcommand{\eq}{\begin{eqnarray}} 
\newcommand{\en}{\end{eqnarray}}
\def\bra#1{\mathinner{\langle{#1}|}}
\def\ket#1{\mathinner{|{#1}\rangle}}
\newcommand{\braket}[2]{\langle #1|#2\rangle}
\newcommand{\nicefrac}[2]{{#1}/{#2}}
\usepackage{hyperref}

\usepackage{microtype}

\begin{document}

\article{}{Tutorial}

\title{Interference of Identical Particles \\ 
from Entanglement to Boson-Sampling}

\author{Malte C. Tichy}
\address{Department of Physics and Astronomy, University of Aarhus, DK--8000 Aarhus C, Denmark}
\address{Physikalisches Institut, Albert-Ludwigs-Universit\"at Freiburg, D-79104 Freiburg, Germany}
\ead{tichy@phys.au.dk}
\begin{abstract}
Progress in the reliable preparation, coherent propagation and efficient detection of many-body states has recently brought collective quantum phenomena of many identical particles 
 into the spotlight. This tutorial introduces the physics of many-boson and many-fermion interference required for the description of current experiments and for the understanding of novel approaches to quantum computing. 
 
The field is motivated via the two-particle case, for which the uncorrelated, classical dynamics of distinguishable particles is compared to the quantum behaviour of identical bosons and fermions. Bunching of bosons is opposed to anti-bunching of fermions, while both species constitute  equivalent sources of bipartite two-level entanglement. The realms of indistinguishable and distinguishable particles are connected by a monotonic transition, on a scale defined by the coherence length of the interfering particles. 

As we move to larger systems, any attempt to understand many particles via the two-particle paradigm fails: In contrast to two-particle bunching and anti-bunching, the very same signatures can be exhibited by bosons and fermions, and coherent effects dominate over statistical behaviour. The simulation of many-boson interference, termed Boson-Sampling, entails a qualitatively superior computational complexity when compared to  fermions. The problem can be tamed by an artificially designed symmetric instance, which allows a systematic understanding of coherent bosonic and fermionic signatures for arbitrarily large particle numbers, and a means to stringently assess many-particle interference. The hierarchy between bosons and fermions also characterises multipartite entanglement generation, for which bosons again clearly outmatch fermions. Finally, the quantum-to-classical transition between many indistinguishable and many distinguishable particles features non-monotonic structures, which dismisses the single-particle coherence length as unique indicator for interference capability. While the same physical principles govern small and large systems, the deployment of the intrinsic complexity of collective many-body interference makes more particles behave differently. 

\end{abstract}

\submitto{\JPB}
\maketitle

\tableofcontents
\clearpage

\section{Introduction:  quantum statistics and many-body coherence}
At the beginning of the twentieth century, three highly successful but seemingly unrelated physical principles urged for a satisfactory connection: Quantum mechanics described discrete atomic spectra, the Pauli exclusion principle \cite{Pauli:1925fk} explained the structure of many-electron atoms, and the Bose-Einstein statistics of photons characterised the spectrum of thermal light \cite{Bose1924,einstein1914quantentheorie}. In 1926, Werner Heisenberg eventually reconciled these three seminal concepts by postulating the permutation (anti-)symmetry of the many-body wavefunction of identical particles \cite{heisenberg1926,PauliNobel}, which allowed him to see that \footnote{``Es scheint mir aber ein wichtiges Resultat dieser Untersuchung, da\ss~Paulis Verbot and die Einsteinsche Statistik den gleichen Ursprung haben, und da\ss~sie der Quantenmechanik nicht widersprechen.'' \cite{heisenberg1926}}  \emph{``the Pauli exclusion principle and the Einstein statistics have the same origin, and they do not contradict quantum mechanics.''}  

Since the advent of quantum physics, the understanding of the permutation symmetry of many-particle wavefunctions facilitated numerous discoveries, ranging from the BCS-theory of superconductivity \cite{PhysRev.106.162} over the description of coherent states of light generated by lasers \cite{PhysRev.131.2766} to the prediction \cite{PhysRev.46.76.2} and observation \cite{Shklovsky:1967hf} of neutron stars. 
These three examples concern macroscopic systems in which the symmetry of the many-fermion or many-boson wavefunction becomes only indirectly apparent through emerging statistical properties.  No control over individual particles is applied, instead, macroscopic effects are ascribed to the microscopic exchange symmetry of the many-particle wavefunction in a top-down approach. In particular, not the \emph{coherence} of the postulated superpositions of permuted many-particle wavefunctions is probed, but rather the \emph{statistical} consequences of the kinematic restrictions on bosons, which favour multiply occupied single-particle states, and on fermions, for which such multiple occupation is impossible. 

Since the 1980s, the development of new experimental technologies allowed a complementary approach to the physics of identical particles, and a direct verification of the intrinsic coherence of many-body wavefunctions. In quantum optics,  many-particle quantum effects became accessible at the level of individual photons, enabling a bottom-up perspective: Sources of correlated light quanta \cite{SinglePhotonSourcesReview}, such as spontaneous parametric down-conversion (SPDC) \cite{PhysRevLett.25.84,Hong:1985ys}, allowed to prepare individual photons, and single-photon detectors permitted their detection, albeit both processes being probabilistic. In particular, the collective interference of two photons that impinge on a beam splitter, observed by Shih and Alley \cite{PhysRevLett.61.2921} and Hong, Ou and Mandel (HOM) in 1987 \cite{Hong:1987mz}, helped to further promote the epistemological questions raised by quantum mechanics \cite{PhysRev.47.777} from the Gedankenexperiment \cite{Bell:1964pt,Clauser:1969qa} to the laboratory \cite{Aspect:1981zr,Aspect:1982ly,PhysRevLett.61.2921}. This development helped to dismiss the interpretation of entanglement as formal idiosyncratic artefact of an incomplete theory, and directed the attention to possible applications \cite{Nielsen:2000fk}.

Although the difficulty of photon generation with SPDC scales exponentially in the number of light quanta, remarkable progress has been achieved. The state-of-the-art quickly progressed from  three \cite{Bouwmeester:1999ys} to four \cite{Ou:1999rr}, six \cite{Xiang:2006uq} and, eventually, to eight interfering photons \cite{Yao:2011uq,Huang:2011fq}; by incorporating measurements and auxiliary particles, interactions can be simulated and first steps towards photonic quantum simulators have been reported \cite{Aspuru-Guzik:2012qm,Schreiber06042012}. In the future, integrated photonic devices \cite{Peruzzo17092010,Peruzzo:2011dq,PhysRevLett.108.010502,Meany:12} that allow re-configuration \cite{J2012ml} promise further   miniaturisation and accessibility of experiments.

Complementing these advances of technologies based on SPDC, other approaches for the generation of individual photons are currently  investigated, such as semi-conductor quantum dots \cite{Flagg:2010ly}, Rydberg atoms \cite{Dudin:2012sy} and even organic molecules \cite{Lettow:2010ve}. Alongside light quanta in the optical regime, also microwave photons can exhibit collective interference and  entanglement \cite{Lang:2013mi,1367-2630-15-10-105025}. In the future, other physical systems may supersede photons, since scientists aim at obtaining full control over the preparation, the coherent evolution, and the number-resolved detection of electrons \cite{Bocquillon01032013,ANDP:ANDP201300181} and atoms in optical lattices \cite{Sherson:2010fk,Bakr:2010fk,Weitenberg:2011vn,Hume:2013cl}.

Although these developments are highly promising, the unrelieved expenditure required for many-particle interference experiments naturally raises the question whether the phenomena that reside in the realm of truly many non-interacting identical particles qualitatively differ from those found within the well-understood two-particle paradigm. Throughout this tutorial, ample clearly affirmative evidence is given in support of such a qualitative difference, we anticipate here that \footnote{We slightly vary P.W.~Anderson's famous quotation \cite{Anderson:1972rc}.} {\emph{more particles behave differently}:} Interference, entanglement and the quantum-to-classical transition between indistinguishable and distinguishable particles feature surprising unexpected phenomena for many particles that cannot be grasped by the intuitive principles  obtained through the two-particle case or via statistical arguments valid in  incoherent environments. This qualitative difference makes quantitative predictions difficult already for moderate numbers of particles, but the complexity can be tamed by imposing artificial symmetries. 

This tutorial presents an illustrative introduction to the physics of many-boson and many-fermion interference, which supplies the reader the necessary tools to describe photonic entanglement manipulation  \cite{Lu:2007ve,Radmark:2009ij,Prevedel:2009ec,Wieczorek:2009ff,Yao:2011uq,Huang:2011fq,Zhao:2004dz} and novel approaches to optical quantum computing \cite{Aaronson:2011kx,Spring15022013,gogolin2013,Leverrier:2013ys,PhysRevA.85.022332,Tillmann:2012ys,Crespi:2012vn,Broome15022013,Shen:2013zr,Motes:2013ys}. Instead of reproducing results in their full generality, we 
 present simplified, intuitive proofs that emphasise the underlying physical and mathematical principles within a  toolbox for the description of many-particle interference. We assume the reader to be familiar with basic concepts of quantum optics (see, e.g., \cite{wallsMilburn}), entanglement theory (for reviews and introductions, see, e.g., \cite{Bengtsson:2006fu,RevModPhys.81.865,Plenio:2006uq,Mintert:2005rc,Tichy:2011fk}) and very basic notions of computational complexity theory (see, e.g., \cite{Nielsen:2000fk,Aaronson:2013kx,Moore:2011fk}); all required vocabulary will, however, be introduced within the text to make it self-contained. We focus on the observable fundamental physical phenomena, and do not tackle particular applications such as quantum metrology \cite{Dowling:2008hc}. Moreover, we do not cover interacting many-fermion and many-boson systems, which  paradigmatically show how the understanding of many-body entanglement can empower numerical techniques \cite{Amico:2008ph,Eisert:2010fk}. 
 
We start with the microscopic two-particle case in Section \ref{twoparticlecaseSec}, in which we discuss the collective interference of bosons and fermions, bipartite entanglement generation with identical particles, and the transition between distinguishable and indistinguishable particles. 
 Our constant leitmotiv in the remaining parts of the tutorial will be the surprising features that arise for large particle numbers.  We consider multi-mode interference in Section \ref{InterferenceSec}, in which we show how coherent many-particle interference dominates over statistical effects, making fermions and bosons exhibit complicated interference patterns. 
  The  scattering problem for bosons is of such great complexity that it constitutes a paradigmatic computational problem of its own, termed Boson-Sampling \cite{Aaronson:2011kx}, for which a classically certifiable instance is presented. The many-particle quantum-to-classical transition features unexpected structures that go beyond the usual coincidence dip or peak found for two particles, as described in Section \ref{DisttransSec}. Multipartite entanglement generation with bosons 
  outmatches the analogous process with fermions, for which we give a proof in Section \ref{MultipartSec}. We discuss open problems and give an outlook on possible future directions at the intersection of coherence and complexity in Section \ref{conclusionsandoutlook}.

\section{Interference of two particles} 
 \label{twoparticlecaseSec}
We introduce coherent many-body effects in a reduced setting with two particles and two modes, following the lines of \cite{TichyDiss}. This elementary and manageable case will, on the one hand, provide us with a reference point to appreciate the impact of the complexity borne by many particles, and, on the other hand, allow us to gently introduce the required formalism and the conventions that we will use throughout the tutorial. 

\subsection{Bunching and anti-bunching in two-particle interference} \label{twopinterSec}
The simplest setup in which the many-particle coherence of a two-boson or two-fermion state manifests itself 
 is a simple beam splitter with two input and two output modes \cite{12387}. Any incoming particle is reflected with probability $R$ and transmitted with probability $T=1-R$. 

\subsubsection{Distinguishable particles: uncorrelated binomial behaviour} \label{distbinob}

\label{quant2phom}
\begin{figure}[ht] \center
\includegraphics[width=1.\linewidth,angle=0]{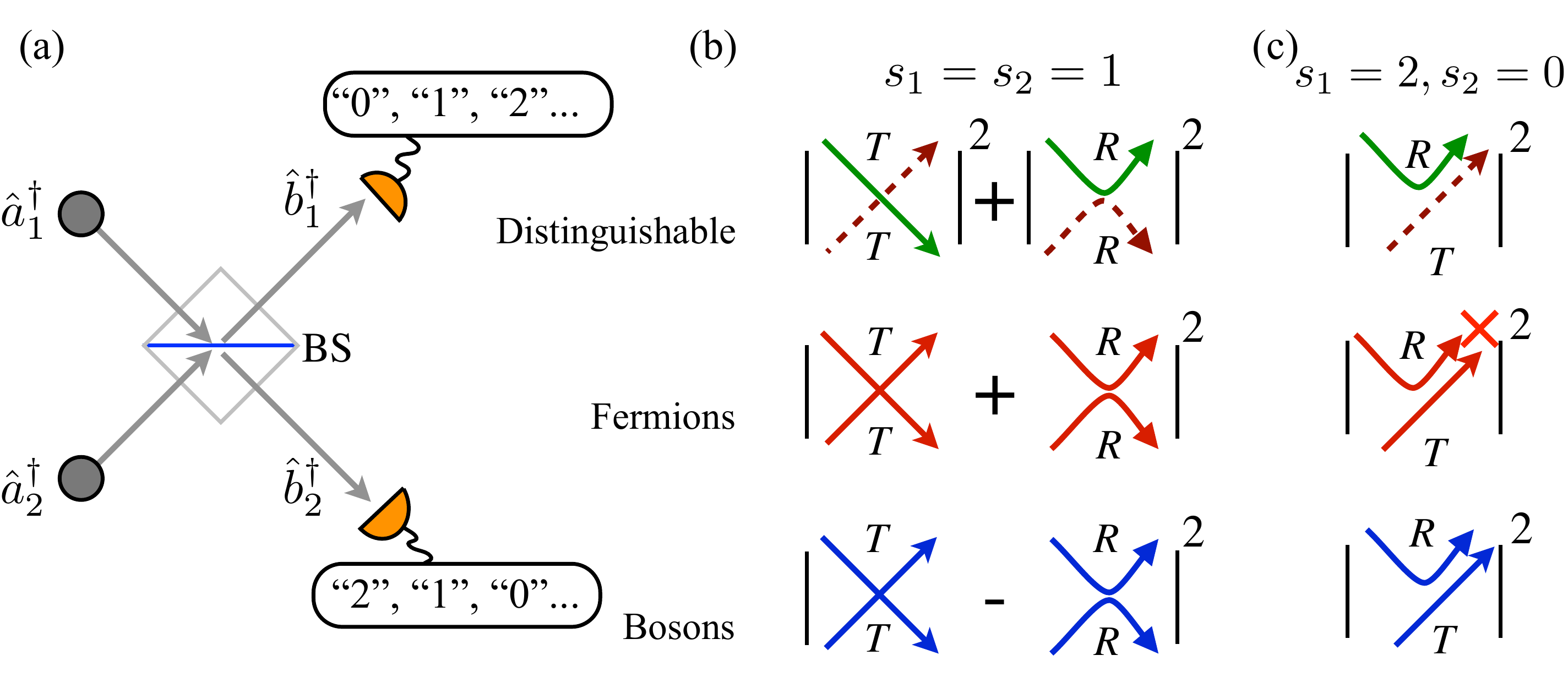}
\caption{Two-particle interference. (a) Two particles are prepared in two input modes $a_1$, $a_2$ and  impinge onto a beam splitter (BS). They are eventually detected in the output modes $b_1$, $b_2$ by the two idealised number-resolving detectors.  (b) To find one particle in each output mode, both particles need to be reflected, or both transmitted. For distinguishable particles (upper row), these processes are added incoherently, while constructive and destructive interference takes place for fermions and bosons, respectively. 
 (c) Only one process contributes to an event with both particles in the same output mode: One particle is transmitted, one reflected. For fermions, the Pauli principle inhibits this process, while it is enhanced for bosons. 
 }  \label{HOMFirstPic.pdf}
 \end{figure}

Let us first consider two distinguishable particles, e.g.~two different species or two photons with different polarisation. Although we assume that the particles are \emph{fundamentally} distinguishable (their wavefunction does not need to be (anti-)symmetric), we take them to be \emph{observationally} indistinguishable, i.e.~we deliberately ignore in our measurements \emph{which} particle is actually measured, but only count only \emph{how many} particles eventually populate the output modes. 
As sketched in Fig.~\ref{HOMFirstPic.pdf}(a), one particle is prepared in each input mode. An \emph{event} with $s_1$ and $s_2$ particles in the first and second output mode, respectively, occurs with probability $P_{\textrm{D}}(s_1,s_2)$. 

In order to observe a \emph{coincident event} with $s_1=s_2=1$, either both particles need to be transmitted or both particles need to be reflected by the beam splitter [see Fig.~\ref{HOMFirstPic.pdf}(a,b)]. These two processes give rise to two fundamentally distinguishable final states. 
 Therefore, the processes do not interfere, and we can simply add their probabilities \cite{PhysRevA.88.012130}: 
\eq
P_{\textrm{D}}(1,1)&=&T^2 + R^2 . \label{distprob1}
\en 
To find two particles in one mode, one particle is reflected, the other one is transmitted, i.e.~only one process contributes, and we find 
\eq 
 P_{\textrm{D}}(2,0)&=& P_{\textrm{D}}(0,2) =T R  \label{distprob2}  .
\en
In particular, for a \emph{balanced} setup, $T=R=1/2$, the distribution of particles in the output modes follows a binomial, and
\eq P_{\textrm{D}}(2,0)=\frac 1 4, \ P_{\textrm{D}}(1,1)=\frac 1 2, \ P_{\textrm{D}}(0,2)=\frac 1 4 . \label{distprob3} \en
This simple Gedankenexperiment with distinguishable particles can be understood in terms of classical probabilities, just like elementary stochastic processes such as two coin tosses. The particles are non-interacting, and they also remain independent and uncorrelated. Finding one particle in one output mode does not reveal any information on the fate of the other: With equal probability will it be in the same or in the other mode.

\subsubsection{Identical particles: correlation without interaction} \label{twoidpa}
The event probabilities for identical particles need to be established on the level of interfering many-body amplitudes, for which our combinatorial intuition breaks down. In particular, although the bosons and fermions under consideration are non-interacting, their behaviour is correlated. 

A treatment in the formalism of second quantisation is beneficial.  The creation operator of a particle in input mode $j$ is denoted by $\hat a_{j}^\dagger$, such that the initial state  reads [see Fig.~\ref{HOMFirstPic.pdf}(a)] 
\eq 
 \ket{\chi_{\text{ini}}^{(2)}} = \hat a^\dagger_{1} \hat a^\dagger_{2} \ket{\text{vac}}  \label{HOMini} ,\en
 where $\ket{\text{vac}}$ denotes the vacuum state. Throughout this tutorial, the letter ${\chi}$ denotes Fock-states, i.e.~$\ket{\chi}=\ket{n_1, n_2, \dots}_{1,2,\dots} $ is interpreted as  mode occupation $n_1$, $n_2$, etc. The letters ${\Psi}$ and ${\Phi}$ are reserved for qudits controlled by observing parties; $\ket{\Phi}=\ket{\phi_1, \phi_2, \dots }$ then denotes the state in which $\ket{\phi_1}$ is observed by the first party, $\ket{\phi_2}$ by the second, and so on. 

The action of the beam splitter amounts to redirecting an incoming particle into a coherent superposition on the output modes. The time-evolution of particle creation operators in the Heisenberg picture  reads  accordingly
\eq 
\hat a^\dagger_{j} &\rightarrow& \hat U \hat a^\dagger_{j} \hat U^{-1} = i \sqrt{R} \hat b^\dagger_{j} + \sqrt{T} \hat b^\dagger_{3-j} ,  \label{singledyn}
\en
where $j=1,2$; $\hat U$ is the unitary time-evolution operator, and the single-particle wavefunction acquires a phase-shift of $i=e^{i \pi/2}$ upon reflection. 
The single-particle time-evolution (\ref{singledyn}) can be inserted into the initial state (\ref{HOMini}), and, for a balanced beam splitter ($R=T=1/2$),   the final state  becomes 
\eq
 \ket{\chi_{\textrm{fin}}^{(2)}}&=&\hat U \ket{\chi_{\text{ini}}^{(2)}} = \hat U  \hat a^\dagger_{1} \hat a^\dagger_{2} \ket{\text{vac}} =  \hat U  \hat a^\dagger_{1} \hat U^{-1} \hat U \hat a^\dagger_{2}  \hat U^{-1} \hat U \ket{\text{vac}}  \nonumber  \\
&=& \frac 1 2 \left(i (\hat b^\dagger_{1})^2 +i (\hat b^\dagger_{2})^2 - \hat b^\dagger_1 \hat b^\dagger_2 + \hat b^\dagger_2 \hat b^\dagger_1 \right) \ket {\text{vac}} , \label{finastateHOM}
\en
where we used $\hat U \ket{\text{vac}}=\ket{\text{vac}}$, and no assumption was made yet on the fermionic or bosonic nature  of the particles. 

\paragraph{Fermions}
The Pauli exclusion principle implies that two fermions cannot co-exist in the same output mode, which is reflected by $(\hat b^\dagger_j)^2=0$ for fermionic creation operators. 
On the other hand, the two processes with both particles transmitted or both particles reflected are now fundamentally indistinguishable, since they lead to the very same final quantum state with one fermion in each output mode. Since fermionic operators anti-commute ($\hat b^\dagger_1 \hat b^\dagger_2 =  - \hat b^\dagger_2 \hat b^\dagger_1$), the  two terms in (\ref{finastateHOM}) that describe the two processes interfere constructively. We term this interference \emph{collective}, since two alternative paths of the whole many-body wavefunction in the many-body Hilbert-space are coherently super-imposed [see Fig.~\ref{HOMFirstPic.pdf}(b)]. 

The final state becomes 
\eq 
\ket{\chi_{\textrm{fin},F}^{(2)}}&=&   \hat b_2^\dagger  \hat b_1^\dagger  \ket {\text{vac}} = \ket{1,1}_{1,2} ,  \label{fermionfinal2}
\en
from which we immediately read off 
\eq 
P_{\text{F}}(2,0)=0, P_{\text{F}}(1,1)=1, P_{\text{F}}(0,2)=0 . \label{FermionsHOM}
\en
In other words, the Pauli principle prohibits the double population of any output mode, which is compatible with the constructive collective interference that enhances the probability to find the particles in different modes. The measurement outcomes  are now anti-correlated: When one fermion is found in one output mode, we immediately know that the second fermion is in the other. 

\paragraph{Bosons} 
Bosonic creation operators commute, $\hat b^\dagger_1 \hat b^\dagger_2 =  \hat b^\dagger_2 \hat b^\dagger_1$, such that the last two terms in (\ref{finastateHOM}) now interfere \emph{destructively}. On the other hand, 
 the two-fold application of a creation operator on the vacuum leads to an over-normalised state: 
\eq   (\hat b^\dagger_1)^2 \ket{\text{vac}} = \sqrt{2} \ket{2,0}_{1,2}. \en 
Thus, the final state for bosons, 
\eq \ket{\chi_{\textrm{fin,B}}^{(2)}}&=&\frac 1 2 \left( (\hat b^\dagger_{1})^2 - (\hat b^\dagger_{2})^2 \right) \ket {\text{vac}} = \frac{1}{\sqrt 2}\left( \ket{2,0}_{1,2}  - \ket{0,2}_{1,2} \right) ,     \label{bosonfinal2} \en
 describes a coherent superposition of both particles in mode 1 and both particles in mode 2, and 
\eq 
P_{\text{B}}(2,0)=\frac 1 2 , P_{\text{B}}(1,1)=0, P_{\text{B}}(0,2)=\frac 1 2. \label{BosonsHOM}
\en
Again, the particles behave in a correlated way: Finding one particle in one mode also reveals the position of the other, which will always be in the same mode. At this stage, it is already possible to anticipate qualitative differences between the simulation of many indistinguishable and many distinguishable particles (which will be discussed in the context of Boson-Sampling in Section \ref{secBosonSampling} below): The simulation of two distinguishable particles can be performed in a Monte Carlo-like approach in which the destiny of each particle is randomly chosen, independently of the other. For correlated bosons and fermions, such efficient Ansatz fails, and a holistic approach is necessary in which the full two-particle probability distribution needs to be sampled.

In a two-particle setup, the signal of identical-particle interference is unambiguous: Constructive interference for fermions turns into destructive interference for bosons (and vice-versa), while bunching of bosons is opposed to the Pauli principle of fermions. This behaviour also agrees with the  intuition gained from incoherent statistical systems, which can be summarised by \emph{bosons bunch, fermions anti-bunch}. For example, the Hanbury Brown and Twiss-effect that is exhibited in the two-point correlation function in thermal light \cite{Hanbury-Brown:1956vn} and thermal Bose \cite{Schellekens:2005bh,Molmer:2008uq} and Fermi gases \cite{Jeltes:2007ly,Kiesel:2002dq,Henny:1999cr} features an analogous enhancement and suppression of the probability to find two particles within their coherence length. The bosonic suppression and fermionic enhancement of the coincident (1,1)-event in Eqs.~(\ref{FermionsHOM},\ref{BosonsHOM}), however, are not  incoherent \emph{statistical} effects, since we start with a well-defined \emph{pure} quantum state, Eq.~(\ref{HOMini}). Instead, these effects are rooted in the \emph{coherent} superposition of two many-particle paths \cite{Tichy:2012NJP}. The (2,0)-event, on the other hand, is only fed by one path, which is why bunching and the Pauli principle are independent of any acquired phases. In other words, bosonic bunching and the Pauli principle are not the consequence  of coherently super-imposed many-particle paths, but they are rather kinematic boundary conditions on the state-space of identical particles. The difference between statistical behaviour and coherent signatures of identical particles will become more apparent below, when the number of particles is increased.

The discussion of collective interference above may, at first sight, seem to contradict Paul Dirac's famous quotation \cite{Dirac:1930vn}:  \emph{``Each photon then interferes only with itself.  Interference between different photons never occurs.''} Dirac, however, refers to expectation values of \emph{single-particle} observables of the form $\hat a^\dagger_k \hat a_j$, which remain unaffected by \emph{many}-particle interference: The expectation value of any single-particle operator, such as the average number of particles $\hat n_1=\hat b^\dagger_1 \hat b_1$, acquires the very same value for the fermionic final state (\ref{fermionfinal2}) and for the bosonic final state (\ref{bosonfinal2}), it also matches the average number of distinguishable particles found in each output mode. In other words, collective \emph{many}-particle interference can only affect \emph{many}-particle observables \cite{Kim:2003nx}. Indeed, Eqs.~(\ref{distprob3}), (\ref{FermionsHOM}), (\ref{BosonsHOM}) effectively describe the full counting statistics including the correlations between the output modes. 

For two particles and two detectors, two -- and only two -- collective paths are possible, such that the enhancement through interference of any transition, $P_{\text{B}}/P_{\text{D}}$ and $P_{\text{F}}/P_{\text{D}}$, can never exceed a factor of two. Any enhancement beyond this limit, e.g.~for thermal light  \cite{PhysRevA.86.013807}, can be attributed to intensity variations.

\subsection{Two-particle distinguishability transition} \label{twopdisttras}
Up to this point, the independent propagation of distinguishable particles [Eqs.~(\ref{distprob1}), (\ref{distprob2}), (\ref{distprob3})] was contrasted to the correlated behaviour of idealised bosons and fermions [Eqs.~(\ref{FermionsHOM}), (\ref{BosonsHOM})]. The latter were assumed to be perfectly indistinguishable in any degree of freedom besides the spatial mode that they are prepared in. In practice, this assumption is rather strong, and, in order to describe  experiments accurately, it is necessary to incorporate mode mismatch, i.e.~\emph{partial} mutual distinguishability.  While the difficulty to mutually match all physical properties of two particles constitutes a great challenge in the laboratory, the deliberate control over such distinguishing degrees of freedom allows one also to quite naturally explore the quantum-to-classical \emph{transition} between indistinguishable, interfering, correlated bosons and fermions, and distinguishable, non-interfering, independent particles. 

\begin{figure}[ht] \center
\includegraphics[width=\linewidth,angle=0]{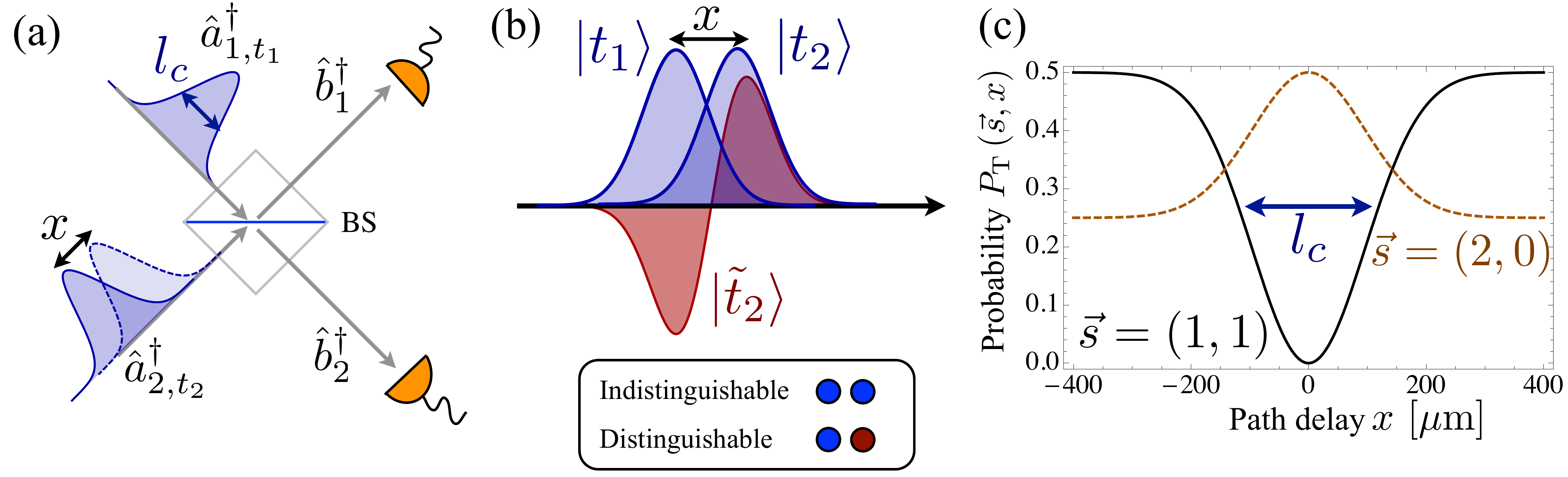}
\caption{(a) Experimental setup for two-particle interference. The ratio of the displacement $x$ between the wave-packets to the coherence length $l_c$ determines whether the incoming photons behave as bosons or as distinguishable particles. (b) The temporal part of the wavefunction of the particle in the second mode, $\ket{t_2}$, can be decomposed into one parallel and one orthogonal component with respect to the temporal distribution in the first mode. (c) The resulting signal for the detection of  $\vec s=(s_1,s_2)=(1,1)$ [blue solid line] and $\vec s=(2,0)$ [orange dashed line] reflects directly the coherence length $l_c$. }  \label{TwoPhotonDistTransition.pdf}
 \end{figure}

\subsubsection{Two partially distinguishable bosons} \label{twopdfsub}
As a realistic example for such transition, we consider photons and choose the time of arrival at the beam splitter as the degree of freedom through which the indistinguishability of the light quanta may be affected. In quantum optics experiments \cite{Hong:1987mz}, the relative delay in the arrival time is manipulated by the spatial displacement of the wave-packets to the beam splitter  [see Fig.~\ref{TwoPhotonDistTransition.pdf}(a)]. 

In order to quantify the distinguishability of the photons,  the discrete mode degree of freedom and the continuous temporal component are treated separately, such that the initial state reads  \cite{Ra:2013kx,younsikraNatComm}
\eq 
\ket{\chi_{\text{ini}}^{\text{HOM}}}=\hat a^\dagger_{1,t_1} \hat a^\dagger_{2,t_2} \ket{\text{vac}} , \label{twodelayed}
\en
where $\hat a^\dagger_{j,t_j} $ creates a photon in the spatial mode $j$ with the temporal wavefunction $\ket{t_j}$. The action of the beam splitter connects the spatial modes, but leaves the temporal component unaffected, such that Eq.~(\ref{singledyn}) is directly inherited,
\eq 
\hat a^\dagger_{j,t_k} &\rightarrow& i \sqrt{R} \hat b^\dagger_{j,t_k} + \sqrt{T} \hat b^\dagger_{3-j,t_k} . \label{linmoregen}
\en

Given the temporal wavefunction of the photon in the first mode, $\ket{t_1}$, the respective wavefunction $\ket{t_2}$ of the particle in the second mode can always be decomposed into an \emph{indistinguishable} contribution, parallel to $\ket{t_1}$, and a \emph{distinguishable} contribution, orthogonal to $\ket{t_1}$, as sketched in Fig.~\ref{TwoPhotonDistTransition.pdf}(b): 
\eq \ket{t_2} =   \ket{t_1} \underbrace{ \braket{t_1}{t_2}}_{=:c_{2,1} } + \ket{\tilde t_2}  \underbrace{ \sqrt{1-|\braket{t_1}{t_2}|^2} }_{=:c_{2,2}} , \label{twoporthog}  \en  where $\ket{\tilde t_2}$ is the projection of $\ket{t_2}$ orthogonal to $\ket{t_1}$, after normalisation,
\eq 
 \ket{\tilde t_2} = \frac{\ket{t_2}-\braket{t_1}{t_2} \ket{t_1} }{\sqrt{1-|\braket{t_1}{t_2}|^2 } } .
\en 
Formally, we have performed a Gram-Schmidt orthonormalisation of $\ket{t_1}$ and $\ket{t_2}$ (which is unnecessary in the case that $\ket{t_1}=\ket{t_2}$, for which $\ket{\tilde t_2}$ is not defined). The initial state becomes 
\eq 
\ket{\chi_{\text{ini}}^{\text{HOM}}}= \left(
c_{2,1} \hat a^\dagger_{1,t_1} \hat a^\dagger_{2,t_1}+
c_{2,2} \hat a^\dagger_{1,t_1} \hat a^\dagger_{2,\tilde t_2} \right) \ket{\text{vac}} ,
\en
where $c_{2,1}$ and $c_{2,2}$ [defined in Eq.~(\ref{twoporthog})]  
 are the respective \emph{weights}: $|c_{2,1}|^2$ quantifies indistinguishability, $|c_{2,2}|^2=1-|c_{2,1}|^2$ is the complementary measure for distinguishability. The two associated components in the wavefunction exhibit  different behaviour: The term $\hat a^\dagger_{1,t_1} \hat a^\dagger_{2,t_1} \ket{\text{vac}}$ exhibits perfect interference, since the particles described by this state are fully indistinguishable. Two particles described by $\hat a^\dagger_{1,t_1} \hat a^\dagger_{2,\tilde t_2} \ket{\text{vac}} $ are perfectly distinguishable and cannot interfere at all, due to the orthogonality of $\ket{t_1}$ and $\ket{\tilde t_2}$. 
 Due to the linearity of the time-evolution (\ref{linmoregen}), the indistinguishable and distinguishable terms can be treated independently. 

The resulting event probability for partially distinguishable particles becomes the average of the probabilities assigned to bosons and to distinguishable particles, weighted by $|c_{2,1}|^2$ and $|c_{2,2}|^2$, respectively:  
\eq 
P_{\text{T}}(\vec s=(s_1, s_2), x) =|c_{2,1}|^2 
 P_{\text{B}}(s_1,s_2) + |c_{2,2}|^2 P_{\text{D}}(s_1,s_2)  \label{twophtondiprob}
\en

To model typical experiments with photons \cite{Cosme:2008uq,Hong:1987mz,Ra:2013kx,younsikraNatComm,Ou:1999lo,Spagnolo:2013fk}, a  Gaussian frequency distribution around a central frequency $\omega_0$ with width $\Delta\omega$ is assumed, 
\eq
 \ket{t_j} =  \int d \omega \frac{1}{\pi^{1/4} \Delta \omega^{1/2}} e^{-\frac{(\omega-\omega_0)^2}{ 2 \Delta \omega^2} } e^{i \omega t_j}  \ket{\omega}. \label{GaussianWF} \en 
The temporal overlap of two photons then becomes 
\eq 
\braket{t_j}{t_k} = e^{i (t_k-t_j) \omega} e^{-\frac{(t_j-t_k)^2}{4} \Delta \omega^2 } , \label{temporaloverlap}
\en
where the phase is not observable in our present two-photon application, since only the absolute-squared values of $c_{1,2}$ and $c_{2,2}$ impact on the probability (\ref{twophtondiprob}). Throughout this tutorial, we consider photons of wavelength $\lambda=800$nm and width $\Delta \lambda =2.5$nm, which corresponds to a coherence length $l_c \approx 226 \mu$m. The spatial displacement $x$ in the experiment is directly related to the arrival times $t_j$ via $x=c (t_2-t_1)$, where $c$ is the speed of light. 

The resulting signals are shown in Fig.~\ref{TwoPhotonDistTransition.pdf}(c). Both the $(1,1)$ and the $(2,0)$ signals constitute direct probes for the indistinguishability $|c_{2,1}|^2=|\braket{t_1}{t_2}|^2$ of the two particles. The width $\sigma_{(1,1)}=\sigma_{(2,0)}$ of the signal equates the single-photon coherence time $l_c$, the shape reflects directly the Gaussian single-photon wavefunction (\ref{GaussianWF}). Therefore, the signal is a \emph{monotonic} function of the absolute spatial displacement $x$ between the photon wave-packets \cite{younsikraNatComm}, and the clear relationship between the displacement and the observed signal can be used to quantify mutual photon indistinguishability.

 Instead of the transition between distinguishable and indistinguishable particles, it is possible to observe  the artificial transition between bosons and fermions \cite{Matthews:2011fk}, or even between different types of \emph{anyons}, exotic particles with unusual exchange phases \cite{Lu:2009zr,Keilmann:2010ys,PhysRevA.85.033823}. 

\subsubsection{Experimental implementation of two-particle interference}
The dip in the (1,1)-signal is named after Hong, Ou and Mandel (HOM), who observed the effect in 1987 \cite{Hong:1987mz}. The visibility of the HOM-dip constitutes today the de-facto standard for ensuring the indistinguishability of two bosonic particles in the experiment. 

In most photonic experiments, the creation of two light quanta relies on 
SPDC \cite{Hong:1985ys}. Photon pairs are generated probabilistically by the passage of a laser pulse through a non-linear crystal. Spectral filtering reduces the bandwidth of the photons, and the quantum state of the field reads \cite{wallsMilburn,al:2010fk}
\eq 
\ket{\chi_{\text{SPDC}}}=\frac{1}{1 - \eta}  \sum_{m=0}^{\infty} \eta^m \ket{m}_s \ket{m}_i  \label{SPDC} ,\en
where $\ket{m}_{s(i)}$ denotes a Fock state of $m$ photons in the signal (idler) mode. The parameter $\eta$ depends on the properties of the non-linear crystal and on the pump power. Typically, $\eta \ll 1$, and the vacuum component $\ket{0}_s \ket{0}_i$ dominates. The exponential suppression of large photon numbers in (\ref{SPDC}) jeopardises SPDC as a scalable system for many-photon generation, but it nevertheless became the indisputable main workhorse of modern single-photon experiments. 

By \emph{post-selecting} events with a well-defined total number of photons $N$, i.e.~by neglecting all final events in which the total number of detected photons does not match the desired $N$, one can effectively project with high fidelity onto the component with a total of $N$ light quanta  in the initial state (\ref{SPDC}). An obstacle is that current ``bucket-detectors'' cannot resolve the number of detected photons, such that events with one and with two or more photons in the same mode give the same experimental signal. Strictly speaking, post-selection  therefore projects on the state that contains \emph{all} components in (\ref{SPDC}) with $2m \ge N$. Due to $\eta \ll 1$, however, one can reliably estimate the rather small impact of such higher photon-number components on the observed events. 

Number-resolving detectors can  be simulated probabilistically by splitting a mode into two and measuring the photon number in each of the output modes. A caveat is that this method is intrinsically probabilistic and decreases the detector efficiency by a factor of two: Two photons that impinge on the same input mode of a beam splitter  exit through the same output mode in 50\% of the runs, in which case only one detector will fire \cite{Brendel:1988ve,Lange:1988qf}. Recent progress in the development of single-photon counting detectors \cite{Sergienko:2008vn,Di-Giuseppe:2003ve,Webb200911799,PhysRevA.86.062113,calkins2013} feeds the hope that photon-counting devices will be available within a few years.

\subsection{Bipartite entanglement generation} \label{bipartiteentsec}
The coherent superposition of the two two-particle paths (both particles reflected and both particles transmitted) considered above can also be used to  generate bipartite entanglement between the particles found in the output modes \cite{PROP:PROP201200079,PhysRevLett.110.140404,Bose:2002vf}. This principle lies at the heart of quantum optics experiments with entangled photons \cite{RevModPhys.84.777}. It can also be adapted to \emph{project} onto entangled states, which allows to detect entangled states and perform protocols such as quantum teleportation \cite{Bennett:1993hc,Bouwmeester:1997dz} and entanglement swapping  \cite{Zukowski:1993ec,Marcikic:2004qr,Lu:2009fk}, as required for quantum repeaters.

\subsubsection{Generation of classical correlations} \label{classcorr}
\begin{figure}[ht] \center
\includegraphics[width=.8\linewidth,angle=0]{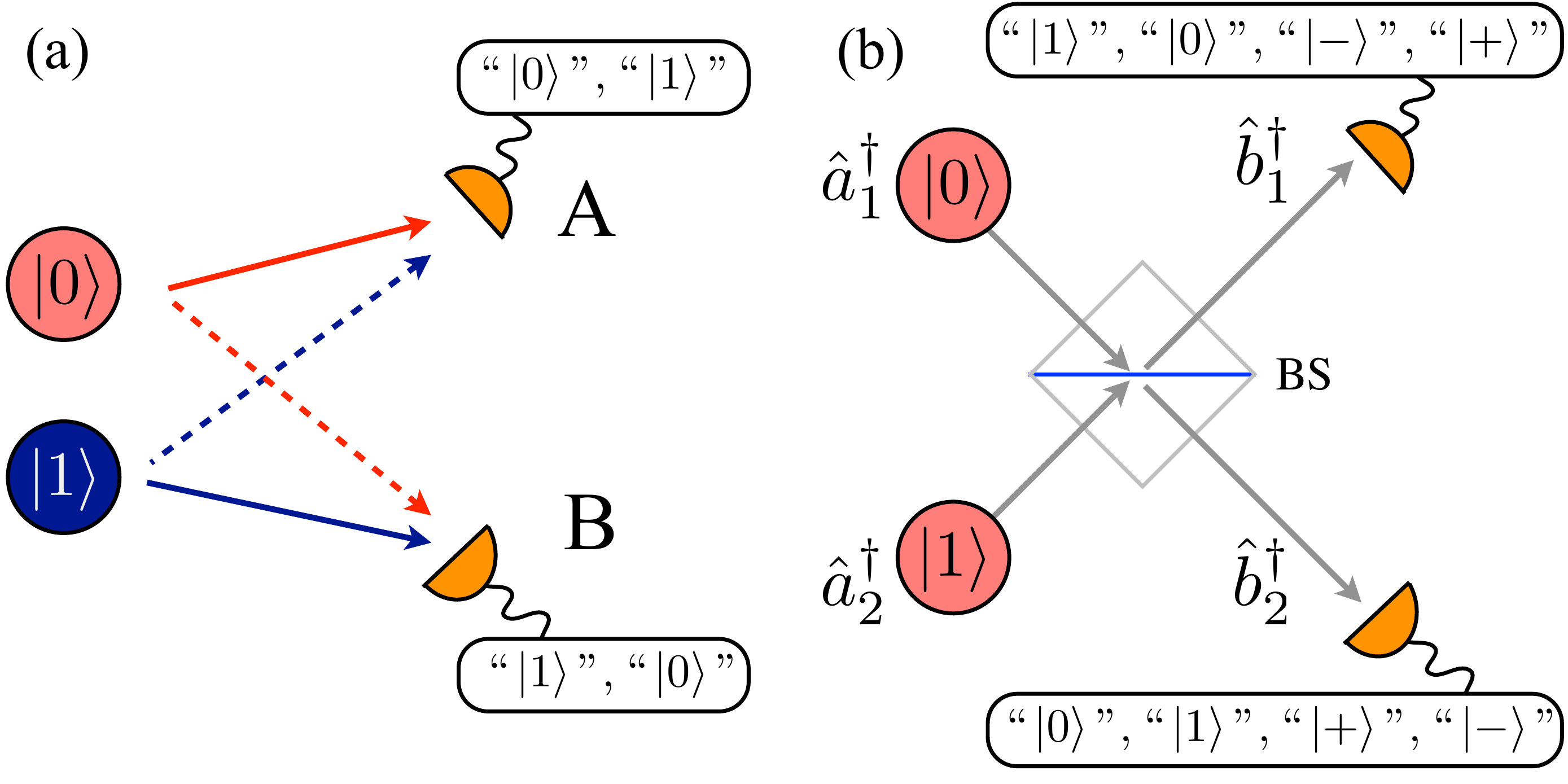}
\caption{(a) Creation of classical correlations. Two two-level systems are attached to two distinguishable objects. The correlations observed at the level of the detectors are purely classical, no coherent superposition of the objects can be measured. (b) A coherent superposition of  two identical particles carrying the two-level systems leads to quantum correlations.  }  \label{BipartitePict.pdf}
 \end{figure}

The origin of quantum correlations generated in experiments with photons can be understood best from an analogy to a classical probabilistic process \cite{Tichy:2011fk}: Consider two macroscopic distinguishable objects that are each attached a two-level quantum system, one prepared in $\ket{0}$, the other in $\ket{1}$ [see Fig.~\ref{BipartitePict.pdf}(a)]. This \emph{internal} state can be any degree of freedom on which a measurement in different bases is possible such that the genuinely quantum nature of correlations can be verified by testing the violation of Bell's inequalities \cite{Bell:1964pt,Clauser:1969qa} or by performing other tests for entanglement \cite{Tichy:2011fk}. 

A probabilistic machine that randomly distributes these two objects to two observers $A$ and $B$ will naturally create correlations between the observed outcomes: When $A$ observes $\ket{0}$, $B$  observes $\ket{1}$, and vice-versa. Since the objects are distinguishable, these correlations are purely classical, and not particularly surprising \cite{Bell:1987uq}, since their origin is manifestly a local-realistic (albeit probabilistic) mechanism: The outcomes of the measurements are determined by the distribution of the objects among the observers, i.e.~before the detection actually takes place. The outcome of one detector does not depend on any action performed at the other and the argument does not rely on $\ket{0}$ and $\ket{1}$ being quantum states, but it can be repeated with coins, socks, or playing cards \cite{Bell:1987uq}. 

The state of the particles as they are measured by the observers is described by a mixed state $\rho_{\text{class}}$, defined in the two-qubit basis $\{ \ket{0,0}, \ket{0,1}, \ket{1,0}, \ket{1,1} \}$:
\eq 
\rho_{\text{class}}=\frac 1 2 \left(\begin{array}{cccc} 
0 & 0 & 0 & 0 \\ 
0 & 1 & 0 & 0 \\
0 & 0 & 1 & 0 \\
0 & 0 & 0 & 0 
\end{array}  \right)  = \frac 1 2 \ket{0,1} \bra{0,1} + \frac 1 2 \ket{1,0} \bra{1,0}  \label{rhoclass}
\en
Since this state can be written as a balanced mixture of two separable states, as explicit in the above equation, it is not entangled, and merely classically correlated. In other words, there are no \emph{coherences} between the two possible outcomes [$\ket{0} (\ket{1})$ observed by $A$ $(B)$ and vice-versa]. Indeed, in the $\ket{\pm}$-basis, defined as 
\eq 
\ket{\pm}= \frac{1}{\sqrt{2}} \left( \ket{0} \pm \ket{1}  \right) ,
\en
the density matrix $\rho_{\text{class}} $ does not exhibit any correlations anymore. We then write 
\eq 
\rho_{\text{class}} = \frac 1 4 \left(\begin{array}{cccc} 
1 & 0 & 0 & -1 \\ 
0 & 1 & -1 & 0 \\
0 & -1 & 1 & 0 \\
-1 & 0 & 0 & 1 
\end{array}  \right) ,
\en
where the diagonal elements correspond to the outcome probabilities of  $\{ \ket{+,+}, \ket{+,-}, \ket{-,+}, \ket{-,-} \}$. 

\subsubsection{Quantum correlations via propagation and detection} \label{correpropdet}
No entanglement is generated by the stochastic classical procedure discussed in the previous section. However, an analogous process that \emph{coherently} distributes \emph{identical} particles to two observers can be used to generate genuinely quantum correlations, i.e.~entanglement. To see this, we again prepare two identical particles in the two input modes of the beam splitter setup [as in Eq.~(\ref{HOMini})], but each particle is initialised in a different internal state, $\ket{0}$ and $\ket{1}$, 
\eq 
\ket{\chi_{\text{ini}}^{\text{2ent}}}=\hat a^\dagger_{1,0}\hat a^\dagger_{2,1} \ket{\text{vac}} , \label{twopent}
\en
that is, the first index of $\hat a^\dagger_{j,k}$ now refers to the external state $j$, the second to the internal state $k$. 
For example, we can prepare two otherwise identical photons in orthogonal polarisation states and identify horizontal (vertical) polarisation with $\ket{0} (\ket{1})$ \cite{Ou:1988jb,PhysRevLett.61.2921}. 
 In analogy to Eq.~(\ref{finastateHOM}), the state after propagation reads 
\eq
 \ket{\chi_{\textrm{fin}}^{\text{2ent}}}&=&\frac 1 2 \left( \underbrace{ \hat b^\dagger_{2,{0}}   \hat b^\dagger_{1,{1}}  -  \hat b^\dagger_{1,{0}}    \hat b^\dagger_{2,{1}} }_{\text{coincident events}}  +i \underbrace{ \left(  \hat b^\dagger_{2,{0}}   \hat b^\dagger_{2,{1}}   + \hat b^\dagger_{1,{0}}   \hat b^\dagger_{1,{1}}     \right) }_{\text{bunched events}}  \right) \ket{\text{vac}} .  \label{bipartiteentgen}
\en
 Following the discussion in Section \ref{distbinob}, the counting statistics observed in the output modes corresponds to the statistics of distinguishable particles. In particular, neither fermions nor bosons prepared in the state (\ref{twopent}) exhibit the HOM dip or peak in (\ref{bipartiteentgen}).

Let us now assume that we \emph{post-select} those events with exactly one particle per spatial output mode, i.e.~coincident events. In an experimental setting, this amounts to neglecting all runs that do not match this requirement, i.e.~discarding all bunched events in which both particles are found in one output mode. When a coincident event is found \emph{without} any observer learning about the internal quantum state of the particles, the state is effectively projected onto 
 \eq
 \ket{\chi^{\text{2ent}}_{\textrm{post}}}&=&\frac {1} {\sqrt{2}} \left( { \hat b^\dagger_{2,{0}}   \hat b^\dagger_{1,{1}}  - \delta   \hat b^\dagger_{2,{1}}  \hat b^\dagger_{1,{0}}   } \right)   \ket{\text{vac}}\label{postselectedstate} , \en
 where $\delta=+1$ for bosons and $\delta=-1$ takes into account the anti-commutation relation of fermionic creation operators. In great contrast to the situation with distinguishable particles in the previous section, the state before the measurement is a \emph{coherent} superposition of both measurement outcomes. Therefore, two observers that control one output mode each will conclude that they are given the state 
 \eq 
\ket{\Psi^+} &= & \frac{1}{\sqrt{2}} \left( \ket{1,0} + \ket{0,1}  \right)= \frac{1}{\sqrt{2}} \left( \ket{+,+} - \ket{-,-}  \right) \label{Psiplu}  ,
 \en
 when the particles are fermions, and 
 \eq 
\ket{\Psi^-} &= & \frac{1}{\sqrt{2}} \left( \ket{1,0} - \ket{0,1}  \right) = \frac{1}{\sqrt{2}} \left( \ket{+,-} - \ket{-,+} \right)  , \label{Psimin}
 \en
 for bosons. These maximally entangled \emph{Bell}-states \cite{RevModPhys.81.865} describe quantum-mechanically correlated particle pairs: Not only will the observers always measure anti-correlated outcomes in the basis $\{ \ket{0}, \ket{1} \}$, but correlations will also prevail for other concerted choices of local measurement bases.  In particular, the singlet state $\ket{\Psi^-}$ yields anti-correlated measurements in every basis [Eq.~(\ref{Psimin})], whereas particles prepared in $\ket{\Psi^+}$ yield perfectly correlated measurement outcomes in the $\{ \ket{+}, \ket{-} \}$-basis [Eq.~(\ref{Psiplu})]. This persistence of correlations in different bases constitutes the qualitative difference between quantum and classical correlations, which allows, e.g.,~to violate Bell's inequality \cite{Bell:1964pt,Tichy:2011fk}. The many-particle quantum coherences exhibited here as entanglement have their origin in the (anti)-symmetry of the bosonic or fermionic wavefunction, as reflected by the parameter $\delta$ in Eq.~(\ref{postselectedstate}). 
  
The states $\ket{\Psi^+}$ and $\ket{\Psi^-}$ can be converted into each other by a \emph{local unitary operation} of the form $\hat U_1 \otimes \hat U_2$, namely a conditional phase-gate on one of the subsystems: 
   \eq 
   \ket{\Psi^+}=- \mathbbm{1}_{2} \otimes \hat \sigma_z \ket{\Psi^-}  ,~~~\hat \sigma_z =\left(\begin{array}{cc} 1 & 0 \\ 0 & -1 \end{array} \right) .
   \en 
From the perspective of entanglement theory \cite{RevModPhys.81.865}, the states generated by bosons and those created by fermions are therefore fully equivalent resources for tasks such as quantum key distribution \cite{PhysRevLett.67.661} or quantum teleportation \cite{Bennett:1993hc}. 

The very assignment of entanglement or separability to a state of identical particles can be conceptually involved, in general, since the formal (anti-)symmetrisation of the many-boson (many-fermion) wavefunction in first quantisation needs to be distinguished from actual physical correlations \cite{Ghirardi-statphys,Ghirardi:2004fk,Tichy:2011fk}. In the present situation, each observing party controls exactly one particle, and assigning entanglement to the quantum state is unproblematic: We quantify precisely those correlations that the parties actually observe; Eqs.~(\ref{Psiplu}) and (\ref{Psimin})  constitute the unambiguous quantum-information abstraction of the physical state (\ref{postselectedstate}). Such quantum correlations carried by particles significantly differ conceptually from the alternative \emph{mode entanglement}: Instead of particles, modes can carry correlations, such that the mode occupation number constitutes the degree of freedom in which the parties can be entangled \cite{PhysRevA.87.022338,Enk:2005gd}. Within this Tutorial, we exclusively focus on the entanglement between particles that are assigned to one detector each, the two-party paradigm considered here will be generalised to $N$ parties in Section \ref{MultipartSec}.

\subsubsection{Partial distinguishability and degradation of entanglement} \label{partialdistentangl}
If the particles prepared in the input modes do not differ only in the degree of freedom in which entanglement is to be created (here: $\ket{0}$ and $\ket{1}$), but also in another physical property, the correlations measured at the output modes will eventually degrade to purely classical ones. In practice, this deteriorating degree of freedom is often the time of arrival, as in Section \ref{twopdisttras}: When the particles can be given the unambiguous labels ``early'' and ''late'', no coherent superposition takes place, and purely classical correlations are observed. 

Combining Eqs.~(\ref{twodelayed}) and (\ref{twopent}), the initial state of two delayed photons reads 
\eq 
\hat a^\dagger_{1,{0}, t_1 }\hat a^\dagger_{2,{1}, t_2 } \ket{\text{vac}} =
\left( c_{2,1} \hat a^\dagger_{1,{0}, t_1 }\hat a^\dagger_{2,{1}, t_1 } + c_{2,2} \hat a^\dagger_{1,{0}, t_1 }\hat a^\dagger_{2,{1}, \tilde t_2 }   \right)\ket{\text{vac}} , \label{enttrans}
\en
where $\hat a^\dagger_{j,k,t_l}$ creates a photon in the external mode $j$, in the internal state $k$, at time $t_l$. 
In Section \ref{twopdfsub} above, the distinguishable component with weight $c_{2,2}$ lead to classical binomial statistics, while the indistinguishable one induced bosonic or fermionic correlated behaviour [see Eq.~(\ref{twophtondiprob})]. In direct analogy, the distinguishable component in (\ref{enttrans}) induces classical correlations, while the indistinguishable one is responsible for entanglement. Consequently, local observers will detect an incoherent \emph{mixture} of a maximally entangled state as given by Eqs.~(\ref{Psimin}) or (\ref{Psiplu}), and the classically correlated, mixed state (\ref{rhoclass}) \cite{TichyDiss,PROP:PROP201200079},
\eq 
\rho =|c_{1,2}|^2 \ket{\Psi^{- \delta}} \bra{\Psi^{- \delta}} + |c_{2,2}|^2 \rho_{\text{class}} = \frac 1 2 \left(\begin{array}{cccc} 
0 & 0 & 0 & 0 \\ 
0 & 1 & - \delta |c_{2,1}|^2 & 0 \\
0 & - \delta |c_{2,1}|^2 & 1 & 0 \\
0 & 0 & 0 & 0 
\end{array}  \right)  , \label{transstate}
\en
where $ \delta =\pm 1$ refers to bosons/fermions. 
That is to say, the indistinguishability $|c_{2,1}|^2$ directly quantifies the many-body coherences of the emerging state and, thus, the quantum nature of the induced correlations. Indeed, a quantitative indicator for entanglement, the \emph{concurrence} \cite{Mintert:2005rc}, 
\eq 
\mathcal{C}=|c_{2,1}| \label{concurrencedist} ,
\en
directly reflects the indistinguishability \cite{PROP:PROP201200079}.

\subsubsection{Limits and applications} \label{limitsapplications}
In summary, the combination of \emph{independent propagation} and \emph{detection} of identical particles leads to the creation of entanglement, \emph{without any direct interaction}. Although the classical analogy in Section \ref{classcorr} may provide an intuitive picture for the mechanism behind the correlations, only the many-particle coherence intrinsic to every state of many bosons or fermions  ensures that the correlations which  are measured between the particles found in the output modes are quantum-mechanical, and, e.g., violate Bell's inequalities. A continuous transition between quantum and classical correlations is observed in (\ref{transstate}), which can be directly compared to the transition between quantum and classical statistics, Eq.~(\ref{twophtondiprob}).

To assess the general potential of bipartite entanglement generation by propagation and detection, it is useful to consider the canonical form of bipartite states, the Schmidt decomposition \cite{Nielsen:2000fk}, 
\eq
\ket{\Psi}= \sum_{j=0}^{d-1} \sqrt{\lambda_j } \ket{\phi_j, \eta_j},  \label{Schmidt}
\en
which exists for every two-party state, i.e.~there are always local bases $\{ \ket{\phi_j} \}$ and $\{ \ket{\eta_j} \}$ such that a single sum index as in (\ref{Schmidt}) is sufficient for a full description of the state. The Schmidt coefficients $\lambda_j$ then contain all information about the entanglement inherent to the state. 

Since two non-interacting particles that are detected by two detectors lead to exactly two physically distinct paths -- independently of how complex the setup may be devised -- qudit-like operations that generate entanglement beyond two levels (i.e.~that involve more than two non-vanishing terms in (\ref{Schmidt})) are impossible for initially unentangled states. One can increase the number of involved levels to up to $d$ when $d-2$ auxiliary particles are used \cite{Calsamiglia:2002dq}.  

Relying on propagation and detection for entanglement manipulation also leads to natural constraints on operations. For example, a measurement that fully discriminates all four Bell states is impossible  \cite{Lutkenhaus:1999bh}, which inhibits deterministic teleportation  \cite{Vaidman:1999qf}, for which non-linear interactions are necessary \cite{PhysRevLett.86.1370}. Although these theorems were formulated for two photons, i.e. bosons, the argument applies for two fermions in an analogous way: The sign change in (\ref{postselectedstate}) does not lead to any conceptual difference between fermions and bosons.

\subsection{Principles governing two-particle interference}
The interference of two particles is a rather manageable situation: Since exactly two distinct paths compete for any event in which particles do not end in the same mode, the emerging behaviour can be understood from  a microscopic perspective. Bosons and fermions exhibit  distinct interference signatures, but they lead to equivalent entangled states, which can be transformed into each other by local unitary operations. For, both, the nature of the correlations [see Eq.~(\ref{transstate})] as well as for the  two-particle interference signal [see Eq.~(\ref{twophtondiprob})], a simple interpolation applies, and the indistinguishability $|c_{1,2}|^2$ quantifies the coherence of the many-body wavefunction. 

Two-particle interference is  also observed in many-mode setups with more involved \emph{single}-particle behaviour, such as quantum walks, possibly with additional disorder \cite{Segev:2013kc,PhysRevLett.105.163905}. The emerging two-particle correlations then remain qualitatively similar to the reduced HOM-setting, since, again, exactly two paths  contribute to an output event with two separated particles \cite{PhysRevA.81.023834,PhysRevA.88.012308}. The amplitudes of these two paths are added for bosons and subtracted for fermions, which leads to characteristic opposite signals in the two-particle correlations, e.g.~bunching and anti-bunching \cite{mayerda,PhysRevA.83.062307,PhysRevLett.108.010502}. In other words, any single-particle interference characteristic for the system at hand is supplemented by the \emph{correlations} induced by the bosonic or fermionic nature of the particles. In the experiment, fermionic behaviour is typically simulated by anti-symmetric entangled states of photons \cite{Matthews:2011fk}; more in general,  \emph{anyons} with unusual exchange phases can be devised \cite{PhysRevA.85.033823}, and different initially correlated states of light give rise to characteristic signatures \cite{Crespi:2013hs,Peruzzo17092010}. 

When not only the number of modes but also the number of particles $N$ is increased, up to $N!$  many-particle paths need to be taken into account for computing an event probability. It is evident that our microscopic treatment of Eq.~(\ref{finastateHOM}) is therefore impractical for many particles, and a more efficient formalism needs to be devised.  From the physical point of view, several questions arises: Are bosonic and fermionic interference signals universal, i.e.~do they persist in the realm of many-body correlations? How can the indistinguishability of truly many particles be ensured? Are the entangled states generated by  many fermions and many bosons equivalent? These three topics will be addressed in the following three sections, in which we give manifold examples for the rich physics of many-particle interference. 

\FloatBarrier

\section{Many-particle interference} \label{InterferenceSec}

\subsection{Many-mode devices}
\begin{figure}[ht] \center
\includegraphics[width=.55\linewidth,angle=0]{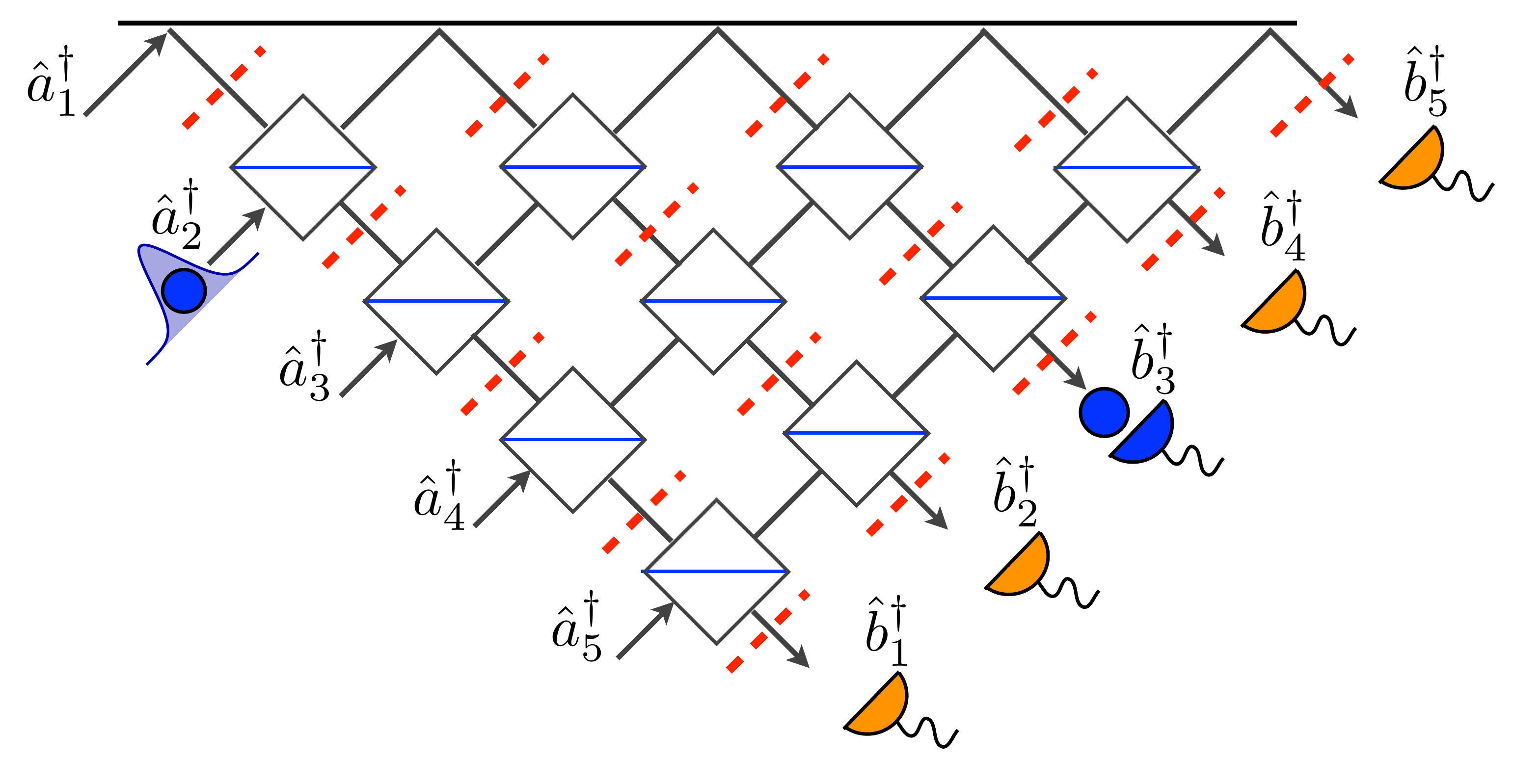}
\caption{Multiport beam splitter with $n=5$ modes in the pyramidal construction \cite{Reck:1994zp}. The matrix element $U_{2,3}$ encodes the amplitude for the particle prepared in mode 2 to exit through mode $3$. It contains all double-slit-like self-interference of the particle. } 
\label{MultiportBS}
\end{figure}

To model  devices for the observation of multi-mode interference of single- \cite{Karski:2009vn,Jeong:2013kq}, two- \cite{Peruzzo17092010,Peruzzo:2011dq,PhysRevA.88.012308,Matthews:2011fk} and many-particle states \cite{Spring15022013,Tillmann:2012ys,Crespi:2012vn,Broome15022013}, we can think of  assembling 
 several elementary beam splitters in a pyramid-like construction, as depicted in Fig.~\ref{MultiportBS}. Any unitary  matrix $U$ of dimension $n\times n$ can be implemented with such a device \cite{Reck:1994zp}, in which $n(n-1)/2$ elementary  beam splitters with certain chosen reflectivities are aligned with $n(n+1)/2$ phase-shifters. The universality still holds even if only one type of beam splitter is available \cite{Bouland:2013uq}. On the other hand, a given unspecified unitary device can be characterised efficiently \cite{Rahimi-Keshari:2013wu}.

  An incoming particle, as the one sketched in mode 2 in Fig.~\ref{MultiportBS}, then interferes with itself within the structure, since several single-particle paths lead to each output mode. 
 Such single-particle interference is already contained in the amplitude $U_{j,k}$; the absolute square, 
  \eq p_{j,k} = |U_{j,k}|^2, \label{singleparticleprob}  \en then denotes the probability for a particle prepared in the $j$th input mode to reach the $k$th output mode. 
 As a convention, the first index of any scattering matrix (the row) refers to the input mode, while the second (the column) denotes the output mode.  
  
 In the laboratory, the constructive procedure sketched in Fig.~\ref{MultiportBS} and described in \cite{Reck:1994zp} is unfeasible: For large optical setups in free space, interferometric stability is  difficult to ensure. Still, the setup shown in Fig.~\ref{MultiportBS} provides us with an illustrative picture for many-mode devices, while recent advances in the fabrication of integrated waveguide structures \cite{J2012ml,Bromberg:2009zr,PhysRevA.87.012309,PhysRevA.87.013842,Metcalf:2013xw,Spagnolo:2012kn,Peruzzo17092010,Peruzzo:2011dq,PhysRevLett.108.010502,Meany:12} makes the experimental realisation of any desired unitary scattering matrix feasible for photons.

For systems of many particles, it is useful to formulate the time-evolution for creation operators, in close analogy to Eq.~(\ref{singledyn}) \cite{Tichy:2012NJP}, 
\eq \hat a^\dagger_j \rightarrow  \hat U \hat a^\dagger_j \hat U^{-1} = \sum_{k=1}^n U_{j,k} \hat b^\dagger_k , \label{timeevocreation} \en
which is independent of the particle species (bosonic or fermionic). This relation can be used to infer the time-evolution of a many-particle state as long as no interaction between particles takes place, which ensures that each particle evolves independently of the presence or absence of others. Formally speaking, the Hamiltonian that governs the system must not contain any terms beyond linear in the creation and annihilation operators. 

\subsection{Many-particle evolution and transition probabilities} \label{mpetp}
To study multi-mode many-particle interference, we consider $N$ particles prepared in the $n$ input modes of a multimode setup and follow the argument of \cite{TichyDiss}. The number of particles in the $j$th mode is denoted by $r_j$, such that $\sum_{j=1}^n r_j=N$. That is, an initial \emph{arrangement} of particles can be characterised by a \emph{mode occupation list} $\vec r=(r_1, \dots , r_n)$.  For indistinguishable particles, the initial quantum state then reads \cite{Tichy:2012NJP}
\eq \ket{\chi^{\textrm{in}}(\vec r)}_{\text{B/F}} = \left[ \prod_{j=1}^n \frac{ \left( \hat a^\dagger_j \right)^{r_j} }{\sqrt{r_j!} } \right] \ket{\text{vac}} .  \label{instatemultip} \en

Generalising the discussion of Section \ref{twopinterSec}, we assume that the number of particles in each output mode is measured after the particles have propagated through the setup. A measurement outcome is characterised by the mode occupation list $\vec s=(s_1,s_2,\dots,s_n)$. Since the number of particles is conserved, by assumption, $\sum_{j=1}^n s_j =N$.

\begin{figure}[ht] \center
\includegraphics[width=.7\linewidth,angle=0]{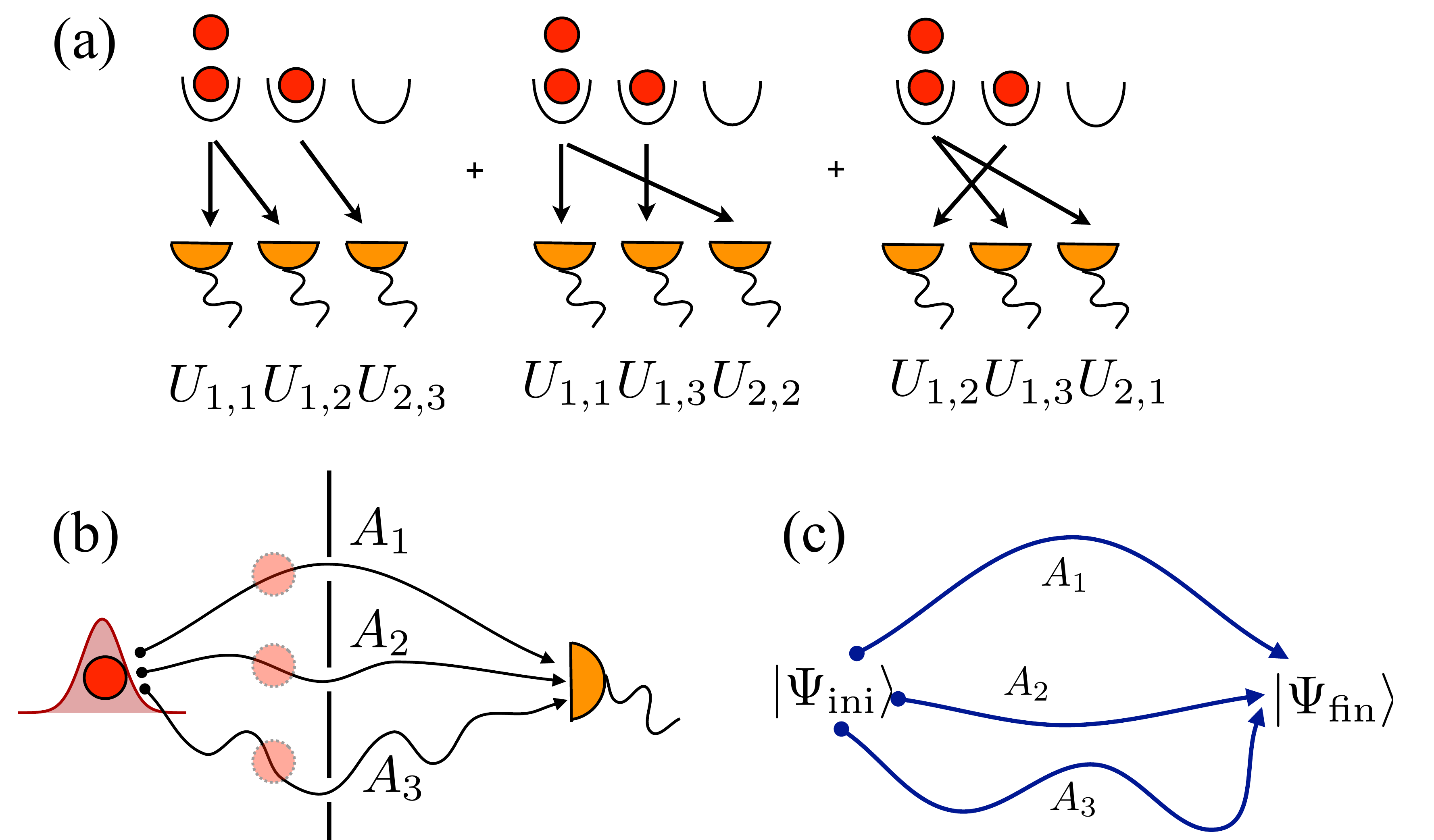}
\caption{(a) Interference of three many-particle paths. The initial arrangement is characterised by $\vec r=(2,1,0), \vec d(\vec r)=(1,1,2)$; the final state is $\vec s=(1,1,1), \vec d(\vec s)=(1,2,3)$.   (b) Single-particle interference of a particle falling on a triple-slit. The amplitudes of the three processes interfere, in close analogy to the interference between the three three-particle paths in (a). (c) In an abstract picture, three interfering processes with amplitudes $A_1, A_2, A_3$ connect the initial to the final state. These states can be single-particle as well as many-particle states. The figure is  adapted from \cite{TichyDiss}.}  \label{Interference.pdf}
 \end{figure}

Instead of a mode occupation list $\vec q$, which states the number of particles in each input or output mode, one can alternatively define the \emph{mode assignment list} $\vec d(\vec q)$, a list of mode numbers such that  $d_k$ denotes the mode in which the $k$th particle is prepared or found in \cite{Tichy:2010ZT,Tichy:2012NJP}. The length of the list reflects the number of particles $N$. Formally, it can be constructed by repeating $q_j$ times the mode number $j$:
\eq \vec d(\vec q) =\oplus_{j=1}^n \oplus_{k=1}^{q_j} (j) = (\underbrace{1,\dots,1}_{q_1},\underbrace{2,\dots,2}_{q_2}, \dots, \underbrace{n,\dots,n}_{q_n}) \label{drepres} \en  
Examples for mode occupation and mode assignment lists can be found in the caption to Fig.~\ref{Interference.pdf}. 
Permutations on a mode assignment list do not effectively change the particle arrangement, i.e.~several mode assignment lists lead to the same mode occupation list. To avoid ambiguity, we therefore always require $d_j \le d_{j+1} $. Due to the Pauli principle, fermions can only be prepared and measured in arrangements $\vec q$ with $0\le q_j \le 1$ for all $j$. Such arrangements (which can also be realised for bosons and distinguishable particles) will be named \emph{Pauli arrangements}. 

Although the mode occupation list is a natural way to describe the \emph{state} of a system, the mode assignment list allows us to formulate a compact and intuitive expression for the probability to find the output event $\vec s$, given the input arrangement $\vec r$. 

\subsubsection{Distinguishable particles}
For distinguishable particles, in close analogy to Eqs.~(\ref{distprob1},\ref{distprob2}), we first consider the possibility that the first particle, in  mode $d_1(\vec r)$, is detected in mode $d_1(\vec s)$; the second particle, prepared in $d_2(\vec r)$, ends in $d_2(\vec s)$, and so on. The probability for this process is the product of the individual transition probabilities, 
\eq p_{d_1(\vec r),d_1(\vec s)} p_{d_2(\vec r),d_2(\vec s)}  \dots p_{d_N(\vec r),d_N(\vec s)} , \en
where $p_{j,k}$ is the single-particle probability given in Eq.~(\ref{singleparticleprob}). 
 The final arrangement $\vec s$ can also be attained when a permutation is applied on the list  $\vec d(\vec s)$. Eventually, all permutations on the particles in the output modes need to be taken into account, and the $N$ particles experience up to $N!$ possibilities to be re-arranged in the output modes.  
Formalising this argument, the total probability becomes 
\eq P_{\textrm{D}} (\vec r, \vec s; U) &=& \sum_{\sigma \in S_{\vec d(\vec s)} }   \prod_{j=1}^N p_{d_{j}(\vec r),\sigma(j)}  ,  \label{generalD}  \en
where the sum over $\sigma \in S_{\vec d(\vec s)}$ runs over all $N!/(\prod_j s_j!)$ permutations of the output mode assignment list $\vec d(\vec s)$. 

\subsubsection{Indistinguishable particles}
For \emph{indistinguishable} particles, the argument can be repeated. Now, however, the exchange of two particles in the output modes does not change the quantum state (up to a global sign change for fermions), such that all possibilities to distribute the particles among the output modes contribute \emph{coherently} to the final state. As a consequence, many-body \emph{amplitudes} of the form $U_{d_1(\vec r),d_1(\vec s)} U_{d_2(\vec r),d_2(\vec s)}  \dots U_{d_N(\vec r),d_N(\vec s)} $ need to be summed, as in Section \ref{twopinterSec}. The  probability $P_{\text{B/F}}(\vec r, \vec s ; U )$ for the transition between an initial state $\vec r$ and a final state $\vec s$ possesses a compact expression:
\eq 
\label{generalampliB}  P_{\textrm{B}}(\vec r, \vec s ; U) &=&  \frac{\prod_j s_j!}{\prod_j r_j!}   \left| \sum_{\sigma \in S_{\vec d(\vec s)}}  \prod_{j=1}^N U_{d_j(\vec r),\sigma(j)} \right|^2 ,   \\
\label{generalampliF}  P_{\textrm{F}}(\vec r, \vec s ; U) &=&    \left| \sum_{\sigma \in S_{\vec d(\vec s)}} \textrm{sgn}(\sigma) \prod_{j=1}^N U_{d_j(\vec r),\sigma(j)} \right|^2 ,  
 \en 
where $\textrm{sgn}(\sigma)$ takes into account the  phase acquired upon exchange of two fermions in the output modes. 

\subsection{Bosonic and fermionic interference pattern}  \label{interferencepattern}
\begin{figure}[ht] \center
\includegraphics[width=.95\linewidth,angle=0]{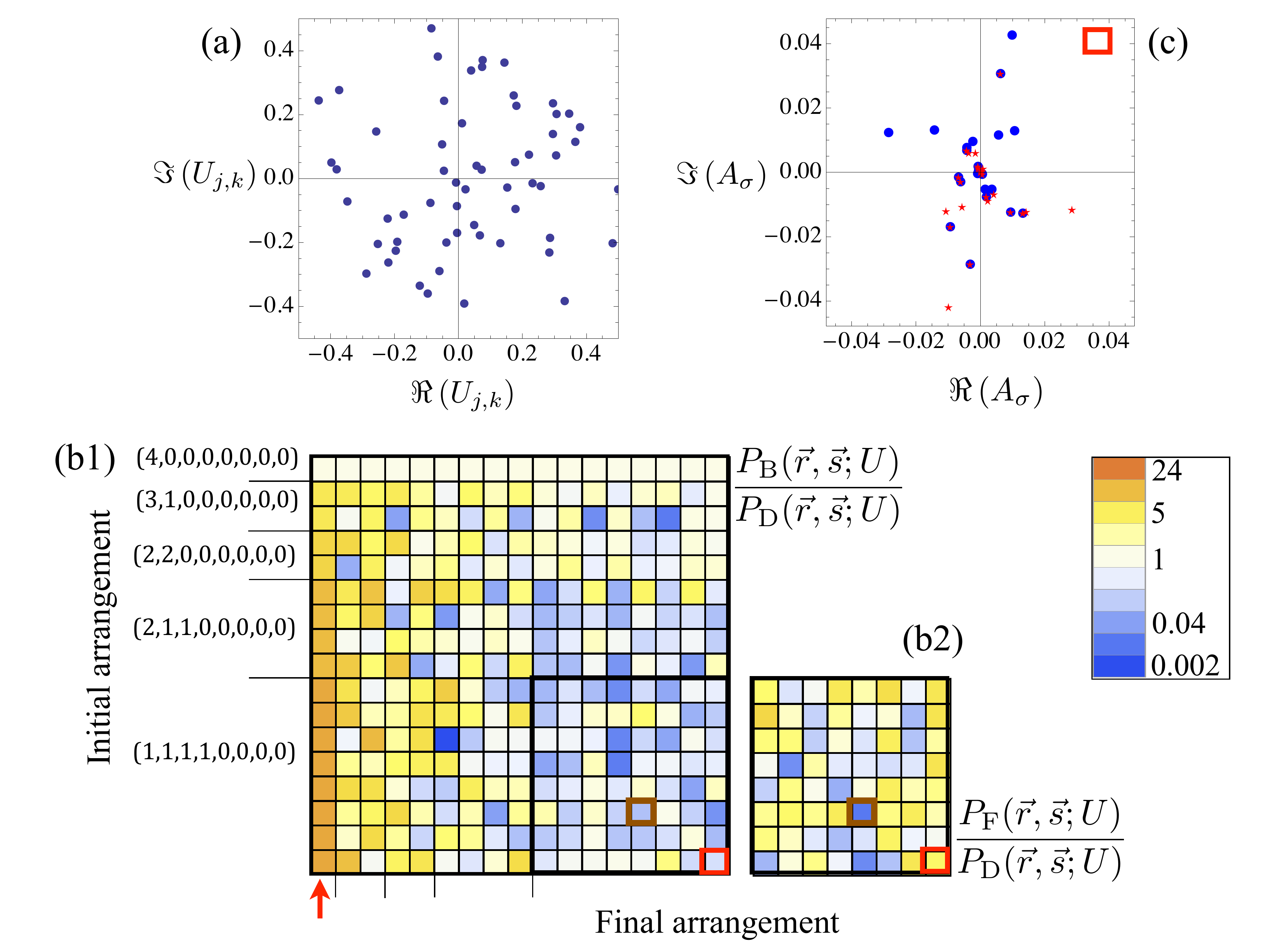}
\caption{Scattering of $N=4$ fermions and bosons on an $n=8$-mode multiport beam splitter. (a) The scattering matrix $U$ is chosen randomly, we show the 64 amplitudes in the complex plane. (b) The transition probability for bosons (b1) and fermions (b2), normalised to the probability for distinguishable particles. The color code indicates whether constructive (reddish colors) or destructive (blueish colors) interference takes place. The arrangements are ordered according to their occupancy, such that high-occupancy final (initial) arrangements are on the left (upper) part of the panel.  We choose 17 out of the 330 possible arrangements, and list one representative for each occupation number on the vertical axis. The other rows in the same grouping are assigned to permutations of the respective list $\vec r$. The order of the arrangements is the same for the final arrangements (horizontal axis). The column denoted with a red arrow represents the bunching final event $\vec s=(4,0, \dots ,0)$, which is gradually enhanced the more the bosons are initially spread out over the modes. For fermions [small panel (b2)], only  Pauli arrangements  can be realised. The interference pattern for fermions can be compared to the black-framed part of the bosonic pattern. The brown-framed transition is an example for a process that is suppressed for bosons as well as for fermions. 
(c) The 24 many-particle amplitudes $A_\sigma$ for the red-highlighted transition in the right lower corner of the interference patterns in (b) are shown in the complex plane; blue circles denote bosonic, red stars fermionic amplitudes. Precisely half of these amplitudes correspond to even permutations and therefore coincide for the two species. We adapt here the visualisation used in \cite{TichyDiss,Tichy:2012NJP}.  }  \label{RandomMatrix.pdf}
 \end{figure}

Before proceeding to a more formal discussion of Eqs.~(\ref{generalD}),(\ref{generalampliB}),(\ref{generalampliF}), it is instructive to obtain some physical intuition with the help of an example with few bosons and fermions, which can be compared to distinguishable particles. For moderate particle numbers $N\le 10$, the sums over all permutations in the above expressions can be evaluated by brute force. We randomly choose  an $n=8$-mode setup out of the unitary Haar ensemble \cite{Zyczkowski:1994uq}, the 64 appearing amplitudes are shown in the complex plane in Fig.~\ref{RandomMatrix.pdf}(a). The transition probabilities for $N=4$ bosons and fermions, normalised to the probability for distinguishable particles,  are shown in Fig.~\ref{RandomMatrix.pdf}(b1,b2).

\subsubsection{Coarse-grained statistical signatures} \label{statisticaleffects}
A global bunching tendency typifies bosons: Events with many particles per mode [on the left-hand-side of (b1)] are privileged, Pauli arrangements [on the right] are rather suppressed. This stands in contrast to the enhancement of all allowed events for fermions: Since any events other than Pauli arrangements (with at most one particle per mode) are excluded for fermions, the average probability of the remaining allowed states is enhanced with respect to distinguishable particles. The average enhancement observed for multiply occupied bosonic states is also a statistical effect: Events that are related to each other by permutations of the output modes are counted several times for distinguishable particles, which favours events with many occupied modes, such as $(1,1,1,1,0,0,0,0)$. For bosons, the mode occupation list also fixes the state, and all states  are a priori equally likely. 

For Pauli arrangements, the difference between the average probability for bosons and fermions dies out when the density of particles $N/n$ is decreased: In the dilute limit $N/n \rightarrow 0$, bunched events with several particles per mode are a priori very unlikely, and their influence on the statistics vanishes \cite{TichyDiss,Aaronson:2011kx}.

For fully bunched events for which one mode receives all particles, $\vec s_b=(0,\dots, 0,N,0 \dots, 0)$, we can formulate an exact relation between the probabilities for distinguishable particles and bosons \cite{TichyDiss,PhysRevA.83.062307},
\eq P_{\text{B}}(\vec r, \vec s_{\text{b}};U) =  \frac{N!}{\prod_{j} r_j !}  P_{\text{D}}(\vec r, \vec s_{\text{b}}; U)  \ge P_{\text{D}}(\vec r, \vec s_{\text{b}}; U) , \label{bunchedfinalstates} \en
 which was recently verified experimentally with photons  \cite{PhysRevLett.111.130503}. 
The probability for a bunched final state is always enhanced for bosons, the more the more broadly the bosons are initially distributed: The factor $\prod_j r_j!$ is then small, such that the \emph{bosonic boost} $N!$ is not jeopardised. This effect can be observed on the left-most column of Fig.~\ref{RandomMatrix.pdf}(b1) [red arrow], where the gradual enhancement of the final arrangement $\vec s_b=(4,0 \dots, 0)$ is apparent. 

Physically speaking, a bunched final state $\vec s_{\text{b}}$ is only fed by one unique many-particle path, since each  particle needs to end in the same output mode. This unique many-particle path is enhanced by the bosonic bunching factor $\nicefrac{N!}{\prod_j r_j!}$, owing to the privilege of highly-occupied states in state space. Since no interference between different paths takes place, the enhancement is independent of the phases of $U$, and can be qualified as a purely \emph{kinematic} effect. 

On the other hand, if all bosons are prepared in the same mode [see the first row of Fig.~\ref{RandomMatrix.pdf}(b1)], the probabilities for bosons and distinguishable particles do not differ at all \cite{Brendel:1988ve,Lange:1988qf,TichyDiss},  
\eq 
P_{\text{B}}(\vec r_b=(0, \dots, 0,N, 0, \dots, 0), \vec s;U) = P_{\text{D}}(\vec r_b=(0, \dots, 0,N, 0, \dots, 0), \vec s;U) .
\en
Here, no interference can take place, because all particles start in the very same initial state, which gives all many-particle paths always the same phase. The bosonic kinematic factors $s_j!$ in (\ref{generalampliB}), which seem to privilege final states with many particles per mode, are then exactly cancelled by the combinatorial factors that, in turn, privilege final states with spread-out distributions, which can be reached via many different many-particle paths.

\subsubsection{Fine-grained granular interference}
Having obtained an understanding of the \emph{statistical} average behaviour of interfering bosons and fermions [panels (b1) and (b2)], the question arises to which extent predictions and systematic statements can be made on the level of \emph{individual} events. The complicated structure in Fig.~\ref{RandomMatrix.pdf}(b1,b2) is  discouraging, and we quickly see that an individual transition can be suppressed for bosons and fermions simultaneously, such that no clearly bosonic or fermionic signal can be identified on the granular level of a single transition $\vec r \rightarrow \vec s$. Considering the red-framed transitions in (b1) and (b2), we see why fermionic and bosonic interferences are not necessarily correlated or anti-correlated: The resulting many-body amplitudes are shown in panel (c) as blue circles for bosons and red stars for fermions. Half of the amplitudes correspond to even permutations $\sigma$, they therefore coincide for bosons and fermions. The other half acquires a factor $(-1)$ for fermions, but not for bosons. For the red-highlighted transition in (b1) and (b2), the fermionic amplitudes interfere constructively, while the bosonic ones interfere rather destructively, due to their more isotropic distribution in the complex plane. Depending on the matrix elements of $U$, however, bosonic and fermionic interference can also be  similar,  e.g., the brown-framed transition in (b1) and (b2) is suppressed for bosons as well as for fermions. That is to say, \emph{interference} is not bound to the \emph{statistics} of the particles, but may act in concert on both species; the clear boson-fermion dichotomy known from the two-particle case breaks down. No systematic prediction can be put forward for an individual bosonic transition knowing the fermionic probability or vice-versa. This can be cast more quantitatively by the correlations between fermionic and bosonic probabilities: While for $N=2$, the fermionic and bosonic suppression/enhancement with respect to distinguishable particles are perfectly anti-correlated, this anti-correlation decreases with increasing particle number $N$ \cite{TichyDiss}. Moreover, when $N/n$, the average number of particles per mode, decreases,  \emph{statistical} effects fade away, while the interference remains: The \emph{average} probabilities for Pauli arrangements for bosons and fermions converge to the value for distinguishable particles in the limit $N/n \rightarrow 0$, since most possible arrangements are then Pauli arrangements \cite{Tichy:2012NJP}. However, a given individual  transition $\vec r \rightarrow \vec s$ can still be strongly enhanced or suppressed for bosons and fermions in comparison to distinguishable particles. 

\subsection{Determinants, permanents and computational complexity} \label{compcomplDet}
The brute-force evaluation of the probabilities in Eqs.~(\ref{generalD},\ref{generalampliB},\ref{generalampliF}) becomes prohibitively difficult for large $N \gtrsim 15$, since up to $N!$ amplitudes need to be summed. For distinguishable particles and fermions, the problem can be simplified considerably, while the complexity in the case of bosons remains unconquerable, in general.  In order to find a more compact expression for the probabilities, a characteristic $N\times N$-matrix  can be defined as 
\eq M_{j,k}=U_{d_j(\vec r),d_k(\vec s)} , \label{definM} \en 
i.e.~$M$ contains those rows and columns of $U$ that correspond to (possibly multiply) occupied in- and output modes, respectively. With $M$ defined as such, we can write 
\eq 
P_{\textrm{D}} (\vec r, \vec s; U) &=& \frac{1}{\prod_{j=1}^n s_j!} \sum_{\sigma \in S(\{1, \dots, N\}) } \prod_{k=1}^N |M_{k,\sigma(k)}|^2 \nonumber \\
&=& \frac{1}{\prod_{j=1}^n s_j!} \textrm{perm}(| M|^2) , \label{distpermdd}
\en
where the function $\textrm{perm}(|M|^2)$ is the \emph{permanent} of the matrix $|M|^2$ \cite{Ryser:1963oa,Scheel:2004uq,Sachkov:2002fk}, for which the absolute-square is taken component-wise. 
For bosons, we find 
\eq
P_{\textrm{B}}(\vec r, \vec s; U) &=& \frac{1}{\prod_{j} r_j! s_j!} \left| \textrm{perm}(M) \right|^2  \label{bosontranspermurep} , 
\en
i.e.~the permanent of the \emph{complex} matrix $M$ needs to be computed before taking the absolute-square to obtain the probability. For fermions, the sign-change allows us to identify the total amplitude as a \emph{determinant}, 
\eq 
P_{\textrm{F}}(\vec r, \vec s; U ) &=& | \textrm{det} (M) |^2 \label{determinantferm} .
\en
Although these three expressions seem formally similar, the computational expenses required for their evaluation differ substantially: The determinant in (\ref{determinantferm}) obeys several algebraic rules and symmetries, e.g. the product rule $\det(A \cdot B)= \det(A)\cdot \det(B)$ \cite{Bernstein:2009uq}. Exploiting these rules, the evaluation of the determinant can be performed in time polynomial in the matrix size $N$, e.g.~the elementary Gaussian algorithm needs $N^3$ operations \cite{Bernstein:2009uq}. 

For distinguishable particles and bosons, the computational tasks in (\ref{distpermdd}) and (\ref{bosontranspermurep}) amount  to calculating the permanent of the real non-negative matrix $|M|^2$ and of the complex matrix $M$, respectively. The permanent has a  similar structural definition as compared to the determinant, but, due 
 to the omission of the sign function $\textrm{sgn}(\sigma)$ in Eq.~(\ref{bosontranspermurep}), all known strategies for an efficient evaluation fail. For example, the determinant product rule does not possess any analogy for permanents. For general matrices, the evaluation of the permanent scales exponentially with the matrix size even using Ryser's algorithm, the best known to date  \cite{Ryser:1963oa}, which poses a serious challenge to any classical computer already beyond a seemingly moderate matrix size of $25 \times 25$. 

It is extremely unlikely that a polynomial-time algorithm exists for the permanent, since the permanent of a ``01-matrix'' that only contains 0 and 1 as entries encodes the number of solutions to a well-defined paradigmatic problem (perfect bipartite matchings in a graph) \cite{Valiant:1979fk}. This relation promotes the permanent of 01-matrices to the complexity class $\# P$-complete \cite{Valiant:1979fk,Aaronson:2013kx}, which is believed to be unconquerable by polynomial-time algorithms. By recursion to results in linear optics quantum computing,  an alternative proof for the complexity of the permanent was recently put forward in \cite{Aaronson:2011uq}. 

For distinguishable particles, the matrix $|M_{j,k}|^2$ 
contains only non-negative entries. For every permutation $\sigma$, the product $\prod_{j=1}^N  |M_{j,\sigma(j)}|^2$ defines by itself a lower bound to the total permanent, since all summands in Eq.~(\ref{generalD}) are non-negative. By sampling over many randomly chosen permutations $\sigma$, the permanent can be estimated, as formalised by the efficient randomised  algorithm described in \cite{Jerrum:2004:PAA:1008731.1008738}. 
This sampling strategy, however, breaks down completely when a complex matrix $M_{j,k}$ is considered, for which interference between pathways makes the sum in Eq.~(\ref{generalampliB}) much less predictable than for non-negative summands, Eq. (\ref{generalD}).

\subsection{Boson-Sampling} \label{secBosonSampling}
It is tempting to state that quantum mechanics makes it possible to \emph{``compute the permanent of a complex matrix with exponential speedup with respect to classical computers''} using bosons that propagate through a multimode setup \cite{PhysRevA.88.013806,Yirka2013}. However, even though the amplitudes of final states characterised by $\vec s$ correspond to the permanent of a certain matrix $M$, these amplitudes are very small, which makes them difficult to extract. Although it is possible to define an observable whose expectation value coincides with  the permanent, its large variance requires an exponential number of measurements to extract the permanent experimentally  \cite{Troyansky:1996ve}. This caveat jeopardises any strategy to directly compute the permanent via the scattering of bosons, although there are current attempts in this direction \cite{Marzuoli}.

\begin{figure}[ht] \center
\includegraphics[width=.7\linewidth,angle=0]{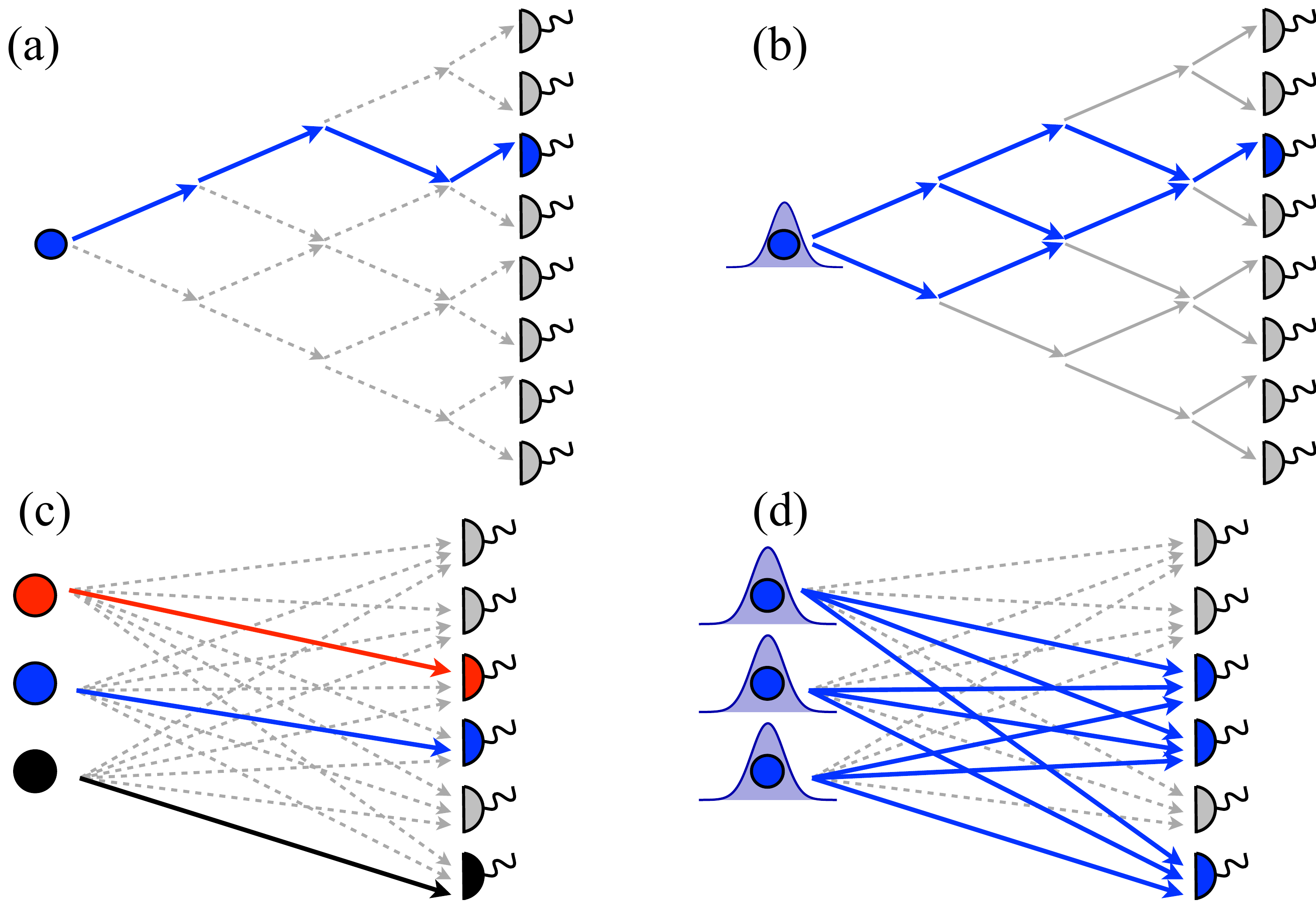}
\caption{Why is Boson-Sampling hard? Simulation of classical and quantum [left and right]  processes with single and many particles  [top and bottom]. (a) A single classical random walker can be simulated  with one path at a time, choosing the next destination according to the graph properties.  (b) A quantum walker requires higher expenditure, since interference needs to be incorporated. Here, the Hilbert-space dimension increases only linearly in the number of steps. (c) The independent stochastic behaviour of many distinguishable particles can be simulated in a Monte Carlo approach, treating one particle at a time. (d) The expenditure required to incorporate the interference of bosons is qualitatively higher, since $N!$ paths need to be taken into account to obtain a single transition probability. The figure is adapted from \cite{Tichy:2013lq}. }  \label{WhyBSisHard.pdf}
 \end{figure}

The complexity of the permanent hinders the calculation or the approximation of an individual transition probability $P_{\text{B}}(\vec r, \vec s;U)$ for typical unitary matrices $U$ drawn according to the Haar measure \cite{Zyczkowski:1994uq}. This complexity is inherited by the computational problem of merely \emph{simulating} a machine in which many bosons interfere \cite{Aaronson:2011kx}, i.e.~generating output events $\vec s$ according to the bosonic probability (\ref{bosontranspermurep}). This hardness holds as long as the number of bosons remains much smaller than the number of modes $N \ll n$. In the other limit, $N \gg n$, semi-classical methods become efficient \cite{Shchesnovich:2013uq,TichyDiss}. 

The difficulty of barely \emph{simulating} the scattering of many bosons, i.e.~producing one event at a time, can be appreciated by analogy to distinguishable particles, and classical and quantum single-particle processes, as sketched in Fig.~\ref{WhyBSisHard.pdf}. In (a), a single random walker on a graph can be simulated via a Monte Carlo approach, such that the next step of the walker is chosen stochastically. Only one path needs to be evaluated at a time. A quantum walker in (b),  however, requires a more holistic approach, since its wave-like nature lets it take several paths simultaneously. The \emph{interference} of all contributing paths needs to be taken into account, which increases the computational expenses. Since the Hilbert-space is typically not exceedingly large (here its dimension is proportional to the number of steps taken), the expenditure  required to simulate a quantum walk is not dramatically greater than for the classical analogy. In the many-particle realm, the argument can be repeated: For distinguishable particles sketched in (c), a Monte Carlo approach in which the output mode $k$ of each particle prepared at the input mode $j$ is  chosen according to the probability $p_{j,k}$ allows one to efficiently generate many-particle events, such that the  mode assignment list of the occurring event is constructed step-by-step-wise. This is possible because the particles are not only non-interacting, but also \emph{independent}, such that each particle can be treated separately (remember the discussion in Section \ref{twopinterSec}) \cite{Aaronson:2013ls}. This strategy makes the actual computation of the permanent in (\ref{distpermdd}) unnecessary, i.e.~the efficient simulability of randomly distributed distinguishable particles does \emph{not} rely on the efficient approximation of (\ref{distpermdd}) by a polynomial-time algorithm \footnote{It is rather the other way around, the polynomial-time algorithm in \cite{Jerrum:2004:PAA:1008731.1008738} is based on a strategy similar to the Monte Carlo approach devised here.}. For many indistinguishable particles, however, such Monte Carlo approach fails,  again, because each event is governed by the \emph{interference} of up to $N!$ many-particle paths that feed it, as sketched in (d). The computational resources that are required to  simulate such process scale \emph{exponentially} with the number of bosons \cite{Gard:2013fk}, since the computation of the full probability distribution is necessary for a simulation \cite{Spagnolo:2013eu}. 

In the language of computational complexity theory, the simulation of many-boson scattering can be re-formulated rigorously as the computational problem to \emph{sample} from a probability distribution in which events $\vec s$ occur with probability $P_{\text{B}}(\vec r, \vec s; U)$, motivating the term \emph{Boson-Sampling} \cite{Aaronson:2011kx}. An algorithm that efficiently solves Boson-Sampling on a classical computer would imply very unlikely reductions in computational complexity theory, which  substantiates the widely shared belief that such efficient simulation is impossible \cite{Aaronson:2011kx}. Remarkably, the analogous problem for fermions, even though it involves the superposition of $N!$ paths just as for bosons, is efficiently simulable, thanks to the benevolent behaviour of the determinant \cite{Aaronson:2013ls}, which emphasises that the \emph{physical} argument for complexity that we have given here cannot replace a rigorous proof \cite{Aaronson:2011kx,Gard:2013fk,Aaronson:2013ls}.

On the one hand, this well-established, rather fundamental restriction rules out that the scattering of bosons be ever simulated reliably and efficiently for more than a dozen particles on a classical computer. It also clearly jeopardises further attempts to characterise the interference of truly many bosons beyond our discussion in Section \ref{interferencepattern}. On the other hand, the difficulties that a classical computer faces upon the simulation of bosons are tantamount for the power that is inherent to any device in which such physical process takes place:  A many-boson interference machine that performs Boson-Sampling is a  powerful quantum computer -- albeit being restricted to one rather abstract  problem without any known application \cite{Ralph:2013xz,Aaronson:2011kx}.

The formidable complexity of Boson-Sampling makes it a serious candidate for a test of the supremacy of quantum devices over classical computers. In comparison to the paradigm of quantum computing, factoring, the simulation of many bosons is \emph{harder} in terms of computational complexity theory, while it may be performed with significantly \emph{less resources} than required for a universal quantum computer \cite{Aaronson:2011kx}. A demonstration of Boson-Sampling in a regime unattainable by classical computers would therefore provide an evident attack on the extended Church-Turing thesis \footnote{The extended Church-Turing thesis states that every computation that can be performed efficiently by a physical device can also be performed efficiently by a classical computer \cite{Aaronson:2013kx}. It is a widely shared assumption among theoretical computer scientists.}. This perspective has ignited further research concerning the scalability of Boson-Sampling to more interfering particles \cite{Rohde:2012ly,Motes:2013ys,Shen:2013zr,Shchesnovich2013,Motes:2014vs}, the generalisation to similar sampling problems with Gaussian states \cite{Lund:2013eb} and entangled states \cite{Rhode:2013vn}. Due to the absence of error-correction in Boson-Sampling, all possible errors, i.e.~in the state preparation and detection \cite{Shchesnovich:2014fz}, in the scattering matrix \cite{Leverrier:2013ys}, and in the partial boson distinguishability \cite{PhysRevA.85.022332,Shchesnovich2013} need to scale appropriately with the system size $N$ to allow a substantial attack on the extended Church-Turing thesis \cite{Rohde:2014uz,Shchesnovich:2014fz}. Boson-Sampling experiments were conducted for three photons that scatter off engineered multimode devices with five or six modes \cite{Broome15022013,Crespi:2012vn,Spring15022013,Tillmann:2012ys}; the outcomes agreed with the theoretical predictions, which are still unproblematic to obtain in this regime.

\subsection{Can we trust a Boson-Sampling device?}
Even if the formidable technical challenges regarding the scalability of concerted photon-sources, unavoidable errors and detector inefficiencies were resolved, a rather fundamental problem remains for a reliable challenge to the extended Church-Turing thesis: the \emph{certification} of an alleged Boson-Sampling device. In the regime of particle numbers $N$ and mode numbers $n$ in which a quantum Boson-Sampling device widely outperforms classical computers, it also becomes unfeasible for the latter to certify by brute force that the quantum device reliably performs the desired sampling operation. This lack of assessability is due to Boson-Sampling being very probably not contained in the complexity class $NP$: Not only is it hard to \emph{simulate} the scattering of many non-interacting bosons, but it also seems unfeasible to reliably \emph{verify} whether a set of sampled events solves the desired  sampling problem, in the limit of large particle numbers  \cite{Aaronson:2011kx,gogolin2013}. A rigorous \emph{proof} for the correct functioning of a Boson-Sampler is therefore out of reach \cite{Aaronson:2013ls}, and it remains a  desideratum to devise methods that yield reasonably convincing evidence that an alleged Boson-Sampling device is trustworthy, such that it can be expected to solve Boson-Sampling with a high degree of confidence \cite{Carolan:2013mj}. 

Under the  strong \emph{symmetric} assumption that no information on the scattering matrix and on the form of the occurring events is used for certification, the resulting distribution of events generated by a Boson-Sampler cannot be distinguished from a uniform distribution (in which every event $\vec s$ is assigned the same probability) after a polynomial number of trials \cite{gogolin2013}. In practice, when the scattering matrix as well as the initial state are  known to the experimenter,  predictions on certain observables are possible, which permits the desired discrimination \cite{Aaronson:2013ls}.  Consider, for example, the average particle number $\langle \hat n_j \rangle$ for output mode $j$, for which the expectation value assumes
\eq 
\langle \hat n_j \rangle  = \sum_{k=1}^N p_{d_k(\vec r),j} , 
\en
where no computation of the permanent is necessary. Based on this observation, the outcome of a Boson-Sampling device that uses a known scattering matrix can be discriminated from the uniform distribution of events \cite{Aaronson:2013ls}: When an event $\vec s$ is recorded, one can immediately compute the value  \cite{Spagnolo:2013eu}
\eq 
\mathcal{P}=\prod_{k=1}^N  \left( \sum_{j=1}^N p_{d(\vec r)_k, d(\vec s)_j} \right) .
\en
If $\mathcal{P}>(N/n)^N$,  one  concludes that the event comes from a Boson-Sampling distribution, otherwise it is assumed to be drawn from the uniform distribution. The experiment is repeated many times, and majority voting is applied to reach a conclusion. This certification of a Boson-Sampling device has been realised experimentally for three photons in a 7-mode as well as in a 9-mode interferometer \cite{Spagnolo:2013eu}.

The drawback of this test is that it also  validates the output of a fully classical sampler that uses distinguishable particles (which can be simulated efficiently, see Section \ref{secBosonSampling} above), or even a sampler that uses fermions \cite{Aaronson:2013ls}. In other words, although some evidence can be obtained that the observed distribution is \emph{compatible} with the one expected for bosons, the characteristic \emph{bosonic} features of the distribution cannot be verified in this manner: Single-particle observables -- as the average number of particles in a mode -- do not witness many-particle interference, in accordance with Dirac's dictum that \emph{``each photon interferes only with itself''} \cite{Dirac:1930vn}, which we discussed for two particles in Section \ref{twoidpa} above. 

One therefore needs to go beyond single-particle observables in order to become sensitive to bosonic  effects. An immediate test is to measure the \emph{correlations} between two output modes, which  exhibit bosonic signals \cite{mayerda,PhysRevA.83.062307}. These correlations are feasible to compute, but they only witness \emph{two}-body interference, which, again, does not prove the genuine interference of $N$ bosons.  Predicting the full $N$-point correlations, however, is equivalent to solving the original Boson-Sampling problem \cite{mayerda}, which entails a dilemma: Computing $M$-body correlations entails a trade-off between the computability of these correlations, unfeasible for large $M$, and the actual stringency of the test, which is only ensured for large $M \approx N$. 

An alternative approach to certifying Boson-Sampling devices is to leave the space of computationally complex problems: Instead of choosing the scattering matrix randomly according to the unitary Haar-measure (which is a crucial assumption for the very hardness of the problem \cite{Aaronson:2011kx}), one can artificially design test-cases for which pertinent properties of the resulting probability distribution are obtained efficiently on a classical computer. In other words,  the Boson-Sampling device is assessed by a task whose exact or approximate solution is known beforehand. The most immediate benchmark is an implementation of the identity matrix $\hat \mathbbm{1}_n$ as a scattering matrix, for which $P_{\text{B/D/F}}(\vec r, \vec s; \mathbbm{1}_n)=\delta_{\vec r, \vec s}$  \cite{Gogolin:2013eu}. This trivial implementation, however, is certainly not satisfactory: Each incoming particle is deterministically redirected to one output mode, and no interference between distinct paths takes place; bosons, fermions and distinguishable particles exhibit the same trivial distribution. 
Therefore, the test-case needs to be physically non-trivial -- in order to assess whether interference indeed takes place -- while it should remain treatable mathematically. 

One approach is based on processes with well-understood average behaviour, such as many-particle quantum walks \cite{PhysRevA.83.062307,mayerda,Gard:13}, for which the observed patterns of distinguishable and indistinguishable particles differ in a predictable way: Bosons that perform a quantum walk  exhibit \emph{bosonic clouding}, which degrades with increased mutual distinguishability of the particles \cite{Carolan:2013mj}. Clouding quantifies to which extent particles are found in the same spatial region at the output, i.e.~how often all particles exit through the same half of the interferometer. Bosons exhibit a larger degree of clouding than distinguishable particles, the question whether this phenomenon can be attributed uniquely to their bunching tendency remains open \cite{MeineckePrivate}. The criterion is based on the \emph{average} behaviour of bosons that propagate in a quantum-walk geometry, i.e.~a coarse-grained statistical observable is used for certification. However, an efficient semi-classical sampling-model \footnote{In the semi-classical model, the destination of each particle is chosen according to a probability distribution that changes from run to run, which formally corresponds to the mean-field limit of many bosons prepared in a Fock-state \cite{Laloe:2010uq,Cennini:2005th,Hadzibabic:2004fk}.} is also validated by any test based on bosonic bunching or clouding \cite{Tichy:2013lq}. In other words, although  bosonic statistical effects  witness features that go beyond the uncorrelated behaviour of distinguishable particles, the observation of these effects is nevertheless insufficient to strictly guarantee a behaviour in accordance to Eq.~(\ref{bosontranspermurep}). Using Fig.~\ref{RandomMatrix.pdf}(b1) to visualise the present argument, the essence of Boson-Sampling does not lie in the average statistical bunching tendency of bosons (the overall red-blue gradient, which is also reproduced by incoherent mixtures of bosons \cite{Tichy:2012NJP} and particles described in a semiclassical approximation \cite{Tichy:2013lq}), but in the granular fluctuations that make each individual event hard to predict. 

A certification protocol based on an efficiently evaluable granular criterion that strictly forbids certain individual well-specified events to occur therefore entails a significantly higher degree of confidence that an alleged Boson-Sampling device exhibits \emph{precisely} the behaviour described by Eq.~(\ref{bosontranspermurep}). Such certification protocol is described  in the following. 

\subsection{A falsifiable instance of Boson-Sampling: the Fourier suppression law}
We argued in Section \ref{compcomplDet} that the superior computational complexity of many-boson scattering in comparison to the analogous problem with fermions is rooted in the absence of symmetries of the permanent  compared to the more benevolent determinant. Our strategy to devise a solvable instance of Boson-Sampling therefore relies on \emph{imposing} artificial symmetries on the scattering matrix $U$, which then allows us to exactly compute the permanent of the pertinent submatrices (\ref{definM}) in a large number of cases. 

In the first place, in order to generate many interfering paths of equal weight, an \emph{unbiased} matrix is chosen, for which all single-particle probabilities are equal, \eq p_{k,l}= |U_{k,l}|^2=\frac 1 n .  \label{classicunbiased} \en 
As a consequence, Eq.~(\ref{generalD}) collapses to a purely combinatorial expression for distinguishable particles: 
\eq 
P_{\text{D}}(\vec r, \vec s; U) = \frac{N!}{n^N \prod_{j=1}^n s_j!} ,  \label{distsimple} 
\en
{\it i.e.}~events occur with probabilities according to a \emph{multinomial distribution} \cite{Ryser:1963oa}, independently of the preparation of the particles $\vec r$, which generalises the binomial distribution (\ref{distprob3}) found for a two-mode beam splitter. 

A \emph{unitary} scattering matrix that fulfils (\ref{classicunbiased}) is called a \emph{complex Hadamard matrix}  \cite{Tadej:2006it}, which can have many different forms. We focus here  on the discrete Fourier transformation between the input and output modes, 
\eq 
\left( U_n^{\text{Fou}} \right)_{j,k}= \frac{1}{\sqrt n} e^{i \frac{2 \pi}{n} j k} \label{foumadef} .
\en
A particle prepared in input mode $j$ thus acquires a phase $ 2\pi j k/n$ when it leaves the setup through the $k$th output mode. The product of such individual single-particle phases always remains a multiple of  $2\pi /n$. Moreover, the symmetries of the Fourier-matrix are translated into a symmetry that relates a transition $\vec r \rightarrow \vec s$ to its time-mirrored inverse, $\vec s \rightarrow \vec r$:
\eq 
P_{\text{B/F}}(\vec r,\vec s; U^{\text{Fou}}_n) = P_{\text{B/F}}(\vec s, \vec r; U^{\text{Fou}}_n) \label{nonsymm} ,
 \en
 which has no analogy for distinguishable particles.

When combined with symmetries of the initial or final states, boson scattering on the Fourier matrix permits the efficient prediction of event probabilities. To better classify symmetries on states, an arrangement $\vec q$ of $N$ particles in $n$ modes is called $m$-\emph{periodic} (for $m$ that divides $n$) when the mode occupation list $\vec q$ can be written as a concatenation of $p=\nicefrac n m$ elementary lists $\vec k$ of length $m$, such that $\sum_{j=1}^{m} k_j=N/p$. Thus, we formally have 
\eq \vec q=( \underbrace{
\underbrace{k_1, k_2, \dots, k_{m}}, \underbrace {k_1, \dots, k_{m}}, \dots, \underbrace {k_1, \dots, k_{m}}}_{p=\nicefrac n m} ) , \label{defiperiodicstate} \en
and the mode assignment list $\vec d(\vec q)$ satisfies
\eq \forall j: d_{j+\frac N p}(\vec q)= d_{j}(\vec q)+m ~, \label{propper} \en
where we identify $d_{N+j} \equiv d_{j}$ and $d_j \equiv  d_j+n$. The list $(1,1,1,1,1,1,1)$ is $1-$periodic ($p=n=7$); $(1,1,0,1,1,0,1,1,0)$ is 3-periodic ($p=3$), while $(1,0,1,0,0,1,0,0,0)$ has no apparent period, since the pattern of length $m=n=9$ is only repeated once ($p=1$). 

\subsubsection{Derivation of the Fourier suppression law for bosons} \label{derivation}
\begin{figure}[ht] \center
\includegraphics[width=.85\linewidth,angle=0]{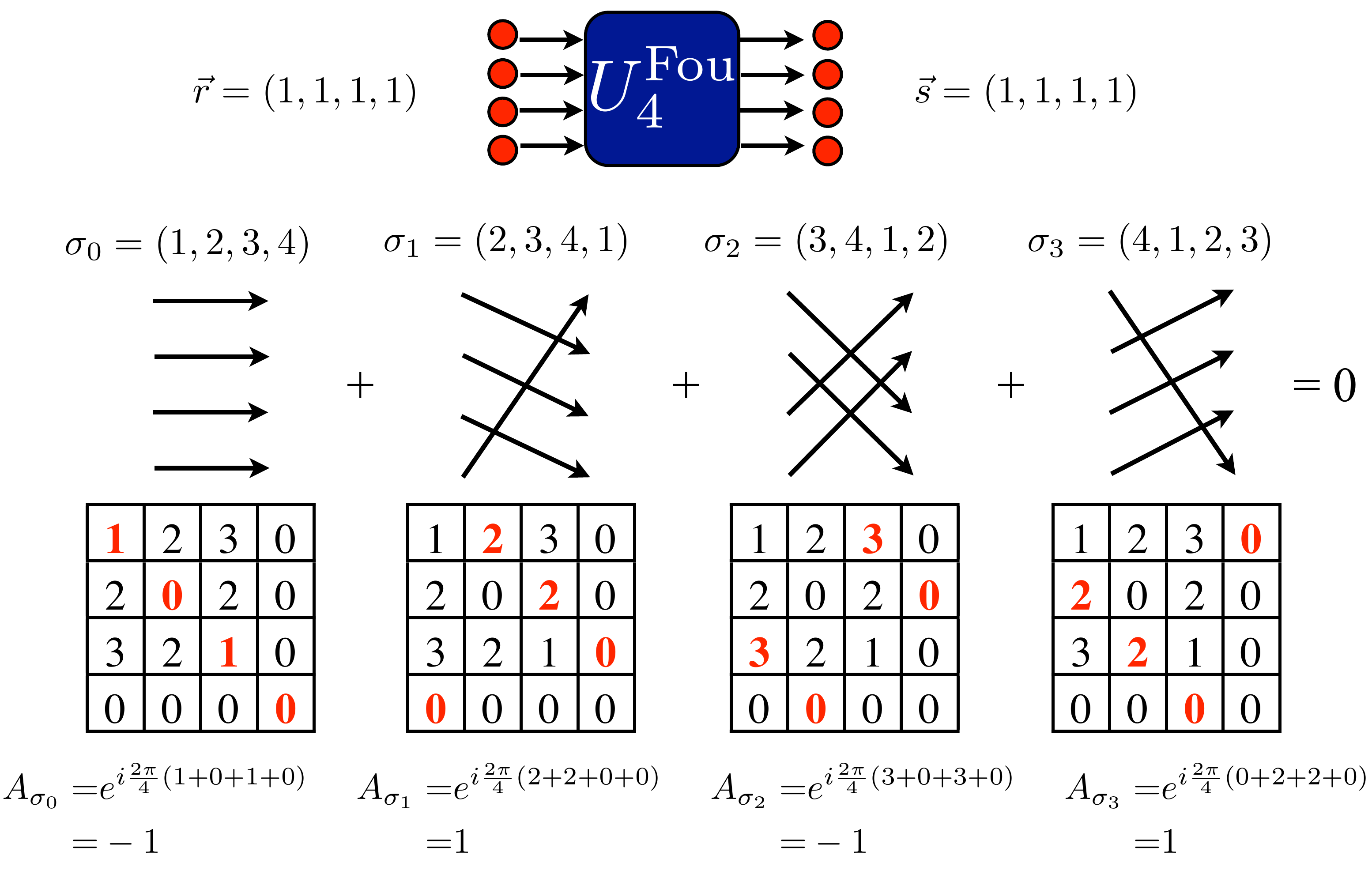}
\caption{Suppression law for the transition $\vec r=\vec s=(1,1,1,1)$ ($N=n=4$). The four permutations $(\sigma_0, \dots, \sigma_3)$ correspond  to four different physical processes in which the incoming particles exit through different output modes, always fulfilling $\vec s=(1,1,1,1)$. The associated many-particle amplitudes can be computed by summing up the corresponding fractions of $ 2 \pi/4$ in the matrices, marked in thick red. Since $Q=1+2+3+4~\text{mod }~4=2$,  the phase that is acquired upon the application of $\Gamma$ is $2\pi/4 Q=\pi$ [see Eq.~(\ref{Qacqui})], and the resulting amplitudes $A_{\sigma_0}\dots A_{\sigma_3}$ cancel.}  \label{SuppressionLawProof}
 \end{figure}

A simplified derivation of the Fourier suppression law for initial and final Pauli arrangements ($r_k, s_j \le 1$) is presented here; the general proof for arbitrary input and output states can be found in Ref.~\cite{Tichy:2012NJP}.  Consider an arrangement $\vec s$ that is $m$-periodic [see Eq.~(\ref{defiperiodicstate})], and an arrangement $\vec r$ with \eq Q=m \sum_{j=1}^N d_j(\vec r) , ~~~\text{mod}(Q, n)\neq 0 \label{defQ}, \en i.e.~$Q$ is not a multiple of $n$. The definition of the Fourier matrix (\ref{foumadef}) is inserted into Eq.~(\ref{generalampliB}), which gives the transition probability for bosons,
\eq 
P_{\text{B}}(\vec s, \vec r; U_n^{\text{Fou}}) =P_{\text{B}}(\vec r, \vec s; U_n^{\text{Fou}}) &=&  \label{BoseFourierApp}  
 \frac {1} {n^N} \left| \sum_{\sigma \in S_N }  \text{exp}\left( i \frac {2\pi}{n} \sum_{j=1}^N  d_j(\vec r) d_{\sigma(j)}(\vec s)  \right) \right|^2 \\
 &=:&  \frac {1} {n^N} \left| \sum_{\sigma \in S_N } A_\sigma \right|^2 \label{BoseFourierApp2}   ,
  \en
where $A_\sigma$ is the many-particle amplitude associated with the process described by the permutation $\sigma$. 
Due to the symmetry of the initial state and of the Fourier matrix, the total sum turns out to vanish. In order to show that, we group the  $N!$ permutations into sets such that each set yields a vanishing amplitude. Let us start with one permutation $\sigma_0$, which contributes the complex summand 
\eq 
A_{\sigma_0}=\text{exp}\left( i \frac {2\pi}{n} \sum_{j=1}^N d_j(\vec r) d_{\sigma_0(j)}(\vec s) \right) .
\en
A mapping between permutations $\Gamma: S_N \rightarrow S_N $ is defined to devise another permutation, $\sigma_1$ \cite{Graham:1976nx}: 
\eq
\sigma_1(k)=\Gamma[\sigma_0](k)=\left( \sigma_0(k)+\frac N p \right) \text{mod}~~ n 
\en
While $\sigma_0$ provides one particular assignment of the input particles onto the output modes, $\sigma_1=\Gamma[\sigma_0]$ shifts the destination for each incoming particle. 
The permutation $\sigma_1$ is connected to the summand 
\eq 
A_{\sigma_1}=\text{exp}\left( i \frac {2\pi}{n} \sum_{j=1}^N d_j(\vec r) d_{\sigma_0(j)+\frac N p}(\vec s) \right) .
\en
Using the periodicity of the final state, Eq.~(\ref{propper}), we find 
\eq
A_{\sigma_1}=\text{exp}\left( i \frac {2\pi}{n} \sum_{j=1}^N d_j(\vec r) (d_{\sigma_0(j)}(\vec s)+m) \right) = 
A_{\sigma_0} \exp\left(  i \frac {2\pi}{n}  Q  \right) .
\en
Further permutations $\sigma_2, \dots, \sigma_{p-1}$ can be constructed using $\sigma_q=\Gamma[\sigma_{q-1}]$. After $p$ applications of the operation $\Gamma$, we come back to our original permutation, $\Gamma[\sigma_{p-1}]=\sigma_0$, i.e. $\Gamma$ possesses a cyclic orbit in $S_N$. 
The assigned amplitudes can be written as 
\eq
A_{\sigma_q}= A_{\sigma_0} \exp\left(i \frac{2\pi}{n} q Q \right)  ,  \label{Qacqui}
\en
and their sum becomes 
\eq 
\sum_{k=0}^{p-1} A_{\sigma_k}  = A_{\sigma_0} \sum_{k=0}^{p-1} \exp\left( i \frac {2 \pi}{n} Q k \right) = 0 , \label{geoser}
\en
where we used  that $Q$ is not a multiple of $n$ in the last step. Therefore, $\exp\left( i \frac {2 \pi}{n} Q  \right) \neq 1$, and  the sum in (\ref{geoser}) can be identified as a truncated geometric series, which vanishes. The proof is illustrated in Fig.~\ref{SuppressionLawProof} for $N=n=4, \vec r=\vec s=(1,1,1,1)$: The four shown permutations that lie on the orbit of $\Gamma$ lead to amplitudes that exactly cancel.  

Since every permutation $\sigma \in S_N$ belongs to some peridic orbit of $\Gamma$, the total transition probability can be written as
\eq 
P_{\text{B}}(\vec r, \vec s; U_n^{\text{Fou}}) &=&  \frac {1}{n^N} \left| \sum_{\sigma \in S_N} A_\sigma \right|^2
=  \frac {1}{n^N}  \left|  \sum_{\sigma_0 \in \tilde S_N } \left(   \sum_{k=0}^{p-1} A_{\Gamma^k[\sigma_0] }  \right)  \right|^2 ,
\en
where $\tilde S_N$ is a set of $N!/p$ permutations that can not be related to each other via the repeated application of $\Gamma$ (they all lie in different orbits of $\Gamma$). Due to (\ref{geoser}), each summand in the above sum vanishes. 

Exploiting the input-output symmetry (\ref{nonsymm}), we can formalise our above observation to state the \emph{suppression law for Fourier matrices} \cite{Tichy:2012NJP,Tichy:2010ZT}, 
\eq \vec s~ m\text{-periodic,}~\text{mod}\left(m  \sum_{j=1}^N d_j(\vec r),n \right)  \neq 0 ~ \Rightarrow ~
 P_{\text{B}}(\vec r, \vec s; U_n^{\text{Fou}}) = P_{\text{B}}(\vec s, \vec r; U_n^{\text{Fou}})  =0 \label{law} . \en
That is to say, given an $m$-periodic initial state, final states are suppressed, unless the sum of the elements of the mode assignment list, multiplied by the period length $m$, can be divided by $n$. The condition in (\ref{law}) is sufficient, but not a necessary criterion for $P_{\text{B}}$ to vanish. The suppression law can also be interpreted as an artificial selection rule: The initial state possesses a symmetry, which remains conserved during the scattering process and manifests itself in the allowed final states.

\subsubsection{Scaling of the suppression law and stringent certification of Boson-Sampling}
 Few properties of the permanent of the Fourier matrix are known \cite{Graham:1976nx}, such that it remains open how to efficiently compute probabilities for bosonic events scattered on the Fourier multiport, in general. The suppression law for bosons, Eq.~(\ref{law}), however, allows us to predict the \emph{strict suppression} of final events $\vec s$ on the basis of the symmetry of the initial state and a simple combinatorial property of $\vec d(\vec s)$, which can be evaluated by pen and paper for photon numbers that are relevant today. In practice, one can efficiently predict suppressed transitions for almost arbitrarily large particle numbers -- the evaluation of the condition in (\ref{law}) remains  unproblematic for $N=10000$, whereas the brute-force computation of the related permanent is absolutely out of question. In other words, there is no known algorithm to efficiently \emph{simulate} the output of Boson-Sampling on the Fourier matrix, but one can easily \emph{verify} whether the suppression law is fulfilled by the output of an alleged Boson-Sampler -- in particular,  semi-classical behavior (inferior to true Boson-Sampling) is clearly revealed as such \cite{Tichy:2013lq}: Events that seem to be favoured by bosonic bunching  can actually be strictly suppressed, such as $\vec r=(1,1, \dots, 1)  \rightarrow \vec s=(N-1,1,0 \dots ,0)$. An overview of different certification protocols and sampling models is given in Table \ref{samplingtab}.

\begin{table}
\begin{tabular}{lcccccc} 
	Test: ~~~~~~~~~~~~~~~~~~~~~~~ Sampling model:  &  Uniform  \cite{gogolin2013} & Classical  \cite{Aaronson:2013ls} & Semi-classical  \cite{Tichy:2013lq} \\ \hline \hline 
Symmetric test \cite{gogolin2013} & {\color{red}$\times$}& {\color{red}$\times$} & {\color{red}$\times$} &  \\
Single-particle observables \cite{Aaronson:2013ls,Spagnolo:2013eu} & $\checkmark$  & {\color{red}$\times$} & {\color{red}$\times$} \\
Bosonic statistical effects \cite{Carolan:2013mj} & $\checkmark$  & $\checkmark$  & {\color{red}$\times$}\\
Fourier suppression law \cite{Tichy:2013lq} & $\checkmark$  &  $\checkmark$ & $\checkmark$  
\end{tabular}
\caption{Success and failure of different tests aimed at ruling out alternative sampling models \cite{Tichy:2013lq}. The columns denote the different samplers that ought to be discriminated from a functional Boson-Sampling device, the rows refer to tests, i.e.~to the certification protocols. Red crosses ({\color{red}$\times$}) indicate that a test protocol fails to efficiently discern the respective sampling model efficiently or erroneously validates it as functional Boson-Sampling device, while checkmarks ($\checkmark$) denote successful and efficient discrimination. }
\label{samplingtab}
\end{table}

A variety of interference effects can be understood via the suppression law. It comprises the two-particle HOM-effect discussed in  Section \ref{twoparticlecaseSec} as a special case: The initial state $\vec r=(1,1)$ has period length $m=1$, the final state $\vec s=(1,1)$ then leads to $\sum_{j=1}^2 d_j(\vec s)=3$, which is not dividable by 2; hence the transition is suppressed for bosons according to Eq.~(\ref{law}). When  the number of particles in the modes is increased, an even-odd effect for $N$ particles that are scattered on a two-mode beam splitter \cite{Campos:1989fk,Laloe:2010uq,nockdipm} arises as another instance: For an initial state of the form $\vec r=(N/2,N/2)$, the final state never  contains any component with an odd number of particles in either mode. Also the suppression of the transition $\vec r=(1, \dots, 1) \rightarrow \vec s=(1, \dots 1)$ for even $N=n$ \cite{Lim:2005qt,Graham:1976nx} can be understood via Eq.~(\ref{law}). 

The suppression law was recently verified for three photons in a three-mode setup (${N=n=3}, {\vec r=\vec s =(1,1,1)}$) \cite{Spagnolo:2013fk} and for four photons in a two-mode setup [${N=4},{ n=2}, \vec r=(2,2), {\vec s=(3,1)}$] \cite{younsikraNatComm},  details will be provided below in Section \ref{DisttransSec}. For these paradigmatic examples with small photon numbers, however, a brute-force calculation of the transition probabilities is still unproblematic, and the suppression law does not (yet) yield a crucial \emph{computational} advantage. 

The intrinsic power of the suppression law becomes apparent  when we move to larger systems, for which the permanent cannot be computed in practice. Not only do then the computational expenses for each individual transition $P_\text{B}(\vec r, \vec s; U)$ grow exponentially with $N$, but also the total number of arrangements (i.e. the number of possibilities to distribute $N$ particles within $n$ modes) explodes: The total number of events $\mathcal{N}_{\text{arr}}^{\text{B}}$ for $N$ bosons reads 
\eq \mathcal{N}_{\text{arr}}^{\text{B}}={ N+n -1 \choose N}=\frac{(N+n-1)!}{(n-1)! N!} \label{numarrbos} . \en 

\begin{figure}[ht] \center
\includegraphics[width=.5\linewidth,angle=0]{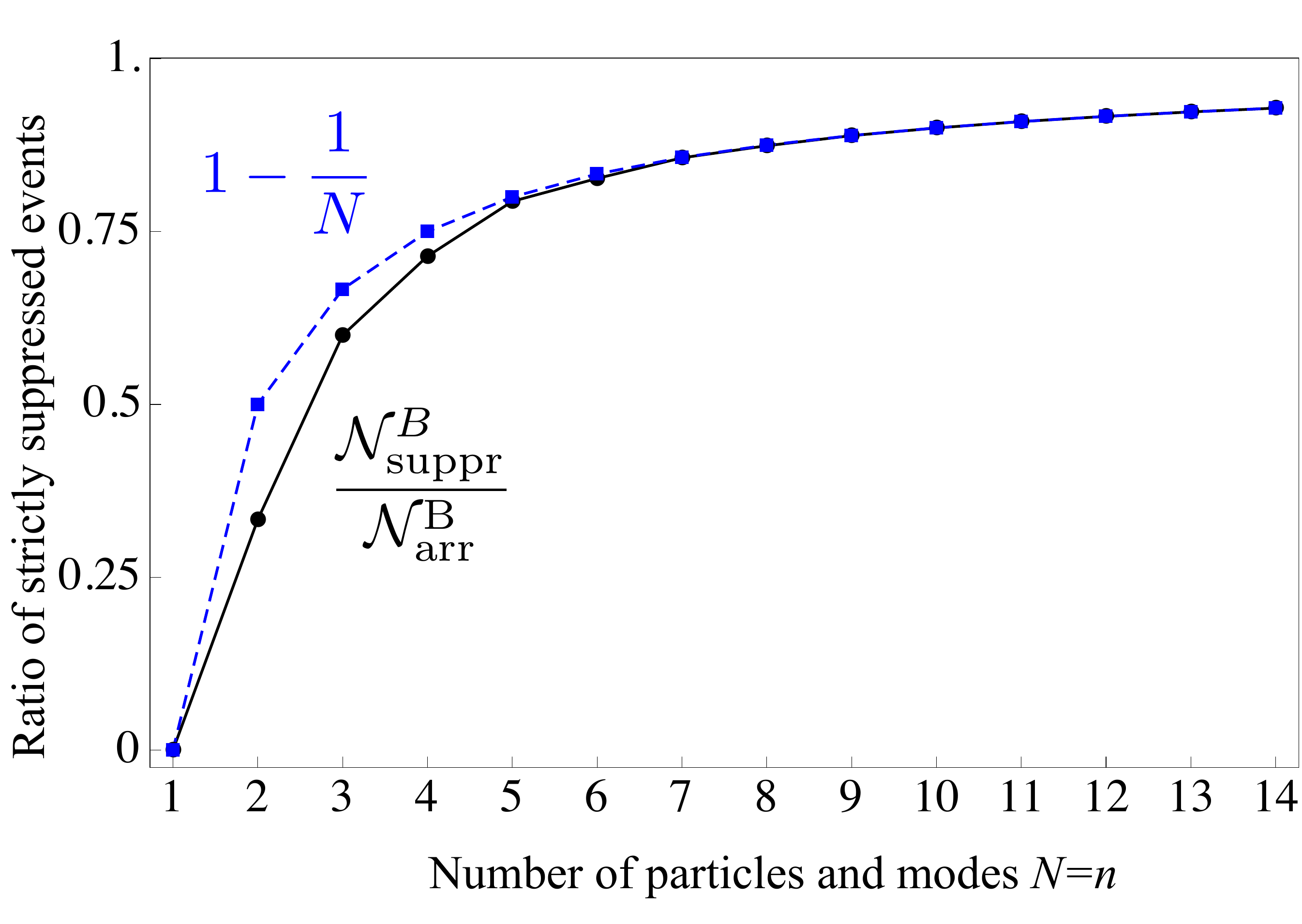}
\caption{Black circles connected by solid black line: Ratio of suppressed events to total number of events,  $\mathcal{N}^{B}_{\text{suppr}}/\mathcal{N}^{B}_{\text{arr}}$, for a setup with $n=N$ and the initial state $\vec r=(1,1,\dots, 1)$. Blue squares, connected by dashed blue line: Estimate $1-1/N$ for the ratio of suppressed events.  Figure adapted from \cite{TichyDiss}.}  \label{NumberSupprEvents}
 \end{figure}

The vast majority of this enormous number of final states, however, is strictly suppressed for the Fourier scattering matrix when an initial state with cyclic symmetry is chosen! To see that, consider $n=N$ and the initial state $\vec r=(1, \dots, 1)$ (i.e.~$m=1, p=n=N$). For each final arrangement $\vec s$, the sum $\sum_{j=1}^{N} d_j(\vec s) $ is an integer between $N$ [all particles in the first mode, $\vec s=(N,0,0,\dots,0), \vec d(\vec s)=(1,1,1\dots, 1)$] and $N^2$ [all particles in the last mode, $\vec s=(0,\dots, 0, N), \vec d(\vec s)=(N, \dots, N)$] \cite{Tichy:2010ZT,Tichy:2013lq}.  The ratio of integers between $N$ and $N^2$ that \emph{cannot} be divided by $N$ can be estimated to be approximately $1-1/N$. Since a transition can only remain unsuppressed when $\sum_j s_j$ is dividable by $N$, the fraction of events that remain unsuppressed is approximately $1/N$. This approximation is confirmed by the data shown in Fig.~\ref{NumberSupprEvents}, in which the fraction of suppressed events is compared to our estimate $1-1/N$. More in general, an initial state with $p$ period repetitions [see Eq.~(\ref{defiperiodicstate})] leads to a fraction of suppressed events of approximately $1-1/p$. 

The large ratio of events to which the suppression law actually applies and  the efficiency of the evaluation of the conditions in (\ref{law}) make the suppression law an exigent and efficient test for Boson-Sampling devices in practice, which ensures that true $N$-body interference occurs. That is, the general complexity of Boson-Sampling can be \emph{tamed} for this particular instance, such as to reliably rule out ill-functioning Boson-Sampling devices and any classical or semi-classical behaviour of many interfering particles that circumvents the true interference of $N!$ many-particle paths \cite{TichyDiss}.  Since the \emph{vast majority} of final events is \emph{strictly} suppressed, a misaligned or otherwise ill-functioning setup will be detected as such as soon as an event is detected that ought to be fully suppressed according to (\ref{law}), which makes this criterion more stringent than a criterion based on average bosonic behaviour accessible by statistical arguments \cite{Carolan:2013mj}.   In particular, the probability that the output of a sampling machine that uses distinguishable particles [Eq.~(\ref{distsimple})], that behaves semi-classically, or that simply generates a uniform distribution \cite{gogolin2013} appears to be compatible with the suppression law is $(1/p)^k$, where $k$ is the number of sampled events and $p$ the number of period repetitions in the initial state \cite{Tichy:2013lq}. As discussed below in Section \ref{mpsemicl}, the suppression law witnesses reliably the partial distinguishability of particles, and becomes more sensitive with increased particle number. Once a Boson-Sampler is certified via the suppression law, it may be used to sample the output of random unitary matrices, for which the classical simulation is truly unfeasible.

\subsubsection{Suppression law for fermions}

\begin{figure}[ht] \center
\includegraphics[width=.95\linewidth,angle=0]{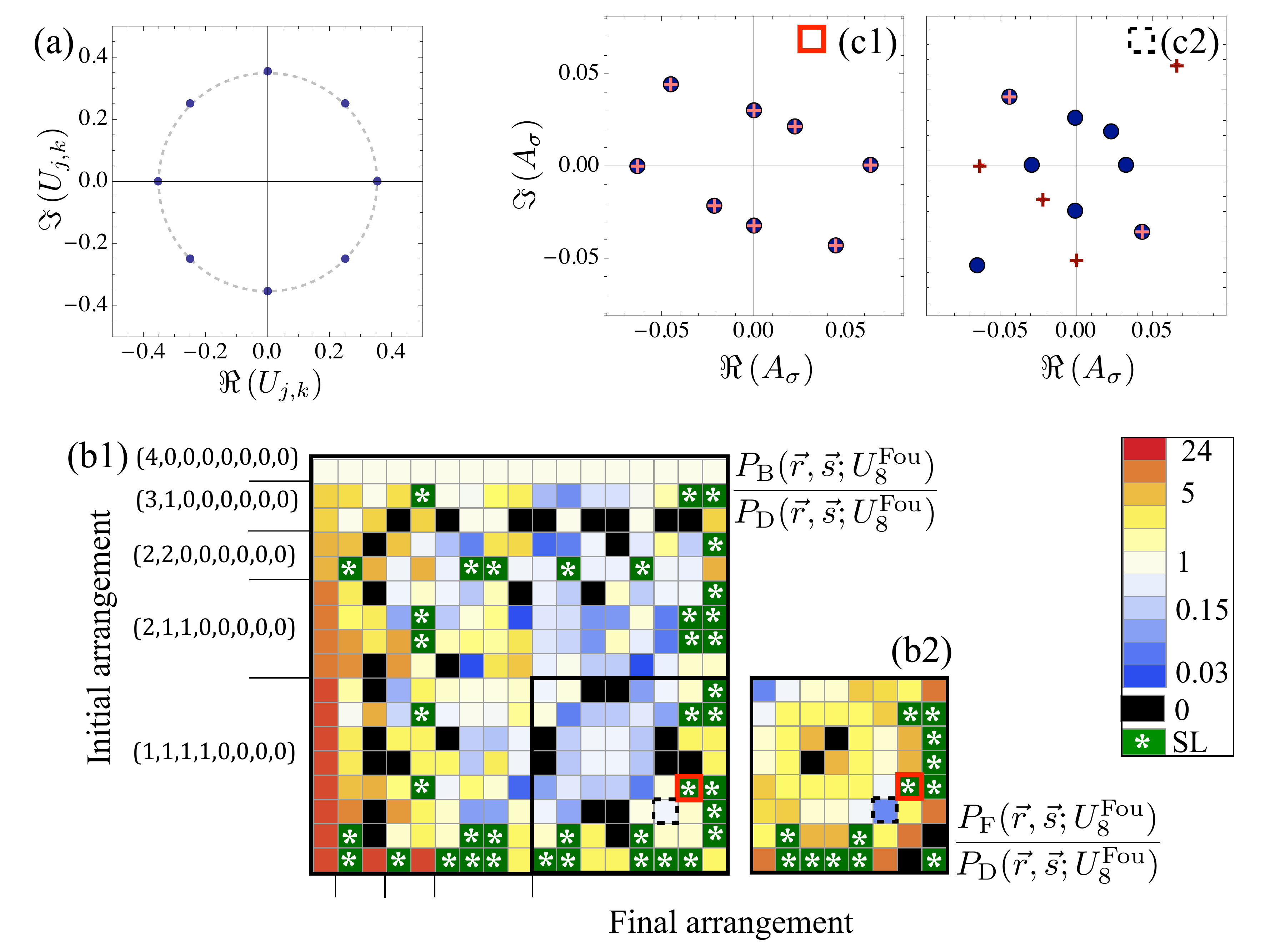}
\caption{Scattering of $N=4$ fermions and bosons on the $n=8$-Fourier multiport. (a) The scattering matrix $U^{\text{Fou}}_8$ contains only eight different amplitudes, which correspond to the eight roots of unity in the complex plane.  (b) Transition probability for bosons (b1) and fermions (b2), normalised to the probability for distinguishable particles. The color code indicates whether constructive (reddish colors) or destructive (blueish colors) interference takes place. The chosen arrangements and axes are the same as in Fig.~\ref{RandomMatrix.pdf}. (c) Many-particle amplitudes $A_\sigma$ in the complex plane. The angle to the horizontal axis reflects the phase, the distance from the origin is proportional to the number of paths $\sigma$ that lead to the same amplitude $A_\sigma$. Blue circles denote bosons, red crosses  fermions. (c1) corresponds to a transition highlighted by a red squares in (b), which is fully suppressed for fermions and for bosons, the resulting amplitudes coincide; (c2) shows the imbalanced amplitudes for the black dotted square in (b), for which the resulting total amplitude does not vanish. We adapt the visualisation used in \cite{Tichy:2012NJP}. }  \label{FourierMatrixPlot.pdf}
 \end{figure}

Although the  suppression law is most useful for bosons, it is instructive to study the analogous problem of many fermions scattered off a symmetric Fourier-multiport. The derivation in Section \ref{derivation} can be repeated in close analogy, but the application of the operation $\Gamma$ on a permutation can change its parity and induce a factor of $(-1)$ in the sum (\ref{geoser}), depending on the parity of $N$ and $N/p$. As a consequence, the suppression law for fermions needs to be adjusted \cite{Tichy:2012NJP}. Eventually, for odd particle numbers $N$ or even $\nicefrac{N}{p}$, we have
\eq 
\label{lawfermi}
 \vec s~ m\text{-periodic},~\text{~mod}\left(m  \sum_{j=1}^N  d_j(\vec r),n \right) \neq 0 ~ \Rightarrow P_{\text{F}}(\vec r, \vec s; U^{\text{Fou}}_n)=P_{\text{F}}(\vec s, \vec r; U^{\text{Fou}}_n)=0 . \label{lawfermi1} \en
For even particle numbers $N$ and odd $\nicefrac{N}{p}$, 
\eq
\vec s~ m\text{-periodic},~\text{~mod}\left(m  \sum_{j=1}^N  d_j(\vec r),n \right) \neq \frac n 2 ~ \Rightarrow P_{\text{F}}(\vec r, \vec s; U^{\text{Fou}}_n)=P_{\text{F}}(\vec s, \vec r; U^{\text{Fou}}_n)=0  \label{lawfermi2} ,
\en
where $n$ is necessarily even when $N$ is even and $N/p$ is odd.

For odd particle numbers $N$ or even $\nicefrac{N}{p}$, the suppression laws for bosons and fermions are \emph{identical}, i.e.~the very same transitions are suppressed for bosons and for fermions. When the laws formally differ (i.e.~even particle numbers $N$ and odd $N/p$), most suppressed events are still shared: The conditions in (\ref{lawfermi1}) and (\ref{lawfermi2}) that an event is suppressed are $\text{mod}(Q,n) \neq 0$ and $\text{mod}(Q,n) \neq n/2$, respectively. For large $n$ and $N$, a large fraction $(n-2)/n$ of the possible final states $\vec s$ will fulfil both criteria simultaneously, i.e.~they will be suppressed for, both, bosons and fermions. For exactly two particles, an arrangement with cyclic symmetry necessarily fulfils $m=n/2, p=2$, and  all bosonically suppressed transitions are enhanced for fermions, and vice-versa. In other words, when we \emph{increase} the number of particles, the signals of bosons and fermions become more and more \emph{similar}! This emphasises again that the suppression of events does not rely on the \emph{statistical} properties of bosons and fermions, but on the coherent superposition of many-particle paths. The suppression law is, in particular, not an instance of the Pauli principle, and it applies, both, in the high-population limit $N \gg n$ as well as for very dilute systems $N \ll n$. 

The  suppression law is compared for $N=4$ fermions and bosons in Fig.~\ref{FourierMatrixPlot.pdf}. The matrix elements of the Fourier matrix are shown in (a), only 8 different elements appear, which lie equidistantly on a circle in the complex plane. Therefore, also the resulting amplitudes in (c1,c2) exhibit symmetries. The resulting interference patterns (b1,b2) exhibit the same \emph{statistical} features as for the randomly chosen matrix [Fig.~\ref{RandomMatrix.pdf}]; however, many fully suppressed arrangements appear (green fields marked with stars). Some transitions are suppressed (black), although they are not predicted to be so by the suppression law. The lowest row of (b1) shows the impact of the suppression law for cyclic initial states: The initial state (not shown) is $\vec r=(1,0,1,0,1,0,1,0)$, such that most events are suppressed, the remaining ones are, consequently, strongly enhanced.  In contrast to random matrices [see Fig.~\ref{RandomMatrix.pdf}],  we can now systematically understand the symmetry properties of the scattering matrix, which leads to the striking similarities between bosons and fermions.

\section{Many-particle distinguishability transition} \label{DisttransSec}
In the previous section, we have explored the remarkable effects that arise for many interfering bosons and fermions in comparison to the two-particle case. At all stages, particles were  treated as either distinguishable or perfectly indistinguishable, which allowed a comparison of fermionic and bosonic interference to the combinatorial behaviour of  distinguishable particles. 

In the experiment, the requirement that all interfering particles be fully indistinguishable is difficult to realise, and even more so for setups with many modes that come with many adjustable variables like mutual path delays. Therefore, the mutual indistinguishability of all particles will require the careful adjustment of many distinct parameters. As in the two-particle case treated in Section \ref{twopdisttras}, \emph{partially} distinguishable particles naturally arise in any realistic description of interference experiments, as we will discuss in the following. 

\subsection{Two-mode setup with few particles}
The simple interpolation between two distinguishable and indistinguishable particles in Eq.~(\ref{twophtondiprob}) may hastily motivate a similar Ansatz for many particles and modes  \cite{al:2009vn}, 
\eq P_{\text{T}}( \vec s, \gamma) = \gamma  P_{\text{B/F}}(\vec s) + (1- \gamma) P_{\text{D}}( \vec s), \label{interpolaaa} \en
where 
$\gamma$ quantifies the indistinguishability in a suitable way, and $P_{\text{D}}$ and $P_{\text{B/F}}$ are the probabilities associated with distinguishable and perfectly indistinguishable bosons/fermions, given by Eqs.~(\ref{generalD}) and (\ref{generalampliB})/(\ref{generalampliF}), respectively. This Ansatz, however, turns out to severely constrain the functional form of the resulting signal, and will be insufficient for a description of the actually observed measurement results.

\begin{figure}[ht] \center
\includegraphics[width=1\linewidth,angle=0]{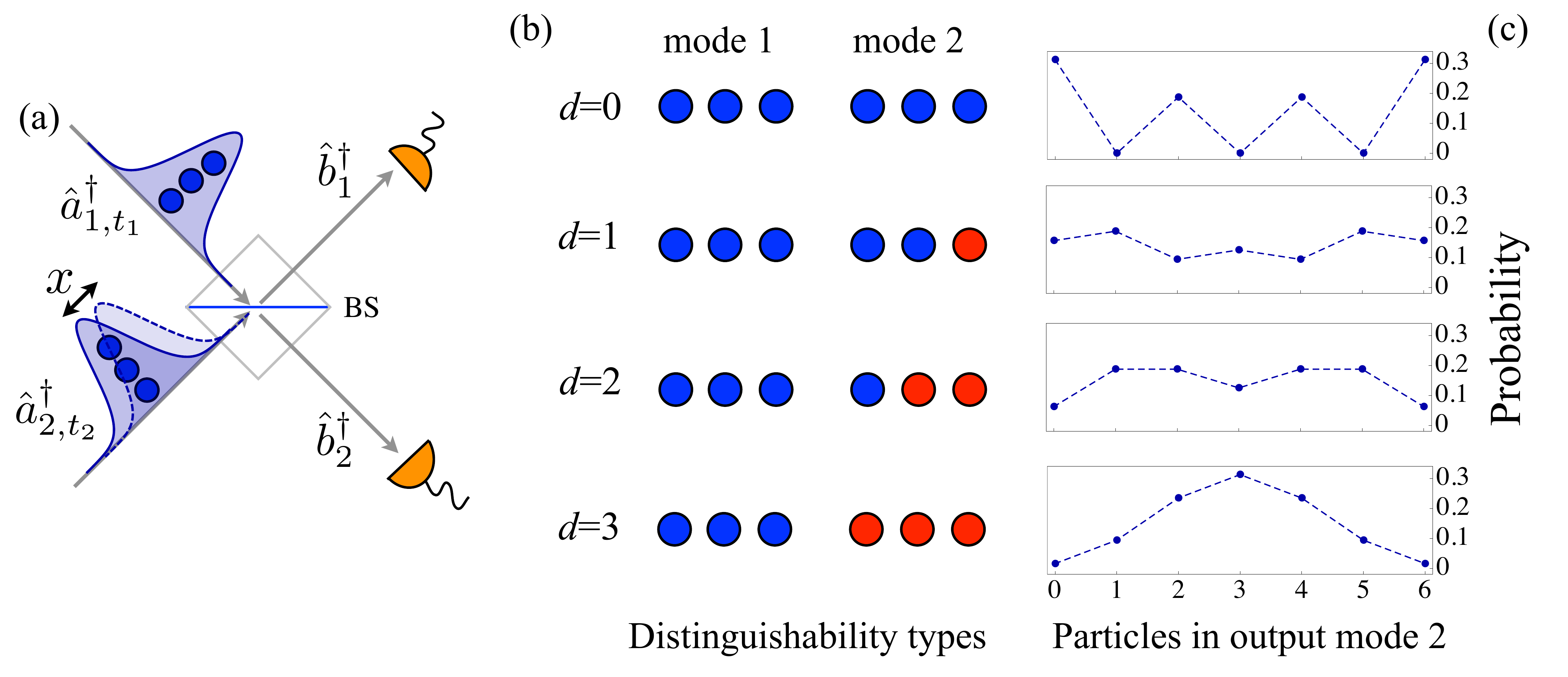}
\caption{(a) Three photons impinge onto each input mode of a beam splitter, $\vec r=(3,3)$. Photons in the same mode $j$ are centered at the same time $t_j$, but photons in mode 2 are delayed with respect to those in mode 1. (b) Decomposition of the wavefunction into 4 different distinguishability types that contain between $d=0$ and $d=3$ distinguishable particles (red circles) in mode 2. (c) Resulting single-mode counting statistics $P_d( s_1, s_2)$ associated to the respective distinguishability type. The signal for $d=0$ corresponds to indistinguishable particles, for $d=3$, a binomial emerges, characteristic for distinguishable particles.  }  \label{TwoModesManyPhotons.pdf}
 \end{figure}

In order to see the origin of such rich behaviour, let us consider the quantum state of $r_1$ and $r_2$ photons in the two  input modes of a balanced beam splitter [see sketch in Fig.~\ref{TwoModesManyPhotons.pdf}(a)], with temporal wavefunction components centered at $t_1$ and $t_2$, 
\eq 
\ket{\chi_{\text{ini}}^{\text{partial}}} =\frac{1}{\sqrt{r_1! r_2!}} \left( \hat a_{1,t_1}^\dagger \right)^{r_1} \left( \hat a^\dagger_{2,t_2} \right)^{r_2} \ket{\text{vac}}  ,
\en
where all particles in the same spatial mode are mutually indistinguishable. This assumption is justified for multi-photon states generated by SPDC  [see Eq.~(\ref{SPDC})]. As in Section \ref{twopdisttras} above, the states $\ket{t_1}$ and $\ket{t_2}$ are the temporal components of the photonic wavefunctions, given by Eq.~(\ref{GaussianWF}). The indistinguishability $|c_{2,1}|^2$ and the distinguishability $|c_{2,2}|^2\equiv 1-|c_{2,1}|^2$ [Eq.~(\ref{twoporthog})] then quantify the degree of collective interference. For $c_{2,1}=0,  c_{2,2}=1$, any two particles in different modes are fully distinguishable with respect to their temporal wavefunction, for $c_{2,1}=1, c_{2,2}=0$, all particles are indistinguishable. 

Following the argument of Section \ref{twopdisttras}, the state $\ket{t_2}$ can be re-written as $\ket{t_2}=c_{2,1} \ket{t_1} + c_{2,2} \ket{\tilde t_2}$, and the initial state becomes \cite{Ra:2013kx,younsikraNatComm,TichyFourPhotons}
\eq 
\ket{\chi_{\text{ini}}^{\text{partial}}} =
\frac{1}{\sqrt{r_1! r_2!}} \left( \hat a_{1,t_1}^\dagger \right)^{r_1} \left( c_{2,1} ~ \hat a_{2,t_1}^\dagger +  c_{2,2} ~ \hat a_{2,\tilde t_2}^\dagger \right)^{r_2} \ket{\text{vac}}  .   \label{iniexact}\en
The binomial expansion of the partially distinguishable term  gives a superposition of $r_2+1$
 orthogonal components, 
\eq 
\ket{\chi_{\text{ini}}^{\text{partial}}}=\frac{1}{\sqrt{r_1! r_2!}} \left(  \hat a_{1,t_1}^\dagger \right)^{r_1} \sum_{d=0}^{r_2} {r_2 \choose d} \left( c_{2,1} ~ \hat a_{2,t_1}^\dagger \right)^d \left( c_{2,2} ~ \hat a_{2,\tilde t_2}^\dagger \right)^{r_2-d} \ket{\text{vac}}  . \label{binoexpdd}
\en
Each summand is associated with a different \emph{type of distinguishability} $d$, as illustrated in Fig.~\ref{TwoModesManyPhotons.pdf}(b). In other words, the wavefunction is a superposition of components with different numbers $d$ ($0\le d \le r_2$) of distinguishable particles in mode 2. Components with different $d$ do not interfere, such that the total probability becomes \cite{Ra:2013kx,younsikraNatComm}
\eq 
P_{\text{T}}(\vec s, x)= \sum_{d=0}^{r_2}  W_d (x)  P_d(\vec s) , \label{sumofpdd}
\en
where $P_d(\vec s)$ is the probability to find the event $\vec s$ associated to the state characterised by $d$ distinguishable particles in mode 2  [see an example in Fig.~\ref{TwoModesManyPhotons.pdf}(b,c)]. The influence of each type of distinguishability is proportional to its weight, which is extracted from (\ref{binoexpdd}),  
\eq 
W_d(x)={r_2 \choose d} |c_{2,1}|^{2d} |c_{2,2}|^{2(r_2-d)}  ,  \label{weights}
 \en which directly generalises   Eq.~(\ref{twophtondiprob}). 

\begin{figure}[ht] \center
\includegraphics[width=1\linewidth,angle=0]{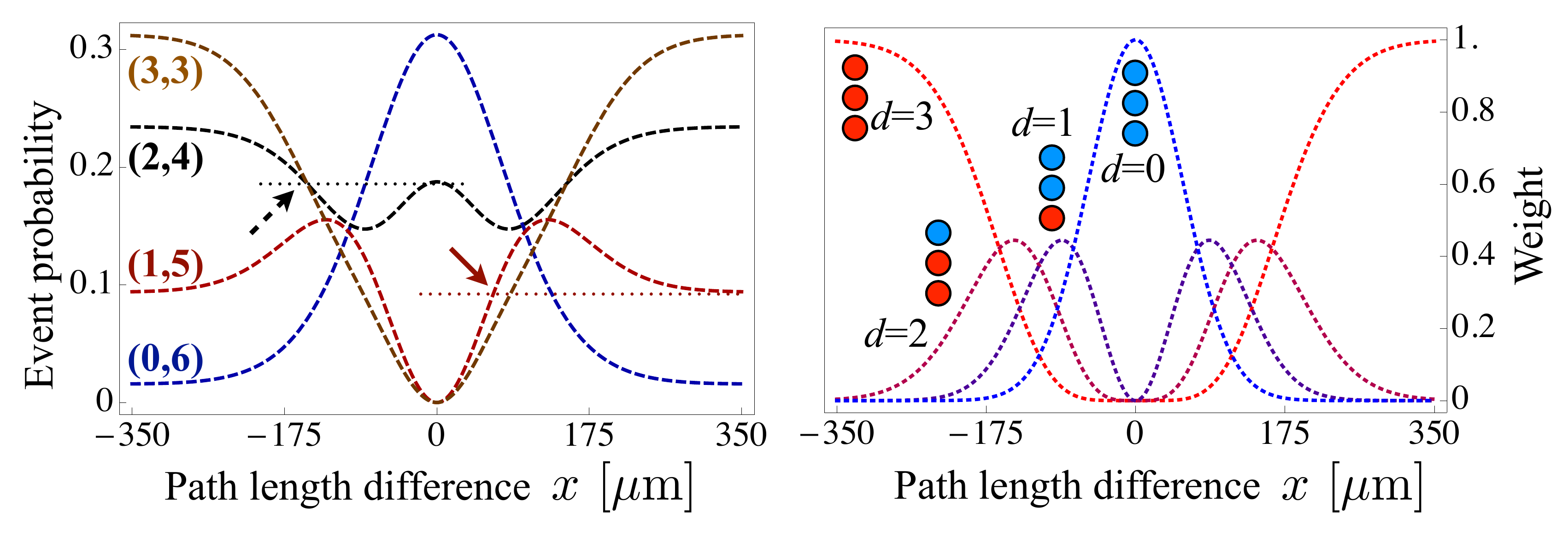}
\caption{(a) Event probability $P_{\text{T}}(\vec s, x)$ for $N=6$ particles prepared in the configuration shown in Fig.~\ref{TwoModesManyPhotons.pdf}(a), as  a function of the path length difference $x$.  The arrows point to values of $x$ which may be mistaken to yield full indistinguishability (black dashed arrow) or perfect distinguishability (red solid arrow), since the signal is compatible with the one for $x\rightarrow 0$ and $x \rightarrow \infty$, respectively. (b) Weights of the different distinguishability types in the wavefunction, Eq.~(\ref{weights}). For $x\rightarrow \infty$ and $x \rightarrow 0$, the distinguishable and indistinguishable components dominate, respectively. For intermediate values $x \sim l_c$, the partially distinguishable types ($d=1,2$) take turns. The figure reproduces data originally published in \cite{Ra:2013kx}. }  \label{TwoModeManyPResults.pdf}
 \end{figure}

Already on this formal level, the expression (\ref{sumofpdd}) is richer than the intuitive Ansatz (\ref{interpolaaa}), which reduces to the extremal components with $d=0$ and $d=r_2$. In principle, the probabilities $P_d(\vec s)$ could assume values that make the resulting signal $P_{\text{T}}$ described by Eq.~(\ref{sumofpdd}) indistinguishable from the one modelled by (\ref{interpolaaa}). In practice, however, this is not the case:  Combining the expressions for $c_{2,1}$ [Eqs.~(\ref{twoporthog}), (\ref{temporaloverlap})] with Eq.~(\ref{sumofpdd}), we show the resulting signal probabilities $P_{\text{T}}(\vec s, x)$ and weights $W_d(x)$ for $\vec r=(3,3)$ in Fig.~\ref{TwoModeManyPResults.pdf}. 
 
Two fully suppressed events for indistinguishable particles can be understood via the rules found in \cite{Campos:1989fk,Laloe:2010uq,nockdipm} and by the suppression law Eq.~(\ref{law}): The events $\vec s=(3,3)$ and $\vec s=(1,5)$ are fully suppressed when the particles interfere perfectly (for  $x=0~\mu$m), since $3+3\times 2=9, 5+1 \times 2=7$, which both cannot be divided by $2$. 

While $P_{\text{T}}((3,3),x)$ features a monotonic dependence on the path length difference $x$, the signal exhibited by $\vec s=(1,5)$ is more intricate: In particular, it is a \emph{non-monotonic} function of $x$, and the intuitive hierarchy $P_{\text{B}}(\vec s) \le P_{\text{T}}(\vec s, x) \le P_{\text{D}}(\vec s)$ is violated. The reason for the  \emph{increase} of the signal for moderate values of $x$ is the constructive interference that enhances $P_{d=1}$ and $P_{d=2}$ with respect to $P_{\text{D}}=P_{d=3}$ and $P_{\text{B}}=P_{d=0}$ [see Fig.~\ref{TwoModesManyPhotons.pdf}(c)]. The operational consequence of this intricate behaviour is that a typical signal $P_{\text{T}}(\vec s,x)$ does not directly reflect the indistinguishability $|c_{1,2}|^2$. For example, there is a value of $x$ (marked with a solid arrow in Fig.~\ref{TwoModeManyPResults.pdf}) for which the probability $P_{\text{T}}(x)$ assumes the value for distinguishable particles, $P_{\text{T}}(x)=P_{\text{D}}$, such that seemingly no interference takes place at all. On the other hand, the non-monotonic signal for $\vec s=(2,4)$ seems to exhibit perfect interference at the delay $x$ marked with a dashed arrow, i.e.~$P_{\text{T}}(x \neq 0)=P_{\text{B}}$. Such non-monotonic signals constitute the \emph{rule} -- and not the exception! -- in many-particle interference: The probabilities $P_d(\vec s)$ for intermediate types of distinguishability $0<d<r_2$ will typically adopt values that are not contained in the interval defined by $P_{\text{B}}(\vec s)=P_{d=0}(\vec s)$ and $P_{\text{D}}(\vec s)=P_{d=r_2}(\vec s)$.

The signals of $\vec s=(0,6)$ and $\vec s=(3,3)$ are monotonic, but their functional dependence on the path length difference $x$ differs from the two-photon HOM-dip shown in Fig.~\ref{TwoPhotonDistTransition.pdf}: The width of the $\vec s=(0,6)$-signal, as quantified by the full width at half maximum (FWHM), amounts to $\sigma_{(0,6)} \approx 0.82 l_c$, while the width of the $\vec s=(3,3)$-signal is increased with respect to the two-photon signal, $\sigma_{(0,6)}=1.14 l_c$. That is to say, although $\sigma_{(1,1)}=\sigma_{(2,0)}=l_c$, the \emph{single}-photon coherence length $l_c$ is not an unambiguous indicator for the shape of \emph{multi}-photon signals. A well-defined quantum state naturally exhibits different degrees of interference in different observables \cite{younsikraNatComm}. The discussed features appear as soon as more than two particles interfere; for four photons prepared in $\vec r=(2,2)$, non-monotonicity \cite{Ra:2013kx} and the reduction of the observed coherence length \cite{younsikraNatComm} were experimentally confirmed. 

The probability of bunched events of the form $\vec s=(N,0)$ always increases monotonically  with the indistinguishability $|c_{2,1}|^2$. Consistent with our argument in Section \ref{statisticaleffects}, bunched events are only fed by one many-particle path, i.e.~no contrasting interference terms compete. Therefore, only the bunching effect for many photons into one mode is observed, without any non-monotonic structure \cite{Ou:1999lo,Ou:1999rr,Xiang:2006uq,Ou:2007ly}. The signal width of $P_{\text{T}}(\vec s=(N,0),x)$  approaches  $ 2 l_c /\sqrt{N}$ in the limit of large particle numbers $N$, which can be shown by analysing the weights in the sum (\ref{sumofpdd}) \cite{younsikraNatComm}.

\FloatBarrier

\subsection{Many particles and semi-classical limit} \label{mpsemicl}
The full counting statistics of many particles in two modes exhibits, both, \emph{granular} and \emph{non-granular} features: At the granular level of individual particles, an even-odd effect suppresses the occurrence of any event with an odd number of particles in either mode, a particular instance of the suppression law (\ref{law}). This granular effect is complemented by a broad characteristic U-like shape, which can be understood via a semi-classical treatment \cite{Laloe:2010uq}. 

\begin{figure}[t] \center
\includegraphics[width=1\linewidth,angle=0]{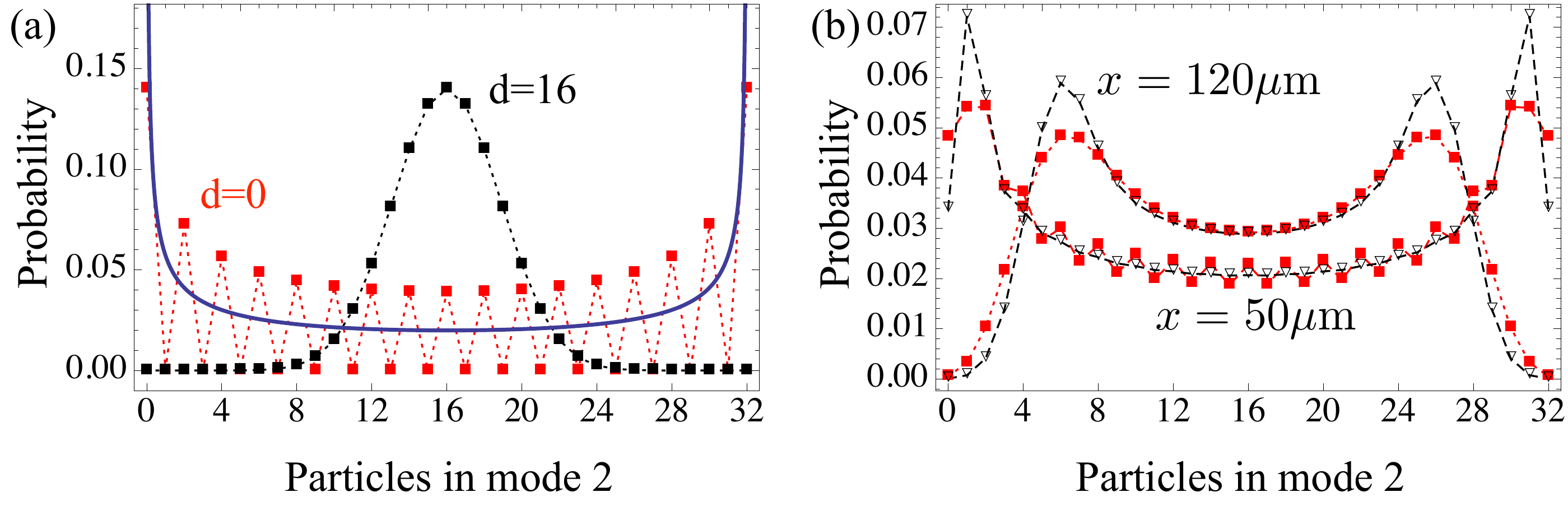}
\caption{(a) Counting statistics for bosons ($d=0$, red squares) and distinguishable particles ($d=16$, black squares, binomial distribution) prepared in $\vec r=(16,16)$, and re-scaled semi-classical probability distribution (solid blue line), given by Eq.~(\ref{semiclassical}) with $x=s_2/N$. (b) Exact counting statistics (red squares) and approximation (black triangles) given by Eq.~(\ref{approxtransi}), for two values of $x$ (marked by dashed arrows in Fig.~\ref{ManyPLimit2Modes.pdf}). For $x=120\mu$m, no even-odd effects are apparent, while they are still visible for $x=50\mu$m. The lines that connect the symbols are intended to guide the eye. }  \label{ApproximationManyP2modes.pdf}
 \end{figure}

In Fig.~\ref{ApproximationManyP2modes.pdf}(a), the counting statistics of $\vec r=(16,16)$ is shown for indistinguishable bosons ($d=0$) and distinguishable particles ($d=16$). A semi-classical approach based on a description of Fock-states by a macroscopic wavefunction \cite{Laloe:2010uq} describes the non-granular features: The semi-classical probability to find the particle fraction $x=s_2/N$ in one output mode is shown as a solid curve in Fig.~\ref{ApproximationManyP2modes.pdf}(a), the analytic expression reads \cite{Laloe:2010uq}
\eq 
P_{\text{semiclass}}(x)=\frac{1}{\pi \sqrt{x(1-x)}}  .  \label{semiclassical}
\en
While the even-odd effects cannot be understood via such coarse-grained semi-classical approach \cite{TichyDiss}, the overall shape of the distribution is reproduced well. 

\begin{figure}[t] \center
\includegraphics[width=.9\linewidth,angle=0]{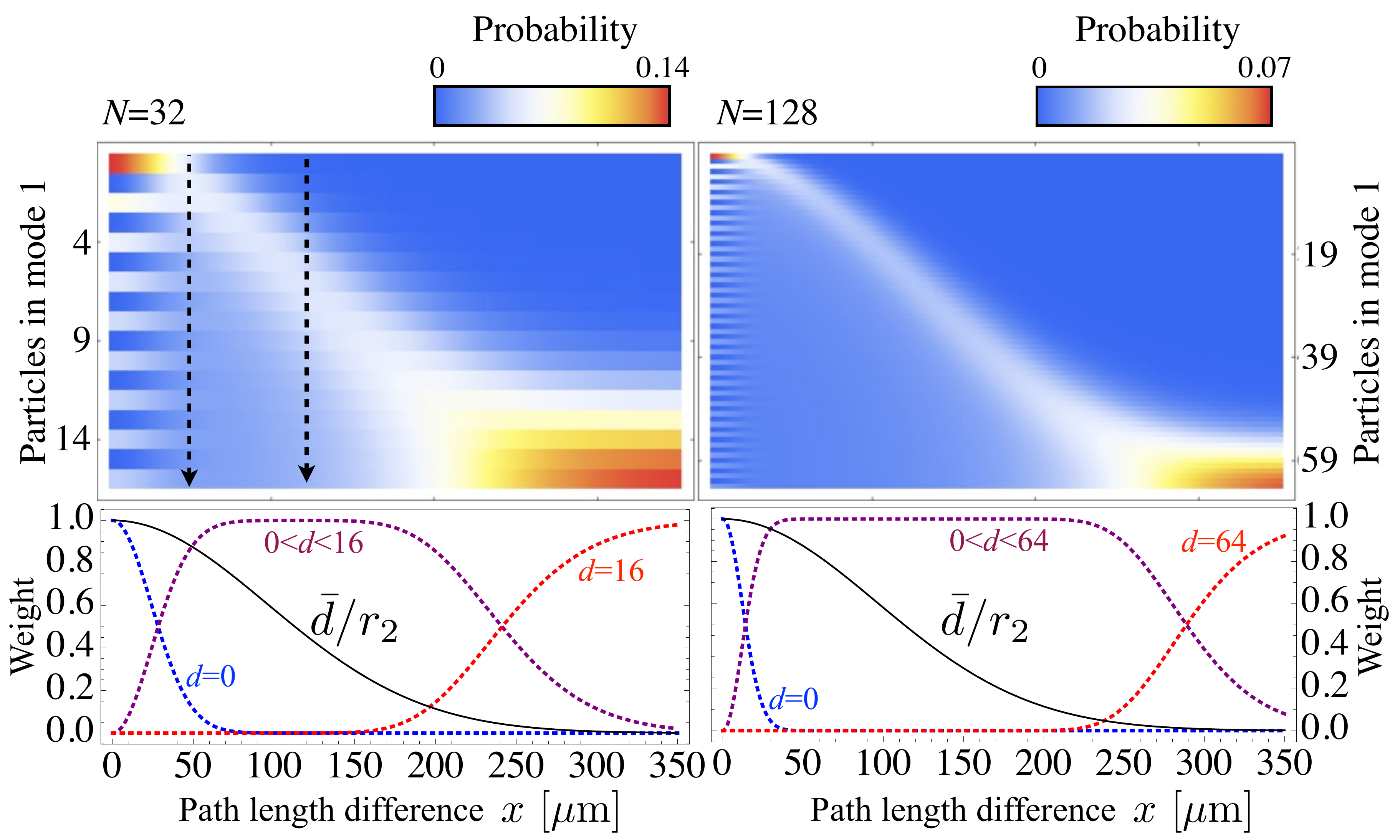}
\caption{Upper panels: Event probability as a function of the path delay $x$ and of the number of particles $s_1$ in mode 1. Due to symmetry, $P(s_1,s_2)=P(s_2,s_1)$, and only the values up to $s_1=N/2$ are shown. Lower panels: Weights of the fully indistinguishable ($d=0$), fully distinguishable ($d=r_2$) and the sum of all partially distinguishable contributions ($0<d<r_2$) to the wavefunction, and expectation value of the most populated type $\bar d/r_2$. Left panels: $N=32$, right panels: $N=128$. The initial state is $\vec r=(N/2,N/2)$. The two black arrows show the path length differences assumed in Fig.~\ref{ApproximationManyP2modes.pdf}.  }  \label{ManyPLimit2Modes.pdf}
 \end{figure}

The question arises to which extent granular and non-granular features are sensitive to partial distinguishability. Since the even-odd effect is borne only by the component with fully indistinguishable particles ($d=0$) in Eq.~(\ref{sumofpdd}), it can be expected to become more and more fragile the larger the total number of particles $N$ is: The weight of the component $W_0(x)$ in (\ref{weights}) is 
\eq 
W_0(x)= |c_{2,1}|^{2 r_2} ,
\en
i.e.~for a fixed displacement $x$, the parameter $c_{2,1}$ remains constant, and $W_0(x)$ decays exponentially in $r_2$ (the number of particles in the second mode). As a consequence, even-odd effects disappear at smaller distinguishability and smaller displacement $x$ when $N$ is increased. For photons with Gaussian frequency distribution [as assumed in Eq.~(\ref{GaussianWF})], the width of $W_0(x)$ in $x$ decreases as $\sigma_1/\sqrt{r_2}$. This effect can be observed in  Fig.~\ref{ManyPLimit2Modes.pdf}, in which the counting statistics for $N=32$ and $N=128$ is shown as a function of the path length difference $x$. The range in which the even-odd effects are still visible corresponds to the interval in which $W_0$ is large enough to support them: For $N=32$, granular effects can still be observed for larger values of $x$ than for $N=128$. 

On the other hand, the characteristic U-like shape of the distribution is rather robust against increases of the total number of particles, the overall distribution for $N=32$ resembles indeed the one for $N=128$, although it is scaled with the total number of particles. Non-granular effects are, thus, more robust against distinguishability, since they are not only borne by the fully indistinguishable component with $d=0$. This similarity suggests an approximate treatment of the  wavefunction (\ref{iniexact}): It can be approximated by the contribution of the distinguishability type $d=\bar d$ associated with the most prominent weight $W_{\bar d}(x)$, $\bar d=r_2 |c_{2,1}|^{2}$. That is,  
\eq
 \ket{\chi_{\text{ini}}^{\text{partial}}} \approx \ket{\chi_{\text{approx}}}= \frac{(\hat a_{1,t_1}^{\dagger})^{r_1}}{\sqrt{r_1!}} \frac{(\hat a_{2,t_1}^{\dagger})^{\bar d}}{\sqrt{{\bar d}!}} \frac{ (\hat a_{2,\tilde t_2}^{\dagger})^{ r_2-{\bar d} } }{\sqrt{ (r_2-{\bar d})!}} \ket{\text{vac}} ,   \label{approxtransi}
  \en {\it i.e.~}the state (\ref{iniexact}) is modelled by $r_1$ particles in the first input mode, \emph{exactly} $\bar d$ particles in the second input mode that interfere with the particles in the first input mode, and $r_2-\bar d$ particles in the second input mode that do not. Since the distribution $W_d$ is binomial in $d$, its relative width in $d$ decreases with $\nicefrac{1}{\sqrt{r_2}}$, and the approximation becomes more accurate for large particle numbers. The most influential distinguishability type, $\bar d$, is shown as solid black line in the lower panels of Fig.~\ref{ManyPLimit2Modes.pdf}, the shape is identical for $N=32$ and $N=128$, the ratio $\bar d/r_2$ being independent of the total particle number $r_2$. It also governs the position of the maximum of the distribution in the upper panels, which moves from $N/2$ (for the binomial distribution characteristic for distinguishable particles) to the extremes $0$ and $N$ (U-like distribution of indistinguishable bosons): The $\bar d$ indistinguishable particles that interfere typically all end in one of the two modes, while the remaining $N-\bar d$ particles are evenly distributed among the modes. 
   The counting statistics of $N=32$ particles computed via the approximation  (\ref{approxtransi}) is compared to the exact calculation based on Eq.~(\ref{iniexact}) in Fig.~\ref{ApproximationManyP2modes.pdf}(b) for two intermediate values of the displacement $x$. The resulting exact distribution resembles the approximated one, while the computational expenses for the latter are significantly lower.

In general, truly quantum features borne by $d=0$,  such as the even-odd effect discussed here, are more sensitive to partial distinguishability than the average, non-granular, coarse-grained shape of the distribution, which can also be described in a semi-classical approach. The more particles are involved, the more sensitive becomes the suppression law (\ref{law}), which turns into a more and more stringent criterion for full indistinguishability when the number of particles increases \cite{Tichy:2013lq}. In other words, the degree of satisfaction of the suppression law is a measure for the impact of the \emph{perfectly} quantum features of the states that are observed in granular observables, while the width of the U-like shape quantifies the \emph{typical} mutual distinguishability between the particles involved in the scattering process.
 
 \FloatBarrier

\subsection{Multi-mode treatment}
Despite the complicated dependence of interference signals on the distinguishability $|c_{2,1}|^2$ inscribed in Eq.~(\ref{sumofpdd}), the parameter $c_{2,1}$ still remains the unique and unambiguous quantifier of the mutual indistinguishability between particles in  two input modes. For three spatial modes, the mutual distinguishabilities of the three different pairings can, in principle, acquire rather independent values, and the concept of distinguishability itself cannot be boiled down to single parameter anymore. For four particles that are prepared in four modes,  features such as non-monotonicity and phase-dependence arising only for non-vanishing values of $x$ were demonstrated theoretically in \cite{TichyFourPhotons}. 

The simplest setup in which these effects arise is discussed in the following, namely three photons that impinge on a three-mode Fourier-beamsplitter \cite{PROP:PROP243}, $N=n=3, U=U^{\text{Fou}}_{3}$. The interference of three photons prepared in the state  $\vec r=(1,1,1)$ has been realised experimentally using an integrated tritter created by femtosecond laser waveguide writing  \cite{Spagnolo:2013fk}.  We discuss the setup of Ref.~\cite{Spagnolo:2013fk} within the formalism of distinguishability types, as introduced above and in Refs.~\cite{TichyFourPhotons,Ra:2013kx,younsikraNatComm}.

\begin{figure}[t] \center
\includegraphics[width=.9\linewidth,angle=0]{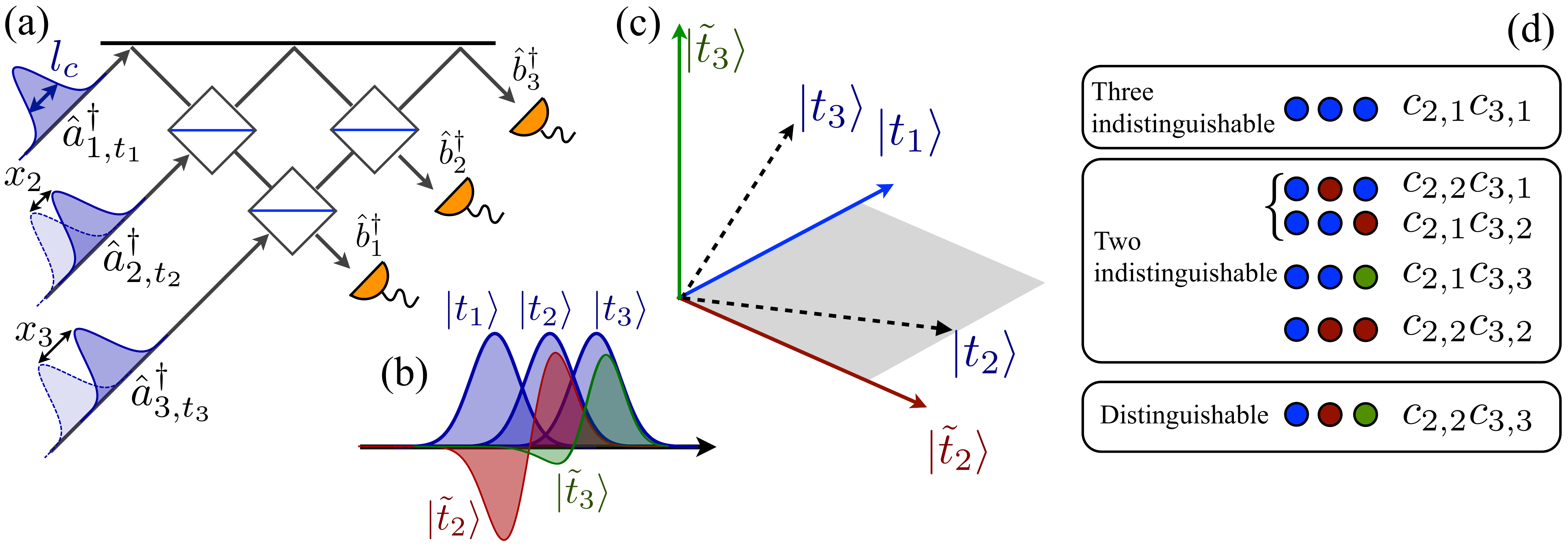}
\caption{(a) Three partially distinguishable particles interfere collectively in a three-mode multiport.  The ratio of the displacements $x_2$,  $x_3$ to the coherence length $l_c$ determines the interference pattern. (b) Decomposition of the temporal components of the wavefunction ($\ket{t_1}, \ket{t_2}, \ket{t_3}$) into orthogonal components ($\ket{t_1}, \ket{\tilde t_2}, \ket{\tilde t_3}$). (c) Visualisation of the three components. The state $\ket{t_2}$ lies in the plane spanned by $\ket{t_1}$ and $\ket{\tilde t_2}$. (d) The total three-photon wavefunction is decomposed into six mutually orthogonal components of different weights. The two components marked by a bracket can feed the same final state, such that their interference needs to be taken into account.}  \label{ThreeParticleDistTrans}
 \end{figure}

The photon in the first mode is centered at time $t_1$; the photons in the second and third mode are then characterised by their arrival times $t_{2}$ and $t_{3}$, respectively: 
\eq 
\ket{\chi_{\text{ini}}^{\text{3trans}}}&=&\hat a_{1,t_1}^\dagger \hat a_{2,t_{2}}^\dagger \hat a_{3,t_{3}}^\dagger  \ket{\text{vac}} 
\en
As in Eqs.~(\ref{iniexact}), the state $\ket{\chi_{\text{ini}}^{\text{3trans}}}$ can be rewritten in an orthonormal basis \cite{TichyFourPhotons},
\eq
\ket{\chi_{\text{ini}}^{\text{3trans}}}&=& \left[ \hat a_{1,\tilde t_1}^\dagger \left( c_{2,1} \hat a_{2,t_1}^\dagger+ c_{2,2} \hat a_{2,\tilde t_2}^\dagger \right) \right. 
\left. \left( c_{3,1} \hat a_{3,t_1}^\dagger + c_{3,2} \hat a_{3,\tilde t_2}^\dagger + c_{3,3} \hat a_{3,\tilde t_3}^\dagger   \right) \right] \ket{\text{vac}} , \label{inic123}
\en
where $c_{j,k} = \braket{\tilde t_k}{t_j}$, and $\hat a^\dagger_{j,\tilde t_k}$ creates a photon in the spatial mode $j$ and in the temporal state $\ket{\tilde t_k}$. The orthonormal components are defined as
\eq 
\ket{\tilde t_1} &=& \ket{ t_1} ,  \\
\ket{\tilde t_2} &=& \frac{ \ket{t_2}- \ket{t_1}\braket{t_1}{t_2} } {\sqrt{1-|\braket{t_1}{t_2}|^2 }} , \\
\ket{\tilde t_3} &=& \frac{ \ket{t_3}- \ket{t_1}\braket{t_1}{t_3} - \ket{\tilde t_2}\braket{\tilde  t_2}{t_3}   }
 {\sqrt{1-|\braket{t_1}{t_3}|^2- |\braket{\tilde t_2}{t_3}|^2 }} ,
\en
such that the temporal wavefunctions can be expanded as follows: 
\eq 
\ket{t_1} &=& \ket{\tilde t_1}   \\
\ket{t_2} &=& \ket{\tilde t_1}c_{2,1} + \ket{\tilde t_2}  c_{2,2}  \\
\ket{t_3} &=& \ket{\tilde t_1}c_{3,1} + \ket{\tilde t_2} c_{3,2}   + \ket{\tilde t_3} c_{3,3}  
\en
Formally, we have performed a Gram-Schmidt orthogonalisation: The three a priori linearly independent vectors ($\ket{t_1}, \ket{t_2}, \ket{t_3}$) yield  three orthonormal states, $\ket{\tilde t_1}, \ket{\tilde t_2}, \ket{\tilde t_3}$ [see sketches in Fig.~\ref{ThreeParticleDistTrans}(b,c)].

Instead of one indicator for distinguishability, $c_{2,1}$, three parameters, $c_{2,1}$, $c_{3,1}$ and $c_{3,2}$, define the physical setting. The restriction to the mutual path delay as distinguishing physical property leaves us here with two independent parameters, the displacements $x_2$ and $x_3$.  When factoring out Eq.~(\ref{inic123}),  six orthogonal terms emerge: 
\eq 
\ket{\chi_{\text{ini}}^{\text{3trans}}}&=& \sum_{d_2=1}^2 \sum_{d_3=1}^3 c_{2,{d_2}} c_{3,{d_3}} \hat a^\dagger_{1, \tilde t_1}\hat a^\dagger_{2, \tilde t_{d_2}} \hat a^\dagger_{3, \tilde t_{d_3}} \ket{\text{vac}}  \label{factoredout}
\en

Three qualitatively different components of the wavefunction can be identified \cite{Spagnolo:2013fk} [see Fig.~\ref{ThreeParticleDistTrans}(b)]: 
\begin{itemize}
\item The term $c_{2,2} c_{3,3}$ dominates when all mutual path delays are much larger than the coherence length $l_c$, one then deals with three distinguishable photons.
\item  If one mutual path delay is much smaller than $l_c$ while the other is much larger, two photons out of the three photons are indistinguishable, and interfere collectively. As observed in Fig.~\ref{ThreeParticleDistTrans}(d), there are four such terms that contain two indistinguishable photons. \item If all mutual delays are much smaller than $l_c$, all photons interfere perfectly, and $c_{2,1}c_{3,1}$ dominates the sum (\ref{factoredout}).  
\end{itemize}
In other words, distinguishability types are now characterised by \emph{two} parameters, $d_2$ and $d_3$, which refer to the  temporal components of the photon in mode 2 and 3, respectively. 

To obtain the total probability of an event $\vec s=(s_1,s_2,s_3)$, we need to incoherently sum over all distinguishability types that can be found in the output modes, and take into account the contribution of each component of the initial wavefunction to the  final state. Formally, we can write 
\eq 
P_{\text{T}}( \vec s , x_2, x_3 ) = \sum_{e_2=1}^2 \sum_{e_3=e_2}^3  \sum_{\sigma \in S_{\{ 1, e_2, e_3 \} } } | \bra{\chi_{\text{fin}}(\vec s, \sigma) }\hat U \ket{\chi_{\text{ini}}^{\text{3trans}}} |^2 ,\label{prob3gamma}
\en
where $e_2$ and $e_3$ are the temporal states of the particles found in the output modes, and the sum over $\sigma$ takes into account all permutations of these particles that are compatible with the final arrangement $\vec s$.

\begin{figure}[t] \center
\includegraphics[width=.9\linewidth,angle=0]{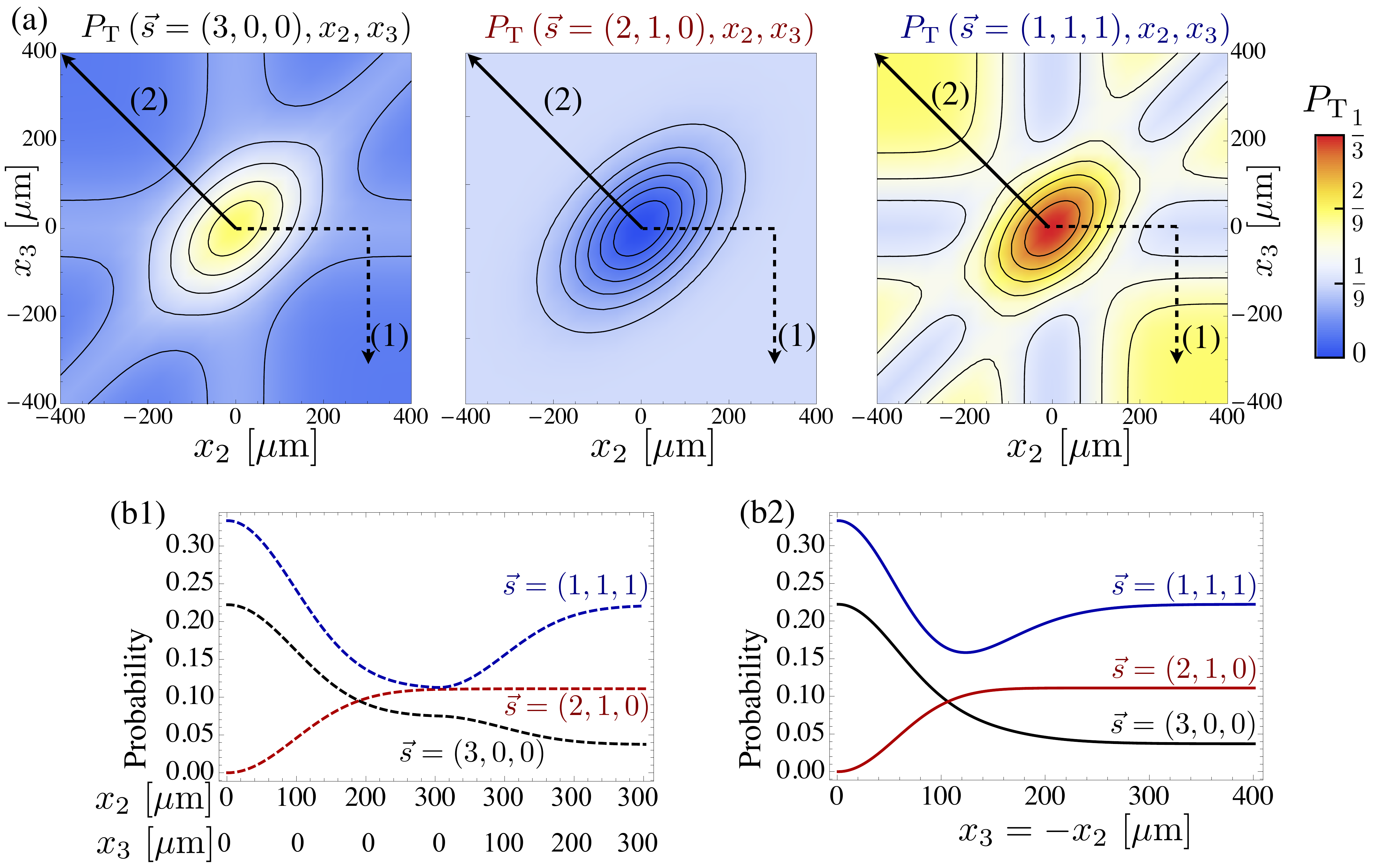}
\caption{(a) Probability for final states $\vec s=(3,0,0)$, $\vec s=(2,1,0)$ and $\vec{s}=(1,1,1) $, as a function of the two path delays $x_2$ and $x_3$, reproducing theoretical calculations originally presented in \cite{Spagnolo:2013fk}. The color-code indicates the probability. The two paths (1) and (2) correspond to two distinct transitions to distinguishability. (b1) Probabilities for the three final events for path (1); first, $x_2$ is increased, subsequently, $x_3$ is increased. (b2) Both displacements, $x_2$ and $x_3$ are increased simultaneously, as for path (2) in (a). }  \label{ThreePhotonsResults.pdf}
 \end{figure}

For most values of $e_2$ and $e_3$, there is exactly one distinguishability type in the wavefunction (\ref{factoredout})  that contributes. For $e_2=1, e_3=2$, however, there are two contributing terms, namely $d_2=1, d_3=2$ and $d_2=2, d_3=1$ [see Fig.~\ref{ThreeParticleDistTrans}(d)]. The interference of these terms can induce fast oscillations of the order of the wavelength of the photons \cite{TichyFourPhotons}, as in $N00N$-state interference \cite{Dowling:2008hc}. In our current setup, however, 
assuming a Gaussian temporal part of the wavefunction as given by Eq.~(\ref{GaussianWF}), the phases of $c_{2,2} c_{3,1} $ and $c_{2,1} c_{3,2}$ are identical, such that no such structures arise.

Using (\ref{prob3gamma}), we compute the event probabilities for $\vec s=(1,1,1),~\vec s=(2,1,0)$ and $\vec s=(3,0,0)$, which are shown in Fig.~\ref{ThreePhotonsResults.pdf}(a) as a function of the path delays $x_2$ and $x_3$. Due to the two independent parameters $x_2$ and $x_3$, there are several inequivalent ways to proceed from full indistinguishability to full distinguishability, two exemplary transitions are shown in (b).  In (b1), we first set $x_3=0\mu$m and increase $x_2$ from 0 to 300$\mu$m $> l_c$, subsequently we increase $x_3$; in (b2), both mutual delays are increased simultaneously, such that the three  wave-packets move away from each other ($x_2=-x_3$). The event $\vec s=(2,1,0)$ remains unaffected by proceeding from full distinguishability to two-particle indistinguishability; it is, therefore, not an indicator for two-particle interference. The event $\vec s=(1,1,1)$ exhibits non-monotonic behaviour: When  two photons are indistinguishable it is partially suppressed due to the elementary two-photon HOM-effect discussed in Section \ref{twoidpa}. Full three-photon interference, however, is constructive, and the signal increases again for  $x_2, x_3 \rightarrow 0$. The bunched event $\vec s=(3,0,0)$ is always enhanced with increasing indistinguishability, consistent with the bunching of four and six photons at a two-mode beam splitter \cite{Ou:1999lo,Ou:1999rr,Xiang:2006uq,Ou:2007ly}, as discussed in Section \ref{statisticaleffects}.

In general, the presence or absence of interference of a many-particle state is not determined by a single length scale as the single-particle coherence length $l_c$, but it can vary considerably, depending on the chosen observable.  The interpretation of any measurement must be done with great care, since the general interference capability of the involved particles is not directly reflected by the emerging signals. The decomposition of the wavefunction into orthonormal components as in Eqs.~(\ref{iniexact}), (\ref{factoredout})  allows us to treat the problem with a discrete finite number of orthogonal terms, although the degrees of freedom that we deal with can be continuous. Alternatively, the permanent and the determinant can be generalised, which allows the expression of transition probabilities in terms of \emph{immanants} \cite{Tan:2013ix,de-Guise:2014yf,Tillmann:2014ye}. For multi-mode many-particle systems, the number of terms in the decomposition (\ref{factoredout}) explodes: For $N$ particles prepared in $n=N$ modes, there are up to  $N$ orthogonal states $\ket{\tilde t_1} \dots \ket{\tilde t_N}$, and a total of $N!$ orthogonal many-particle wavefunction components, many of which will feed the same final states, such that interference between them needs to be accounted for. The problem requires even more computational expenses than idealised Boson-Sampling, and approximations, such as the one performed with Eq.~(\ref{approxtransi}), become necessary in practice.

\FloatBarrier

\section{Multipartite entanglement generation} \label{MultipartSec}
Generalising the argument of Section \ref{correpropdet}, in which we dealt with bipartite entanglement generation using two identical particles, multi-mode devices with many particles can generate \emph{multipartite} entanglement, i.e.~entangled states of $N$ parties that have access to a $d$-dimensional subsystem \cite{RevModPhys.84.777}. Multipartite entanglement largely outmatches the microscopic two-particle case, both, regarding the complexity required for its characterisation as well as through its more profound negation of a local-realistic worldview \cite{Greenberger:1998ff,Bancal:2012fk}. 

The generation of multipartite entanglement with photons has had remarkable success, and a variety of quantum states have been created in the laboratory using propagation and detection \cite{Bouwmeester:1999ys,Lu:2007ve,Radmark:2009ij,Prevedel:2009ec,Wieczorek:2009ff,Yao:2011uq}. The SPDC process that generates two photons  can leave these in a bipartite entangled state, e.g.~the polarisation degree of freedom omitted in Eq.~(\ref{SPDC}) can be correlated. To achieve \emph{multi}partite entanglement of many photons, bipartite entangled pairs need to be manipulated further. In the absence of interaction between photons, linear operations such as beam splitters and phase shifters are the only means to perform physical operations, with consequent restrictions \cite{Piotr-Migdal:2014zl}, whereas non-linearities can be induced by auxiliary photons, detection and feedback. 

Given the no-go-theorems that apply for entanglement generation and projection with two photons (Section \ref{limitsapplications}), the question naturally arises to which extent multipartite entanglement can be generated with many bosons and many fermions. For bipartite quantum states, the Schmidt decomposition [see Eq.~(\ref{Schmidt})]  provides  an unambiguous well-defined canonical form. All properties related to entanglement can then be read off the Schmidt coefficients $\{ \lambda_0 , \dots, \lambda_{d-1} \}$, since these remain invariant under local unitary operations that cannot generate or destroy entanglement. In the case of $N \ge 3$ subsystems, however, there is no constructive analogy to the Schmidt decomposition, which renders the classification and characterisation of multipartite entangled states a demanding field with numerous open questions,  despite recent advances  \cite{PhysRevA.74.052303,PhysRevA.83.022328,Bastin:2009ye,PhysRevLett.104.020504,Bin-Liu1:2011fk,salwey,PhysRevLett.106.180502}. For mixed states, the situation is aggravated further, and the very question whether a multipartite state is entangled or not becomes a challenge, which requires intricate strategies \cite{Huber:2010kx,PhysRevLett.108.020502,PhysRevLett.110.150402}.

In the context of entanglement manipulation with identical particles, the \emph{generalised Schmidt rank} \cite{Eisert:2001uq} is a beneficial entanglement quantifier. It is assigned to a pure state $\ket{\Psi}$ via its representation as a sum of separable states, 
\eq 
\ket{\Psi} =\sum_{r=1}^R  \lambda_r \ket{\Phi^{\text{sep}}_r} . \label{SchmidtRank}
\en
The Schmidt rank of $\ket{\Psi}$ is the smallest integer $R$ for which such representation exists; for bipartite states,  $R$ is the number of nonvanishing Schmidt coefficients in the representation (\ref{Schmidt}).  Given a multipartite state, it is often rather difficult to compute the Schmidt rank; in the following, however,  bounds on the Schmidt rank for multipartite entangled states generated by propagation and detection will be formulated, following \cite{tichy:2013PRA}.

\subsection{Framework for many-particle entanglement generation}
Experiments for the generation of multipartite entanglement with identical particles are based on linear setups of aligned beam splitters, phase-shifters and elements that manipulate the internal degree of freedom of the particles. The linear propagation of $N$ identical particles that carry a  $d$-dimensional internal degree of freedom (a qudit) can be accommodated formally in a scattering setup characterised by an $n=N d$-dimensional matrix $W$ that contains all amplitudes between input and output states. The framework \cite{tichy:2013PRA} described in the following allows us to describe a variety of experiments and proposals, an exhaustive overview over the state-of-the-art of such methods can be found in Ref.~\cite{RevModPhys.84.777}. 

As in Section \ref{correpropdet}, the creation operator for a particle in the $k$th spatial mode and in the $l$th internal state is denoted by $\hat a^\dagger_{k,l}$. Elements that connect different spatial modes without affecting the internal degree of freedom (i.e.~\emph{non-polarising} components) will connect modes via their first index. Local operations that act only on the internal degree of freedom of a particle in one mode will affect  the second index. Such elements can be combined, i.e.~the internal state of a particle may be changed \emph{and} the particle may be redirected into another spatial mode conditioned on the internal state. Therefore, for convenience of notation,  the doubly indexed modes $a_{k,j}$ are identified with single-indexed modes 
 \eq 
\hat a^\dagger_{d(k-1)+j} := \hat a^\dagger_{k,j} , \label{identific}
 \en
 i.e.~the mode $a_{m}$ with $0 \le m \le dN-1$ is associated with the qudit in the external state $k=\lfloor m/d \rfloor$ and in the internal state $m~\text{mod}~d$. 

For example, for $d=2$, the qubit case is recovered, which can be realised via photons that carry a qubit encoded in their polarisation. A non-polarising beam splitter is then described by the scattering matrix 
\eq 
U_{\text{BS}} = \frac 1 { \sqrt{2} } \left(\begin{array}{cccc} 
 i & 0 & 1 & 0  \\
 0 & i & 0 & 1 \\
 1 & 0 & i & 0 \\
 0 & 1 & 0 & i 
 \end{array} \right)  ,
\en 
which precisely induces the time-evolution (\ref{singledyn}), independently of the internal state of the particle. 

Another essential element in photonic experiments is a  polarising beam splitter. This element transmits all horizontally polarised photons (identified with $\ket{0}$) but it reflects all vertically polarised photons ($\ket{1}$), as  described by 
\eq 
U_{\text{PBS}} = \left(\begin{array}{cccc} 
 0 & 0 & 1 & 0 \\
 0 & 1 & 0 &0 \\
 1 & 0 & 0 & 0 \\
 0 & 0 & 0 &1 
 \end{array} \right)  .
\en 
Half-wave-plates in the first and second spatial modes act as beam splitters in polarisation space, they are described by 
\eq 
U_{\text{HWP}} = \left(\begin{array}{cccc} 
  \cos 2 \theta_1  & i \sin 2 \theta_1 & 0 & 0 \\
 i \sin 2 \theta_1 &  \cos 2 \theta_1 & 0 &0 \\
 0 & 0 &   \cos 2 \theta_2 & i\sin 2 \theta_2 \\
 0 & 0 & i \sin 2 \theta_2 &  \cos 2 \theta_2
 \end{array} \right)  ,
\en 
where $\theta_j$ is the respective orientation of the plate in mode $j$. Via these elements, every thinkable transformation on incoming modes can be realised \cite{Reck:1994zp}. 

Closely following the formalism used in Refs.~\cite{TichyDiss,tichy:2013PRA}, we treat the following question: Given a desired $N$-qudit-state defined by the $d^N$ entries of a tensor $c_{j_1, \dots , j_N}$, 
\eq \ket{\Phi}& =&\sum_{j_1=0}^{d-1} \dots \sum_{j_N=0}^{d-1} c_{j_1,\dots, j_N} \ket{j_1, \dots , j_N}  ,  \label{initstates} \en
and $N$ indistinguishable bosons or fermions prepared in a certain initial state, is there a linear setup $W$ that generates $\ket{\Phi}$? 

\begin{figure}[ht] \center
\includegraphics[width=.67\linewidth,angle=0]{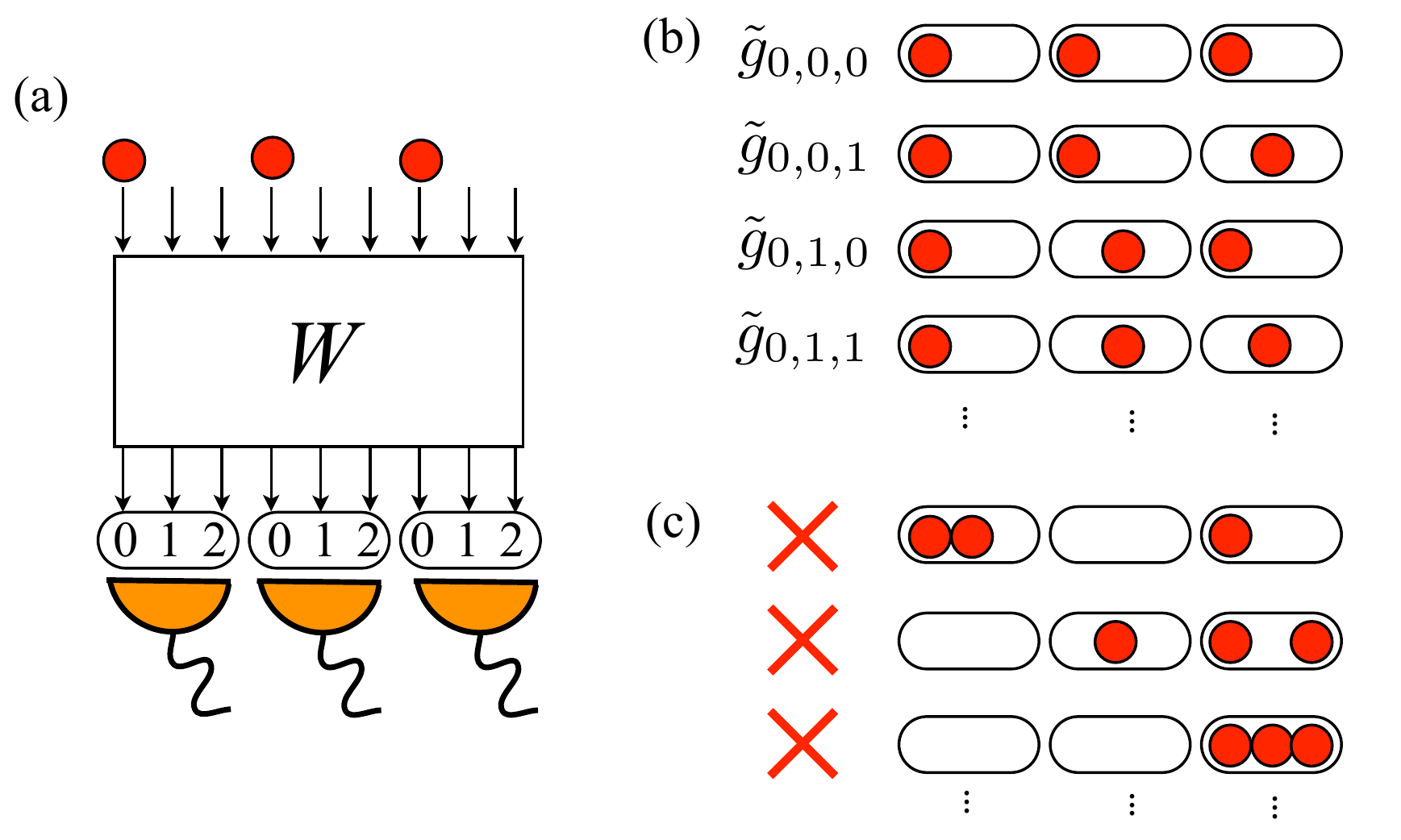}
\caption{Multipartite entanglement generation. (a) Interference setup for three three-dimensional qudits (here: the internal degree of freedom is a qutrit, $d=3$, and there are three particles, $N=3$, which makes a total of $n=d N=9$ states). The particles are initially prepared in the state $\hat a^\dagger_{1,0}a^\dagger_{2,0}a^\dagger_{3,0} \ket{\text{vac}} \equiv \hat a^\dagger_{0}\hat a^\dagger_{3}\hat a^\dagger_{6} \ket{\text{vac}}.$ (b) Components of the post-selected wavefunction that contain one particle in each external mode. These components contribute to the state measured by the observing parties. (c) Components that are not eigenvectors of the post-selection operator are neglected by the post-selection procedure. Figure inspired from \cite{tichy:2013PRA}.}  \label{SuperpositionsMultipent.pdf}
 \end{figure}

To answer this question, an explicit representation of the state that is generated upon the scattering of $N$ particles on such general multi-mode setup is required. Here, we assume that we are given the separable initial state 
\eq
\ket{\chi_{\text{ini}}^{\text{multi}}} = \prod_{j=1}^N \hat a^\dagger_{j,0} \ket{\text{vac}} \equiv  \prod_{j=1}^N \hat a^\dagger_{d(j-1)} \ket{\text{vac}} ,  \label{initmultipaent}
 \en
 i.e.~each qudit $j=1\dots N$ is initialised in the first qudit state, $\ket{0}$, as illustrated in Fig.~\ref{SuperpositionsMultipent.pdf}. The qudit \emph{label} then corresponds to an \emph{external} degree of freedom, e.g.~the spatial mode the particle is found in. A general multimode-setup induces a transformation as given by Eq.~(\ref{timeevocreation}), 
\eq 
\hat a^\dagger_{m} \rightarrow \sum_{l=0}^{N d -1}  W_{m,l} \hat b^\dagger_{l} ,
\en
i.e.~a particle initially prepared in mode $m$ exits the setup in a superposition of all modes $0\le l \le Nd-1$, with amplitude $W_{m,l}$, and the identification of external and internal states for the single-indexed $b_k$ is done in analogy to Eq.~(\ref{identific}). The matrix $W$ can take any form, i.e.~manipulations of the internal degrees of freedom are allowed and it is not restricted to be unitary: Any non-unitary matrix can be embedded in a (larger) unitary matrix, which physically corresponds to incorporating particle loss \cite{He:2007tg,Bernstein:2009uq}.  

As for the two-particle case treated in Section \ref{bipartiteentsec},  post-selection is applied. Thus, we can assume to find exactly one particle in each external output mode. Formally speaking, the state is projected onto the subspace with one particle in each group of $d$ modes $b_j$, ${(0,\dots, d-1)}, {(d, \dots, 2d-1)}, \dots, {((N-1)d, \dots, Nd-1)}$. Particles measured at the output modes can be identified with the qudits under control by the respective party, as sketched in Fig.~\ref{SuperpositionsMultipent.pdf}. 
 The quantum state that these parties control at the output modes then reads \cite{tichy:2013PRA}
\eq \ket{\Phi}  = \sum_{j_1=0}^{d-1} \dots \sum_{j_N=0}^{d-1} \tilde g_{j_1, \dots ,j_N} \ket{j_1, j_2 ,  \dots j_N} , \label{gep1}
 \en
 where the amplitude $\tilde g_{j_1, \dots ,j_N}$ for each component of the wavefunction (\ref{gep1}) is given by the total scattering amplitude for bosons (\ref{generalampliB}) or fermions (\ref{generalampliF}). For the initial state (\ref{initmultipaent}), we have 
\eq 
 \label{gexpress} 
  \tilde g_{j_1, \dots ,j_N}    =  \sum_{\sigma \in S_N} \text{sgn}_{\text{B/F}}(\sigma)  \prod_{k=1}^N W_{d (\sigma[k]-1),d(k-1)+j_k}  ,
\en
where $\text{sgn}_{\text{B}}(\sigma)=1$ and $\text{sgn}_{\text{F}}(\sigma) = \text{sgn}(\sigma)$. The post-selection procedure is probabilistic (not always does one successfully find exactly one particle in each spatial output mode, but also unwanted events sketched in Fig.~\ref{SuperpositionsMultipent.pdf}(c)), hence $\tilde g$ is typically sub-normalised. 

Given a separable state (\ref{initmultipaent}) of bosons or fermions,  a desired entangled state specified by a tensor $c_{j_1, \dots, j_N}$ can be created if the equations 
\eq 
\tilde g_{j_1, \dots , j_N} = \eta c_{j_1, \dots, j_N} , \label{equationsfmpps}
\en
possess a solution for the matrix $W$ with $\eta>0$. The success probability for state generation is then $|\eta|^2$. 

The difficulty to find a scattering matrix $W$ that solves the set of $d^N$ equations (\ref{equationsfmpps}) lies in the non-linear dependence of the $\tilde g_{j_1, \dots, j_N}$ on the matrix elements $W_{k,l}$. In practice, one typically approaches the creation of entangled states by heuristic methods, and  severely restricts the form that $W$ may take. Also the initial state  might be distinct from (\ref{initmultipaent}), e.g.~it might  be already entangled in the first place. If this is taken into account, every linear-optics experiment in which entanglement between photons is created by propagation and detection can be described in the current framework. 

Using a rough estimate of the number of physical parameters, we can quickly see that not every thinkable state can be realised by propagation and detection: The total number of controllable parameters  is bound by $dN^2$, which is the number of relevant entries in the scattering matrix $W$. This polynomial growth in $N$ is dominated by the exponential increase of the size of the Hilbert space, $d^N$. Hence, there are quantum states that cannot be generated without interaction,  given the separable initial state (\ref{initmultipaent}).

\subsection{Combinatorial bound} \label{combibounds}
A first explicit restriction on states of the form (\ref{gexpress}) can be understood from a combinatorial argument: When $N$ distinguishable classical objects are distributed among $N$ observers, there are precisely $N!$ observable combinations. If  there are indistinguishable replica among the objects, the effective number of combinations is reduced accordingly. In particular, for $N$ identical objects, only one collective outcome is possible.  

Keeping this classical analogy in mind, the final state $\ket{\Phi}$ in (\ref{gep1}) is re-written as  
\eq 
\ket{\Phi}=\sum_{j_1=0}^{d-1} \dots \sum_{j_N=0}^{d-1}  \sum_{\sigma \in S_N} \text{sgn}_{\text{B/F}}(\sigma)  \left( \prod_{k=1}^N W_{d(\sigma[k] -1), d(k-1) +j_k} \right) \ket{j_1, \dots , j_N}  , \label{initstatecomb}
\en
where the definition of $\tilde g_{j_1, \dots, j_N}$ [Eq.~(\ref{gexpress})] was inserted. This last expression equates 
\eq
\ket{\Phi}= \sum_{\sigma \in S_N} \text{sgn}_{\text{B/F}}(\sigma)  \left[ \otimes_{m=1}^N  \left( \sum_{j_m=0}^{d-1}  W_{d(\sigma[m] -1), d(m-1) +j_m} \ket{j_m}_m \right) \right] , \label{combreprese}
\en
where $\ket{j_m}_m$ is the state of the qudit in the $m$th external mode prepared in the $j_m$th internal state. This representation emphasises that there are at most $N!$ separable terms in (\ref{combreprese}). Therefore, 
the Schmidt rank $R$ [see Eq.~(\ref{SchmidtRank})] of any state $\ket{\Phi}$  is bound by $N!$. 

When two bosons initially populate the very same mode $a_j$, the number of distinct permutations in (\ref{combreprese}) is reduced: Formally, we can use the same initial state (\ref{initmultipaent}), but the two pertinent rows of the matrix $W$ need to be set equal. Then, transpositions acting on these two modes leave the resulting summand in (\ref{combreprese}) invariant. Therefore, given the initial mode occupation list $\vec r$, the total number of distinct terms in (\ref{combreprese})  is $N!/\prod_{j=1}^n r_j!$. In particular, a Bose-Einstein condensate [$\vec r=(N,0, \dots, 0)$] cannot be used as a source to generate entanglement via propagation and detection \cite{Sasaki:2011uq}. 

For $d=2$, the bound is not tight, since one can write any entangled $N$-qubit state as a sum of at most $2^{N-1}$ separable terms, and $2^{N-1} \le N!$. In other words, the bound does not imply any restriction on qubit-like entanglement manipulation, e.g.~in the photon polarisation.  For $d>2$, however, the bound severely restricts the set of accessible states. Such higher-dimensional degrees of freedom can be realised by the spatial path \cite{PhysRevLett.94.220501,PhysRevLett.61.2921}, the time of arrival \cite{PhysRevLett.82.2594,Marcikic:2004qr,Halder:2007th,Kwon:2013qf}, or the orbital angular momentum \cite{Vaziri:2002nx,Inoue:2009dq,Mair:2001zr,PhysRevA.83.012306} of photons. A means to circumvent the combinatorial bounds is, e.g., to realise a single qudit  by several photons \cite{PhysRevA.78.042321,Baek:2007mi}. 

\subsection{Bosonic and fermionic bounds} \label{Bosonfermionsent}
The combinatorial bound can be understood from a rather intuitive, combinatorial perspective, and it applies to bosons in the very same way as to fermions. When the expression (\ref{gexpress}) is  analysed in more detail, however, a relic of the complexity gap between  bosons and fermions discussed in Section \ref{compcomplDet} is recovered. 

To aid notation, the matrix $\tilde W$ is defined as the $N\times dN$-submatrix of $W$ that contains the rows corresponding to the initially occupied input modes, i.e.~the rows labelled with $j=0,d,2d,3d, \dots, (N-1)d$. 
Closely following the argument in Section \ref{compcomplDet}, the coefficient related to the state (\ref{gep1}) is re-written as the permanent or the determinant of a sub-matrix $V_{(j_1, \dots, j_N)}$ of $\tilde W$, 
\eq 
 \tilde g_{j_1, \dots ,j_N}^{(\text{F})}   & =&  \text{det}\left( V_{(j_1, \dots, j_N)}  \right) ,   \label{fermmulti} \\
 \tilde g_{j_1, \dots ,j_N}^{(\text{B})}   & =& \text{perm}\left(     V_{(j_1, \dots, j_N)}    \right) ,
\en
where $ V_{(j_1, \dots, j_N)}$ is the $N \times N$-matrix that contains the columns $j_1, d+j_2, 2d+j_3, \dots $ of $\tilde W$. Remember that the columns are assigned to the output modes, i.e.~each possible outcome $(j_1, \dots, j_N)$ is assigned one matrix $V_{(j_1, \dots, j_N)} $. 

The symmetries of the fermionic determinant that appears in (\ref{fermmulti}) will allow us to constrain the number of parameters required to describe states generated by fermions. These symmetries are absent for bosons, which leads to a larger  number of  controllable parameters. 

\subsubsection{Reduction of parameters of the fermionic determinant}
We restrict ourselves to the treatment of qubits ($d=2$), the argument can be extended to higher-dimensional systems in close analogy \cite{tichy:2013PRA}. A matrix is defined for the even column numbers, $E=V_{(0,0,\dots, 0)}$ and one for the odd columns, $F=V_{(1,1,\dots, 1)}$. Without restrictions to generality, we assume that $\tilde g^{(\text{F})}_{0,\dots, 0}= \text{det} \left( E \right)\neq 0$, i.e.~$E$ is not singular. Thus, the column vectors of $E$ are linearly independent, and constitute a basis. Each column vector of $F$ can therefore be written as a linear combination of column vectors of $E$, 
\eq 
\vec F_{j} = \sum_{k=1}^N C_{j,k} \vec E_{k} , \label{EFdef}
\en
where  $C$ is an $N\times N$-matrix. Similarly, each column vector of every $V_{(j_1, \dots, j_N)}$ can be written as a superposition of column vectors of $E$. 

Using the rule that the determinant of a matrix remains unchanged when a multiple of a column vector is added to another column, we can write the coefficient $\tilde g^{(\text{F})}_{j_1, \dots, j_N}$ as the product of two determinants, 
\eq 
  \tilde g^{(\text{F})}_{j_1, \dots ,j_N} &=& \text{det} \left( E \right) \text{det}\left( \bar C(j_1, \dots, j_N) \right) = \tilde g^{(\text{F})}_{0, \dots , 0} \text{det}\left( \bar C(j_1, \dots, j_N) \right) , \label{determinantproduct}
\en
where $\bar C(j_1, \dots, j_N)$ is the matrix constructed from $C$ as follows: For each possible index value $k$ $(1 \le k \le N)$, delete the $k$th row and the $k$th column of $C$ if $j_k$ is even. The resulting matrix $\bar C(j_1, \dots , j_N)$ is an $m\times m$-submatrix of $C$, where $m$ is the number of even coefficients $j_k$. For convenience, we set the determinant of the empty matrix $\bar C(0,\dots 0)$ to unity. 

The determinant of $\bar C$ appearing in (\ref{determinantproduct}) is called a \emph{principle minor}  \cite{Bernstein:2009uq}. There are $2^N$ principle minors, i.e.~one for every possible coaxial submatrix of $C$, including the empty matrix associated to $\tilde g_{0,\dots,0}$ and the full matrix $C$ related to $\tilde g_{1,\dots,1}$. 

At this stage, we have re-written the coefficient $\tilde g_{j_1,\dots,j_N}$ as a product of $\tilde g_{0,0,\dots, 0}^{(\text{F})}$ and a function of the matrix $C$ that depends on the $N^2$ matrix elements. Thus, there are at most $N^2+1$ parameters in total (remember that we initially started with the $2N^2$ matrix elements of $\tilde W$). The $2^N$ principle minors of $C$, however, do not depend on all $N^2$ matrix elements of $C$, but they can be expressed by exactly $N(N-1)+1$ independent variables \cite{MacMahon:1893ff,Stouffer:1924lh,Muir:1894fu}. The total number of independent parameters for fermions therefore amounts to  \cite{tichy:2013PRA}
\eq
\text{dim}_{\text{F}} = N(N-1)+2 , \label{fermiondim}
\en
where one of these parameters only contributes to the normalisation and the global phase of the emerging state. 

\subsubsection{Reduction of parameters of the bosonic permanent}
For bosons, the argument in the last section breaks down, since the product rule for determinants has no analogue for permanents; in general,  $\text{perm}(A \cdot B) \neq \text{perm}(A) \text{perm}(B)$. 

The total number of independent parameters can still be reduced by identifying those variables that only contribute to the total normalisation and phase of the emerging state \cite{tichy:2013PRA}. Each row and each group of two columns in $\tilde W$ can be rescaled with a complex factor, which  contributes a global factor to all amplitudes. Out of these $2N-1$ parameters,  $2N-2$ can be eliminated, such that the total number of effective parameters is bounded by  \cite{tichy:2013PRA}
\eq 
\text{dim}_{\text{B}} =2N (N -1)+2  \ge \text{dim}_{\text{F}} . \label{bosondim}
\en
Strictly speaking, the values given in (\ref{fermiondim}) and (\ref{bosondim}) constitute only upper bounds on the total number of physical parameters. Numerical studies on random states show that for small systems ($N\le 8$), the dimensionality of the manifold of states indeed coincides with the upper bounds \cite{tichy:2013PRA}. 
 
In other words, experiments that generate photonic  entanglement use the bosonic superiority of light quanta over fermions, which reflects the superior complexity of  Boson-Sampling in comparison to the simulation of fermions [see Section \ref{compcomplDet}].  Note that the smaller dimension of the manifold of states generated by fermions is not a simple consequence of the Pauli-principle that exclude certain states, since we only deal with Pauli arrangements in the first place. It is not immediate, however, which are the precise non-local characteristics that states generated by bosons possess in contrast to those created with fermions. In particular, no entanglement measure is known which clearly demarcates the two sets of states.  

Partial distinguishability in uncontrollable degrees of freedom deteriorates the quantum character of multipartite correlations, just like for two particles (Section \ref{partialdistentangl}). When $k$ particles become fully distinguishable, they cannot contribute to quantum correlations, and the resulting state becomes $(k+1)$-separable, i.e.~the  density matrix that describes the measurement outcomes at the level of the detectors can be expressed as a mixture of $(k+1)$-separable pure states that do not carry genuine $N$-body entanglement \cite{TichyDiss}. How entanglement measures for multipartite states \cite{doi:10.1080/09500340210121589}  behave quantitatively as a function of partial distinguishability -- e.g.,~as defined in Section \ref{DisttransSec} -- constitutes an open question, and a relationship analogous to Eq.~(\ref{concurrencedist}) remains a desideratum. 

\FloatBarrier

\section{Conclusions and outlook} \label{conclusionsandoutlook}
Throughout this tutorial, the coherent evolution of many non-interacting identical particles was found to be virtually subjugated by collective interference. Our physical leitmotiv, the qualitatively \emph{different} behaviour of \emph{many} particles in comparison to two, appears in many different contexts. The coherent conspiration of many-particle paths  leads to the independence of bosonic and fermionic interference (Section \ref{InterferenceSec}), to the superiority of bosons over fermions (Sections \ref{secBosonSampling} and \ref{Bosonfermionsent}) and to the non-monotonicity of the quantum-to-classical transition of many particles (Section \ref{DisttransSec}). The consequences of coherent many-body effects thus reach far beyond statistical bosonic (fermionic) (anti)bunching familiar from thermal, incoherent environments. The latter do not require any multi-particle coherence, which is why they became noticeable experimentally much earlier \cite{Hanbury-Brown:1956vn} than genuinely quantum coherent effects that  elude classical descriptions. The complexity that we encountered upon scaling up the system is not only borne by the large number of particles, but also by the coherence that inhibits the application of statistical methods for many particles, which implies a forceful distinction between coherent many-body effects and statistical features \cite{Tichy:2012NJP}. 

Established experimental techniques for photons \cite{Aspuru-Guzik:2012qm,RevModPhys.84.777}  and emerging technologies for ultracold atoms that could extend photonic experiments in the near future \cite{Kierig:2008fk,Hume:2013cl,Sherson:2010fk,Weitenberg:2011vn,Cheinet:2008ly,Bakr:2010fk} feed the hope that many-particle interference might become a workhorse for a variety of applications, since this phenomenon lies at the roots of quantum metrology \cite{Dowling:2008hc},  simulation \cite{Aspuru-Guzik:2012qm} and computation \cite{RevModPhys.79.135}. 

In particular, the domain of Boson-Sampling  is progressing at a remarkable pace \cite{Spagnolo:2013eu,Carolan:2013mj}. Scaling to larger particle numbers, however, entails  serious technological challenges, which might be tackled by slightly simplifying the formulation of the problem while keeping its formidable hardness \cite{Lund:2013eb}. On the conceptual side, a rigorous certification  comes with difficulties without precedent. 
The \emph{generation} of random events that are compatible with any a priori known property such as bosonic clouding \cite{Carolan:2013mj} or the suppression law (\ref{law}), and the \emph{verification} whether a set of events satisfies the test criterion can both be done efficiently. Hence, a malicious adversary who pretends to possess a Boson-Sampling machine can easily generate a distribution that passes the test \cite{Gogolin:2013eu}. 
A remedy to this caveat is a ``one-way'' function by which a scattering matrix is implanted a ``hidden'' structure. For this matrix, it would remain classically hard to \emph{predict} which events are suppressed (i.e.~the actual test-criterion remains unknown to the experimenter even if the matrix itself is known), whereas the assessor who generates the matrix can do such prediction efficiently and certify the output of the Boson-Sampling machine. Such scenario of asymmetric complexity (the simulation is hard, the verification is easy, just like for factoring and multiplication) is not ruled out, but believed to be unlikely \cite{Aaronson:2011kx,Aaronson:2013dq}. If we set aside the skeptic scenario (which becomes relevant only if ever any application of Boson-Sampling in cryptography is conceived), the suppression law provides a feasible means to generate reasonable evidence that a many-particle Boson-Sampling machine works correctly: The criterion is virtually unforgivable with respect to partial distinguishability or deviations from the ideal unitary matrix, and offers a strict \emph{granular} criterion which is violated by classically or semi-classically behaving particles  \cite{Tichy:2013lq}. First certification experiments have already been reported \cite{Spagnolo:2013fk,Spagnolo:2013eu,Carolan:2013mj}, the suppression law for many particles may provide a more stringent benchmark in the future \cite{Tichy:2013lq}. A functional certified Boson-Sampling machine might then constitute the first quantum device that effectively outperforms classical computers.

Multipartite entanglement generation via propagation and detection can be regarded as Boson-Sampling with post-selection. The superior complexity of bosons translates here into a larger manifold of generable states as compared to fermions, while it is widely open to date whether every state that can be generated by fermions can also be created by  bosons. Although the manipulation of entanglement with non-interacting fermions is currently not being pursued experimentally, further studies of these questions  may contribute to the general classification of multipartite entangled states and enable a fruitful cross-fertilisation between the two exciting fields of multipartite entanglement and many-particle interference. 

While unwanted partial distinguishability is the main obstacle to the generation of entanglement and to the observation of  many-particle interference as in the context of Boson-Sampling, the functional dependence of  experimental signals on the indistinguishability \cite{Ra:2013kx,younsikraNatComm} could be used to characterise different types of decoherence processes. For example, processes that keep the interfering particles indistinguishable but which scramble the acquired phases will lessen any coherent many-particle interference, while the typical statistical behaviour with average bosonic and fermionic signatures will remain. A process that jeopardises the mutual indistinguishability of the involved particles, however, will also wipe out such statistical effects. On the other hand, first steps have been performed to study the interplay of disorder acting on single particles  and many-particle interference effects, such as Anderson localisation \cite{Segev:2013kc} of many particles  \cite{PhysRevA.86.040302,PhysRevLett.105.163905,Crespi:2013hs}, which opens the door to study disorder in the many-particle space.

The symmetrisation postulate for identical particles and quantum entanglement make, in principle, comparable fundamental statements on physical reality \cite{TichyDiss}: It is impossible to assign a particle label to one photon in a system of two  indistinguishable light quanta; likewise, no spin-state can be assigned to an electron that is entangled to a photon prior to the measurement. Both concepts stem from the early days of quantum physics, their appreciation, however, could not have evolved more differently: The undisputed acceptance of the symmetrisation postulate is borne, first of all, by its  impressive success, which today encompasses tests on systems ranging from hadrons at the highest artificially maintainable energies to molecules at the lowest ever measured temperatures. Its fast reception was  supported further by its very satisfactory conceptual appeal: It translates a  \emph{physical} postulate (no measurement can distinguish any two particles of the same kind, all intrinsic physically measurable properties of each particle of a species are identical) into a formal mathematical framework (the permutation symmetry of all admissible many-particle operators and the (anti-)symmetry of fermionic and bosonic wavefunctions). The roles are typically reversed when it comes to a discussion of entanglement: One then starts with the linear structure of the Hilbert space that  \emph{formally} permits to write down a state that describes a coherent superposition of \emph{physically distinguishable} many-particle states, and one is then left to find a meaningful interpretation of such formal construct. The challenge to distill the physical principles that underlie the observed measurement outcomes is ongoing \cite{dowereally2012}, but it has been greatly alleviated by the possibility to physically generate entangled states in the laboratory. Almost a century after the first attempts to describe the quantum mechanics of many particles, the intrinsic connection between indistinguishability and entanglement has now become apparent and tangible experimentally. In the future, either phenomenon may  be understood best from a perspective that encloses both.

\section*{Acknowledgements}
I would like to thank Klaus Mayer, Klaus M{\o}lmer and David Petrosyan for carefully reading the manuscript and for providing very valuable feedback, Andreas Buchleitner for continuous guidance and stimulating debates, Fernando de Melo, Florian Mintert and Markus Tiersch for illuminating discussions, and Yoon-Ho Kim, Hyang-Tag Lim and Young-Sik Ra for insight into photonic experiments.   Financial support by the German Academic Merit Foundation and by the Alexander von Humboldt-Foundation through a Feodor Lynen Fellowship are gratefully acknowledged. 


\providecommand{\newblock}{}

\end{document}